\documentclass{aastex631}
\usepackage{graphicx}
\usepackage{natbib}
\usepackage{babel}
\usepackage{float}
\usepackage[utf8]{inputenc}

\usepackage[toc,page]{appendix}

\usepackage[caption=false]{subfig}

\usepackage{txfonts}
\usepackage{xspace,color}
\usepackage{soul,ulem}

\defcitealias{2018ApJ...863..134H}{H18}
\defcitealias{2023ApJ...951..138L}{L23}

\shorttitle{Searching for metal-poor dwarf galaxies in the S-PLUS survey}
\shortauthors{M. Grossi, D. R. Gonçalves  et al.}

\begin{document}

\title{The Southern Photometrical Local Universe Survey (S-PLUS): searching for metal-poor dwarf galaxies}

\correspondingauthor{M. Grossi}
\email{grossi@astro.ufrj.br}

\author[0000-0003-4675-3246]{M. Grossi}
\affiliation{Observat\'orio do Valongo, Universidade Federal do Rio de Janeiro, Ladeira Pedro Ant\^onio 43, Sa\'ude, Rio de Janeiro, RJ, 20080-090, Brazil}

\author{D. R. Gon\c{c}alves}
\affiliation{Observat\'orio do Valongo, Universidade Federal do Rio de Janeiro, Ladeira Pedro Ant\^onio 43, Sa\'ude, Rio de Janeiro, RJ, 20080-090, Brazil}

\author{A. C. Krabbe}
\affiliation{Departamento de Astronomia, Instituto de Astronomia, Geof\'isica e Ci\^encias Atmosf\'ericas da USP, Cidade Universit\'aria, 05508-090 São Paulo, SP, Brazil}

\author{L. A. Guti\'errez Soto }
\affiliation{Instituto de Astrofísica de La Plata, UNLP-CONICET, Paseo del Bosque s/n, B1900FWA, LA PLATA, Argentina}

\author{E. Telles}
\affiliation{Observat\'orio Nacional, Rua General Jos\'e Cristino, 77, S\~ao Crist\'ov\~ao, Rio de Janeiro, 20921-400, Brazil}

\author{L. S. Ribeiro}
\affiliation{Observat\'orio do Valongo, Universidade Federal do Rio de Janeiro, Ladeira Pedro Ant\^onio 43, Sa\'ude, Rio de Janeiro, RJ, 20080-090, Brazil}

\author{T. S. Gon\c{c}alves}
\affiliation{Observat\'orio do Valongo, Universidade Federal do Rio de Janeiro, Ladeira Pedro Ant\^onio 43, Sa\'ude, Rio de Janeiro, RJ, 20080-090, Brazil}


\author{A. E. de Araujo-Carvalho}
\affiliation{Observat\'orio do Valongo, Universidade Federal do Rio de Janeiro, Ladeira Pedro Ant\^onio 43, Sa\'ude, Rio de Janeiro, RJ, 20080-090, Brazil}

\author{A. R. Lopes}
\affiliation{Instituto de Astrofísica de La Plata, UNLP-CONICET, Paseo del Bosque s/n, B1900FWA, LA PLATA, Argentina}
\affiliation{Departamento de Astronomia, Instituto de Astronomia, Geof\'isica e Ci\^encias Atmosf\'ericas da USP, Cidade Universit\'aria, 05508-090 São Paulo, SP, Brazil}

\author{A. V. Smith Castelli}
\affiliation{Instituto de Astrofísica de La Plata, UNLP-CONICET, Paseo del Bosque s/n, B1900FWA, LA PLATA, Argentina}
\affiliation{Instituto de Astrofísica de La Plata, UNLP-CONICET, Paseo del Bosque s/n, B1900FWA, LA PLATA, Argentina}

\author{M. E. De Rossi}
\affiliation{Universidad de Buenos Aires, Facultad de Ciencias Exactas y Naturales y Ciclo B\'asico Com\'un. Buenos Aires, Argentina}
\affiliation{CONICET-Universidad de Buenos Aires, Instituto de Astronom\'ia y F\'isica del Espacio (IAFE). Buenos Aires, Argentina}

\author{C. Lima-Dias}
\affiliation{Departamento de Astronom\'ia, Universidad de La Serena, Avda. Ra\'ul Bitr\'an 1305, La Serena, Chile}


\author[0000-0002-9269-8287]{Guilherme~Limberg}
\affiliation{Kavli Institute for Cosmological Physics, University of Chicago, 5640 S. Ellis Avenue, Chicago, IL 60637, USA}

\author{C. E. Ferreira Lopes}
\affiliation{Instituto de Astronom\'{i}a y Ciencias Planetarias, Universidad de Atacama, Copayapu 485, Copiap\'{o}, Chile}
\affiliation{Millennium Institute of Astrophysics, Nuncio Monse\~{n}or Sotero Sanz 100, Of. 104, Providencia, Santiago, Chile}

\author[0000-0002-2925-1861]{J. A. Hernandez-Jimenez}
\affiliation{Universidad de Investigaci\'on y Desarrollo, Departamento de Ciencias B\'asicas y Humanas, Grupo FIELDS, Calle 9 No. 23-55, Bucaramanga, Colombia}

\author{P. K. Humire}
\affiliation{Departamento de Astronomia, Instituto de Astronomia, Geof\'isica e Ci\^encias Atmosf\'ericas da USP, Cidade Universit\'aria, 05508-090 São Paulo, SP, Brazil}

\author{A. L. Chies-Santos}
\affiliation{Instituto de F\'isica, Universidade Federal do Rio Grande do Sul, Av. Bento Gon\c{c}alves 9500, Porto Alegre, RS, 90040-060, Brazil}

\author{L. Lomel\'i-N\'u\~nez }
\affiliation{Observat\'orio do Valongo, Universidade Federal do Rio de Janeiro, Ladeira Pedro Ant\^onio 43, Sa\'ude, Rio de Janeiro, RJ, 20080-090, Brazil}

\author{S. Torres-Flores }
\affiliation{Departamento de Astronom\'ia, Universidad de La Serena, Avda. Ra\'ul Bitr\'an 1305, La Serena, Chile}


\author{F. R. Herpich}
\affiliation{Laborat\'orio Nacional de Astrof\'isica (LNA/MCTI), Rua Estados Unidos, 154, Itajub\'a 37504-364, Brazil}

\author{G. B. Oliveira Schwarz}
\affiliation{Departamento de Astronomia, Instituto de Astronomia, Geof\'isica e Ci\^encias Atmosf\'ericas da USP, Cidade Universit\'aria, 05508-900 São Paulo, SP, Brazil}
\affiliation{Universidade Presbiteriana Mackenzie, Rua da Consola\c{c}\~ao, 930, Consola\c{c}\~o, S\~ao Paulo, 01302-907, Brazil}

\author{A. Kanaan}
\affiliation{Departamento de F\'isica, Universidade Federal de Santa Catarina, Florian\'opolis, SC, 88040-900, Brazil}

\author{C. Mendes de Oliveira}
\affiliation{Departamento de Astronomia, Instituto de Astronomia, Geof\'isica e Ci\^encias Atmosf\'ericas da USP, Cidade Universit\'aria, 05508-900 São Paulo, SP, Brazil}

\author{T. Ribeiro}
\affiliation{Rubin Observatory Project Office, 950 N. Cherry Ave., Tucson, AZ 85719, USA}

\author{W. Schoenell}
\affiliation{Giant Magellan Telescope, 251 S. Lake Ave, Suite 300, Pasadena  91101, USA}

\begin{abstract}
The metal content of a galaxy's interstellar medium reflects the interplay between different evolutionary
processes such as feedback from massive stars and the accretion of gas from the intergalactic medium. Despite
the expected abundance of low-luminosity galaxies, the low-mass and low-metallicity regime remains
relatively understudied. Since the properties of their interstellar medium resemble those of early galaxies,
identifying such objects in the Local Universe is crucial to
understand the early stages of galaxy evolution.
We used the DR3 catalog of the Southern Photometric Local Universe Survey (S-PLUS) to select low-metallicity dwarf galaxy candidates based on color selection criteria typical of metal-poor, star-forming, low-mass systems.
The final sample contains approximately 50 candidates. Spectral energy distribution  fitting of the 12 S-PLUS bands reveals that $\sim$ 60\% of the candidates are best fit by models with low stellar metallicities. We obtained long-slit observations with the Gemini Multi-Object Spectrograph to follow-up a pilot sample and confirm whether these galaxies have low metallicities.
We find oxygen abundances in the range $7.28<$ 12 + log(O/H) $< 7.82$ (4\% to 13\% of the solar value), confirming their metal-poor nature. Most targets are outliers in the mass-metallicity relation, i.e. they display a low metal content relative to their observed stellar masses. In some cases, perturbed optical morphologies might give evidence of dwarf-dwarf interactions or mergers.
These results suggest that the low oxygen abundances may be associated with an external event causing the accretion of metal-poor gas, which dilutes the oxygen abundance in these systems.
\end{abstract}


\section{Introduction}

The metallicity of the interstellar medium (ISM)   is a fundamental tracer of the evolutionary process of a galaxy  \citep{1979A&A....80..155L,2004ApJ...613..898T}. It depends on several factors, such as the star formation rate and the gas-mass fraction, but it also reflects the interplay between stellar feedback and the accretion of metal-poor gas from the intergalactic medium \citep{2013ApJ...772..119L,2014MNRAS.438..262P}, 
driven by the processes that drive and regulate the baryon cycle within galaxies \citep[see][for a review]{2019A&ARv..27....3M}.

Low-luminosity galaxies are less chemically evolved than more massive systems, likely due to a less efficient star formation activity and a higher fraction of metals lost in supernova events \citep{2009A&A...505...63G,2013ApJ...779..102K,2015ApJ...815L..17M,2019IAUS..344..161G}. However, despite the expected large population of low-luminosity galaxies, the low-mass and low-metallicity regime is still relatively little studied among the galactic populations.
In the standard cosmological scenario, dwarf galaxies play a crucial role in the hierarchical formation of structures, since the first objects to form and contribute, through mergers, to the growth of more massive systems were low-mass and metal-poor galaxies \citep{2009ApJ...693.1859B, 2011ARA&A..49..373B}.
\citet{2024ApJ...976L..15C}  have recently reported the discovery
of eight faint galaxies
in the redshift range between 6 and 8 with the  James Webb Space Telescope. These galaxies exhibit stellar masses between $10^6$ and $10^7$ M$_{\odot}$ and have oxygen abundances of   12 + log(O/H) $\sim$ 7, approximately 2\% the solar value\footnote{In this work we adopt a solar oxygen abundance of 12 + log(O/H)$_{\odot}$ = 8.69 $\pm$ 0.04 \citep{2021A&A...653A.141A}.}. 
Nearby gas-rich, low-metallicity dwarfs are expected to be analogous to such primitive galaxies both because of their low abundances of heavy elements and the presence of intense bursts of star formation \citep{2016ApJ...830...67B}. 
However, due to their low intrinsic luminosity, detailed investigations of these objects at great distances is extremely complex. 
Therefore, it is important to combine the most detailed observations of the closest analogous objects with those of younger and more distant galaxies. 

Dwarf galaxies in the Local Universe can be studied in different galaxy environments, from large under-dense regions -- the voids -- to rich clusters \citep{2019IAUS..344..319G}. Dwarf galaxies in low-density regions often show peculiar
gas morphology and kinematics, likely associated with external gas accretion, and they can experience strong bursts of star formation \citep{2012MNRAS.426.3041H}. The most metal-poor star-forming dwarfs in the Local Universe are found in such large underdense regions \citep{2016IAUS..308..390S}. 

Progress in identifying new metal-poor systems has been relatively slow.
Detecting these faint galaxies requires that they are relatively nearby, and/or that they are experiencing a recent star formation episode.
Robust abundance determinations are
challenging to derive for faint galaxies, and because these systems are harder to ﬁnd due to their low masses, the current metallicity relations are substantially underpopulated at the low-luminosity/low-mass end.
Recent wide-field surveys have detected only a small number of dwarf galaxies with metallicities below one tenth of solar (commonly referred to as metal-poor dwarfs, \citet{2000A&ARv..10....1K}) in the Local Universe
\citep[$z \lesssim 0.02$;][]{2012A&A...546A.122I,2018ApJ...863..134H,2020ApJ...891..181M,2023ApJ...951..138L}. A few extremely metal poor (XMP) objects (i.e. with 12 + log(O/H) $<$ 7.35, approximately 5\% of the solar value, \citet{2020ApJ...891..181M}) have been found at low redshifts 
\citep{2009A&A...503...61I,2016ApJ...822..108H,2018MNRAS.473.1956I,2019MNRAS.483.5491I,2020ApJ...898..142K,2024MNRAS.527.3486I}. These galaxies are characterized by having low stellar mass M$_* <  10^7$ M$_{\odot}$ and very compact structures. Discovering and studying XMP galaxies is fundamental to understand the process of star formation in a metal poor ISM 
and to provide constraints for theoretical models of
chemical evolution of low-mass galaxies.


Metal-poor galaxies can be identified using selection criteria based on either spectroscopic \citep{2012A&A...546A.122I,2017A&A...599A..65G,2022ApJ...925..131H,2024ApJ...961..173Z} or photometric diagnostics 
\citep{2015MNRAS.448.2687J,2017MNRAS.465.3977J,2018ApJ...863..134H,2020ApJ...898..142K,2023ApJ...951..138L}.
Wide-area photometric surveys can provide a more efficient tool in finding low-metallicity targets, since they allow to preselect a larger number of candidates that can be subsequently confirmed or rejected with a targeted spectroscopic campaign.
\citet{2017MNRAS.465.3977J} identified low metallicity systems in the Sloan Digital Sky Survey (SDSS) using a combination of $gri$ color cuts.
Oxygen abundances obtained via the direct method
varied within the range 7.45 $<$ 12 + log (O/H) $<$ 8.0,
and about 20\% of the subset of galaxies with spectroscopic observations  had less than 10\% of the solar value. 
\citet[][hereafter H18]{2018ApJ...863..134H} applied $ugriz$ color selection criteria typical of low-metallicity star-forming dwarf galaxies, such as IZw18 or Leo~P to galaxies selected in the SDSS.
Follow-up spectroscopic studies revealed that approximately 45\% of the galaxies selected using these color cuts had oxygen abundances below 1/10th of the solar value, and only 6\% had lower values,  $\lesssim 5$\% Z$_{\odot}$.
\citet[][hereafter L23]{2023ApJ...951..138L} performed a similar study in the Dark Energy Survey (DES) using FUV+$griz$ color cuts. The oxygen abundances of the spectroscopic subsample were found to vary between 7.4 $<$ 12+log(O/H) $<$ 8.2 with a mean of 12+log(O/H) = 7.8 ($\sim$ 0.13 Z$_{\odot}$).


\begin{table*}
\begin{center}
\caption{Observational and measured properties of the S-PLUS galaxies observed with GMOS (1-4) and with spectroscopic observations from previous studies (5-6). $g$, $r$, $i$ magnitudes extracted from the S-PLUS DR3; spectroscopic redshifts measured from the brightest emission lines in the GMOS spectra; stellar masses estimated from the $(r-i)$ colour using Eq. (1) \citep{2003ApJS..149..289B}. }
\begin{tabular}{ccccccclc}
\hline \hline
\noalign{\smallskip}
ID & S-PLUS ID & RA & DEC & $g$ & $r$ & $i$ & $\: \: \,$z$_{spec}$ &  log(M$_{*}$) \\
&          & (J2000)   & (J2000)    & [AB mag] & [AB mag] & [AB mag] &         &        [M$_{\odot}$]  \\ 
\noalign{\smallskip}
\hline \hline
\noalign{\smallskip}
1 & DR4\_3\_SPLUS-s27s07\_0011404              & 00:37:13.1 & -35:08:48 & 16.99 $\pm$ 0.02   & 16.79$\pm$0.01 & 16.79 $\pm$ 0.03 & 0.0138 & 8.50$\pm$0.11 \\ 
2 & DR4\_3\_SPLUS-n14s01\_0028263              & 10:00:27.6 & -17:28:25 &  17.84 $\pm$ 0.05  & 17.54$\pm$0.02 & 17.51 $\pm$ 0.04 & 0.0162 & 8.40$\pm$0.11 \\ 
3 & DR4\_3\_SPLUS-n13s01\_0024584              & 10:01:16.3 & -16:13:49 &  18.14 $\pm$ 0.05   & 17.76$\pm$0.03 & 17.80 $\pm$ 0.06 & 0.0075 & 7.53$\pm$0.11 \\ 
4 & $\: \:$DR4\_3\_SPLUS-n02n27\_0037849$^a$   & 12:23:23.3 & 01:48:52  &  17.18 $\pm$ 0.03   & 17.07$\pm$0.02 & 17.00 $\pm$ 0.04 & 0.0067 & 7.88$\pm$0.11 \\ 
\noalign{\smallskip}
\hline
\noalign{\smallskip}
5 & DR4\_3\_SPLUS-s29s37\_0046039              & 04:05:20.4 & -36:48:59 &   16.39 $\pm$ 0.01   &  16.29$\pm$0.01 & 16.26 $\pm$ 0.02 & 0.0027$^b$ & 7.34$\pm$0.11 \\ 
6 & $\: \:$DR4\_3\_SPLUS-n03s22\_0021537$^{c}$ & 11:57:12.0 & -02:41:12 &   17.28 $\pm$ 0.03   &  17.10$\pm$0.02 & 17.20 $\pm$ 0.04 & 0.0046$^d$ & 7.27$\pm$0.11 \\
\noalign{\smallskip}
 \hline\hline
\noalign{\smallskip}
\end{tabular}
\label{tab:properties}
\end{center}
\footnotesize{$^a$ other ID: SDSSJ122323.40+014854.1, $^b$ \citet{2009A&A...505...63G}, $^{c}$ other ID: SDSS J115712.37-024111.2, $^d$ \citet{2016ApJ...819..110S}.}
\end{table*}

\begin{figure}
\centering
\includegraphics[bb=12 5 600 380, scale=0.65, clip]{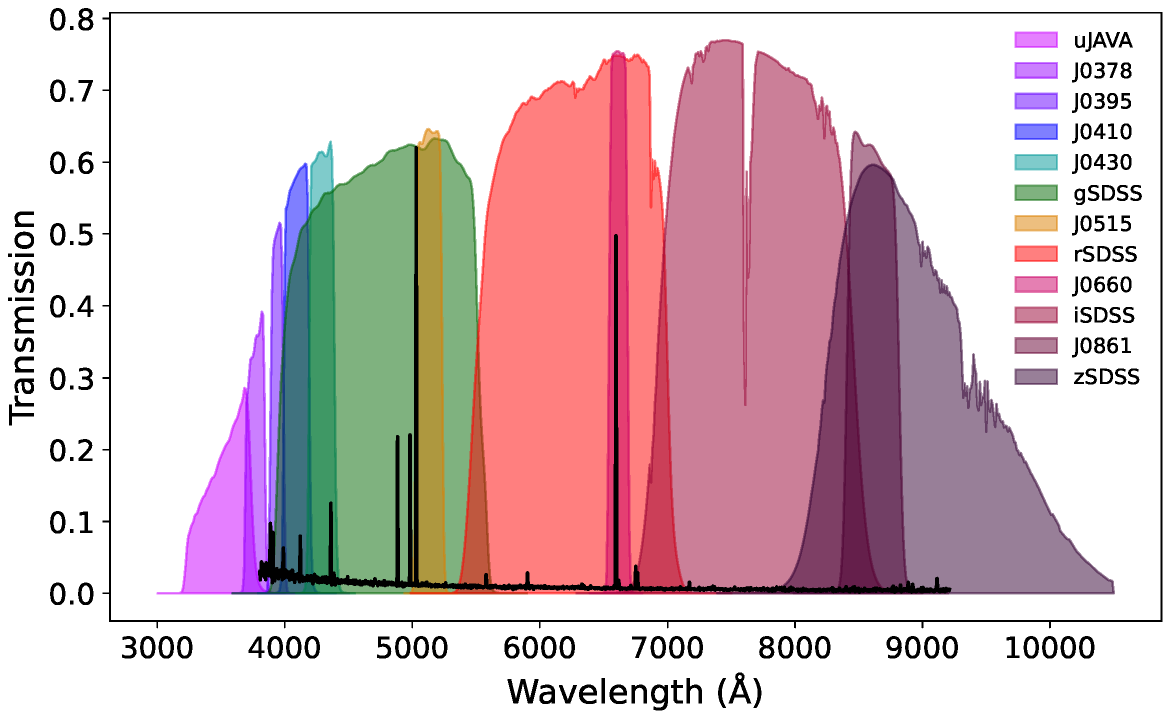}
\caption{Transmission curves of the 12 S-PLUS filters overlaid on the SDSS spectrum of one of the targets of this work (DR4\_3\_SPLUS-n03s22\_0021537). The H$\alpha$ emission line falls within the wavelength range of the J0660 filter, illustrating the importance of the narrow-band filters in identifying star-forming dwarf galaxy candidates.}
\label{fig:img01a}
\end{figure}

\begin{figure}
\centering
\includegraphics[scale=0.65]{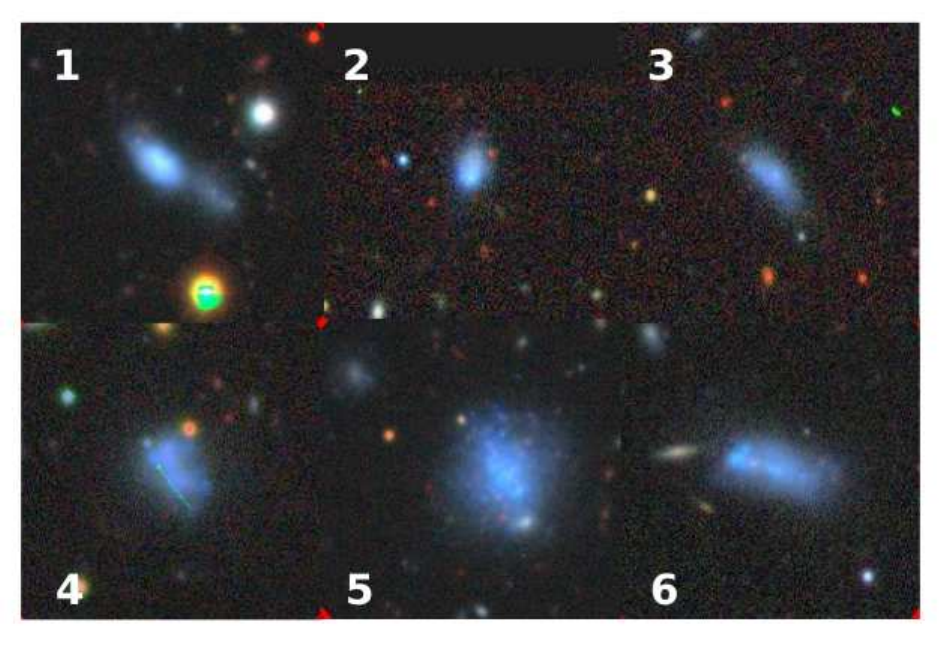}
\caption{Legacy $grz$ cutouts of the six metal-poor dwarf galaxy candidates selected in this work. Legacy images are deeper than S-PLUS allowing to infer more details about the galaxies morphology and structure. The cutout size is 55$^{\prime\prime} \times 55^{\prime\prime}$. See Table 1 for the definition of the ID numbers given at the top-left and bottom-left corners of each postage stamps.}
\label{fig:img}
\end{figure}

The Southern Photometric Local Universe Survey\footnote{https://splus.cloud/} (S-PLUS) is a 12-band optical survey conducted using the T80-South 0.83-m robotic telescope (T80S) located at Cerro Tololo Interamerican Observatory \citep{2019MNRAS.489..241M}. S-PLUS uses the Javalambre 12-band magnitude system \citep{2019A&A...622A.176C}, which includes four SDSS-like $griz$ broad-band, a Javalambre intermediate-band $u$ filter, and seven narrow-band filters centered on prominent stellar spectral features such as [O {\sc ii}]$\lambda\lambda$3727,3729, Ca{\sc ii} H+K, H$\delta$, H$\alpha$, and the Ca{\sc ii} triplet. The
panoramic camera features a single charge-coupled device
(CCD) with 9.2k $\times$ 9.2k pixels, a ﬁeld of view of
$1\fdg4 \times 1\fdg4$, and a pixel scale of 0$\farcs$55 pix$^{-1}$ \citep{2014arXiv1403.5237B}. The survey will cover an area of $\sim$ 9300 deg$^2$ of the southern sky.
Although there are other ongoing wide-area
photometric surveys in the southern hemisphere, the S-PLUS unique 12 multi-band filter system enables a better sampling of the wavelength range between 3000 $\AA$ and 10000 $\AA$ to perform detailed analysis of the spectral energy distribution and stellar population of galaxies. Specifically, the availability of the $u$ and J0660\footnote{This narrow-band filter is centered at $\lambda =$ 6614 $\AA$ and is suitable for investigating the H$\alpha$ emission line.} filter imaging over a large-area of the sky makes S-PLUS a unique survey for studying star forming galaxies. Moreover, it has been shown that an improved accuracy in physical parameter estimation from spectral energy distribution fitting (SED) can be attained when combining medium- and narrow-band filters with broad-band ones \citep{2015A&A...582A..14D, 2017MNRAS.471.4722M, 2019A&A...622A.181S, 2020MNRAS.499.3884M, 2021A&A...649A..79G, 2023MNRAS.526.1874T, 2024MNRAS.531..327N}.

Our aim is to start building a sample of low-metallicity dwarf galaxy candidates in the southern hemisphere using the S-PLUS survey and to perform a spectroscopic follow-up observations to confirm their chemically unevolved nature. Once we have constrained their abundances we can derive scaling relations and study their global properties in order to understand whether their low metallicity are due to either a slow secular evolution or to some external events related to their local environment. Here we present the preliminary results of a pilot spectroscopic follow-up campaign of the first metal-poor candidates with the  Gemini
Multi-Object Spectrograph (GMOS) attached to the 8-m Gemini
South telescope (Gemini-S).

This article is organized as follows. In Section \ref{sec:2} we present the sample selection, in Section \ref{sec:3} we describe the Gemini-S/GMOS observations and the data reduction. We discuss our results in Section \ref{sec:4}, and we present our Summary and Conclusion in Section \ref{sec:5}. In this work, we adopt the cosmological parameters $H_0$ = 70 km s$^{-1}$ Mpc$^{-1}$, $\Omega_m$ = 0.3, and $\Omega_{\Lambda} = 0.7$.

\section{Sample Selection}\label{sec:2}

The search for low-metallicity objects in the S-PLUS survey was performed on the third data release (DR3)
which covers an area
of $\sim$2000 deg$^2$. The photometric 5$\sigma$ depths range from 19.1 to 20.5 mag depending on the filter \citep[see][for details]{2022MNRAS.511.4590A}.

\subsection{color cuts}\label{sec:2.1}

Following \citetalias{2018ApJ...863..134H}, 
we applied color selection criteria typical of low-metallicity star-forming dwarf galaxies, such as IZw18 or Leo~P.
These color cuts were adapted from the analysis of galaxies in the SDSS survey to identify low redshift ($z \sim$ 0.01–0.02),  blue compact sources with colors consistent with H {\sc ii} regions:


\[ \: \: \: 0.2 \leq u - g \leq 0.8, \]
\[ -0.2 \leq g - r \leq 0.4, \\ \]
\[ -0.7 \leq r - i \leq 0.1, \\ \]
\[ -0.4 - 2\Delta z \leq i - z \leq 0.3, \]
\[ \: \: \: r - J0660 > -0.5 \]

\vspace{0.2cm}


\noindent where $\Delta z$ is the uncertainty of the $z$ band magnitude, and $J0660$ is the filter covering the H$\alpha$+[N {\sc ii}] lines.
The fourth condition is defined to include also objects with a poorly constrained $z$-band magnitude (\citetalias{2018ApJ...863..134H}).
Taking advantage of the S-PLUS multiband filter set, we add an additional constraint,  $r - J0660 > -0.5$, to ensure that galaxies exhibit detectable H$\alpha$ emission\footnote{We note that, given the characteristics of the narrow-band J0660 filter,
the H$\alpha$ emission can only be detected within the redshift range $z < 0.02$ \citep{2019A&A...622A.180L,2019MNRAS.489..241M}.} \citep{2020A&A...633A.123G}.
Figure \ref{fig:img01a} shows the SDSS spectrum of one of the galaxies of our sample, with the S-PLUS photometry and transmission curves of the 12 filter system overlaid. 
We applied these criteria to the S-PLUS DR3 catalog including only objects with $r < 18$ mag ($\sim$~200000 objects).
This selection returned a total of 380 candidate objects.
We visually inspected the S-PLUS images to reject spurious candidates
such as stars or star-forming regions located in the spiral arms of
larger galaxies.
After applying an absolute magnitude cut of $M_g > -18$ mag\footnote{To derive the absolute magnitudes at this stage of the selection, we used the photometric redshifts derived by \citet{2022A&C....3800510L} for the S-PLUS survey.}, which defines dwarf galaxies \citep{1981ApJ...247..823T,2014A&A...562A..49M,2019A&A...625A.143V}, we obtained a final sample of 47 metal-poor dwarf galaxy candidates (see Fig. \ref{fig:img} as an example).

\subsection{Spectral energy distribution (SED) fitting}\label{subsec:SED}

As a first test to confirm the metal-poor nature of our candidates, we performed SED fitting of the 12 S-PLUS bands with the Code Investigating GALaxy Emission \citep[CIGALE,][]{2019A&A...622A.103B}.
CIGALE combines a stellar SED with dust attenuation and emission components.
The code fits the spectrum (from 3000$\AA$ to 10000$\AA$) of each galaxy and  derives global properties such as star formation rate, stellar mass, bolometric luminosity, the minimum intensity value of the stellar radiation field. The stellar component is described by providing
the relative contribution of both the young and the old stellar components to the total emission of the galaxy.
Once the predicted theoretical SEDs are modeled, they are fitted to the observed SED.
The quality of the fit is assessed by the best $\chi^2$ \citep[and a reduced best $\chi^2$ defined as $\chi_r^2 = \chi^2/(N-1)$, with $N$ being the number of data points,][]{2019A&A...622A.103B}.

CIGALE allows for a number of star formation histories
(SFH) such as exponentially declined or delayed.
We used a delayed exponential star formation history with e-folding times
varying between $\tau = 2$ and 7 Gyr for the older stellar populations and an exponential burst with  an age of 20 Myr and e-folding time $\tau = 50$ Myr  for the latest star formation event. We left the stellar metallicity as a free parameter, ranging between 1/200th Z$_{\odot}$ and 1 Z$_{\odot}$, adopting the Small Magellanic Cloud (SMC)  
attenuation law \citep{2019A&A...622A.103B}.
The majority of the candidates (29 out of 47) have SEDs that are best-fitted by stellar metallicity below Z = 0.002 (1/10th Solar).  
The stellar masses obtained from population synthesis models range between 10$^7$ and few times 10$^9$~M$_{\odot}$, and they are comparable to those obtained using the relation between $r - i$ color and mass-to-light ratio derived by stellar population synthesis models \citep{2003ApJS..149..289B}

\begin{equation}\label{eq:mstar}
\log(M_*/\textrm{M}_{\odot}) = 1.672 + 1.114\,(r - i)_0 - 0.4\,M_{i,0}
\end{equation}

\noindent where $(r - i)_0$ is the galaxy color corrected for the Milky Way extinction, $M_{i,0}$ absolute $i$ magnitude in AB mag units, also corrected for Milky Way extinction, and a \citet{2001MNRAS.322..231K} initial mass function (IMF) is assumed (see Table \ref{tab:properties}). This calibration has been used to derive the stellar masses of other samples of low-mass galaxies \citepalias[e.g.][]{2018ApJ...863..134H}. Therefore, we will use it throughout this work to facilitate comparison with other sets of low-metallicity dwarf galaxies
(see Sections \ref{sec:2.3},  \ref{subsec:4.5}, and \ref{subsec:4.6}).
Following \citet{2003ApJS..149..289B} we adopt a systematic uncertainty of 25\% for the mass-to-light ratios obtained with this method.

Four objects were selected to be observed with Gemini-S/GMOS between 2022 and 2023 (ID 1 - 4 in Table \ref{tab:properties}). 
DR4\_3\_SPLUS-s29s37\_0046039 (ID 5 in Table \ref{tab:properties}) has oxygen abundances derived with the direct method in the literature (see Table \ref{tab:properties}). Another object -- DR4\_3\_SPLUS-n03s22\_0021537 (ID 6 in Table \ref{tab:properties}) -- has emission line fluxes available from the SDSS. Its metallicity was estimated by \citet{2016ApJ...819..110S}, although not with the direct method. Therefore we downloaded the SDSS spectrum and calculated the metallicity using the same approach as for the four galaxies observed with GMOS.  We display the images of these six targets in Fig. \ref{fig:img}. The images are taken from the DESI Legacy Imaging Surveys\footnote{https://www.legacysurvey.org/}  \citep{2019AJ....157..168D}. Being deeper than S-PLUS, Legacy images allow us to better identify the morphology and structure of our targets.

\begin{figure}
\includegraphics[bb=12 5 460 360, scale=0.55, clip]
{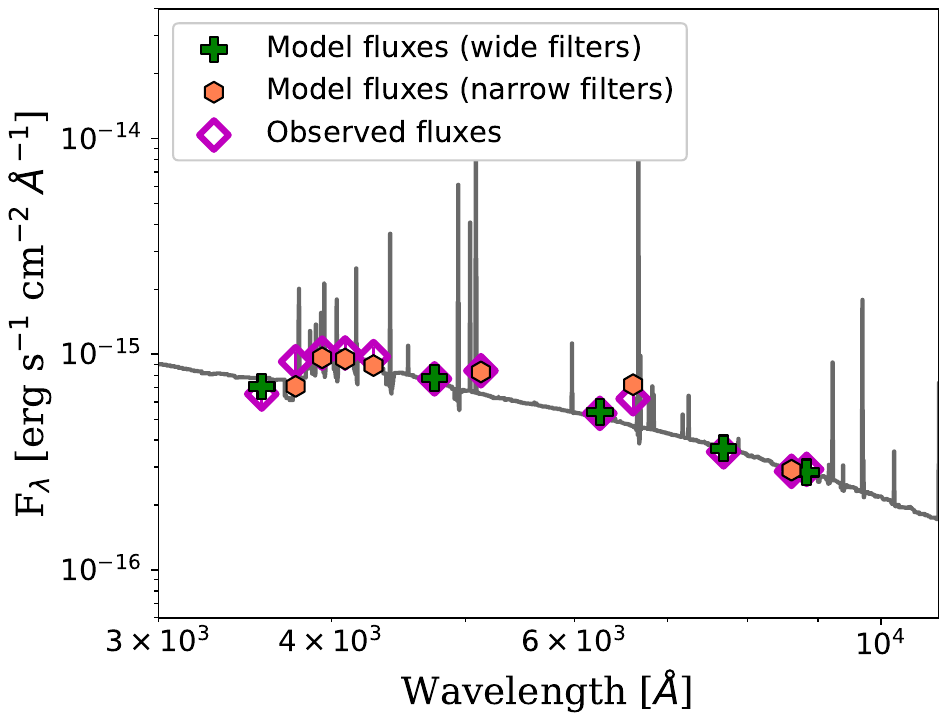}
\includegraphics[bb=12 5 460 360, scale=0.55, clip]
{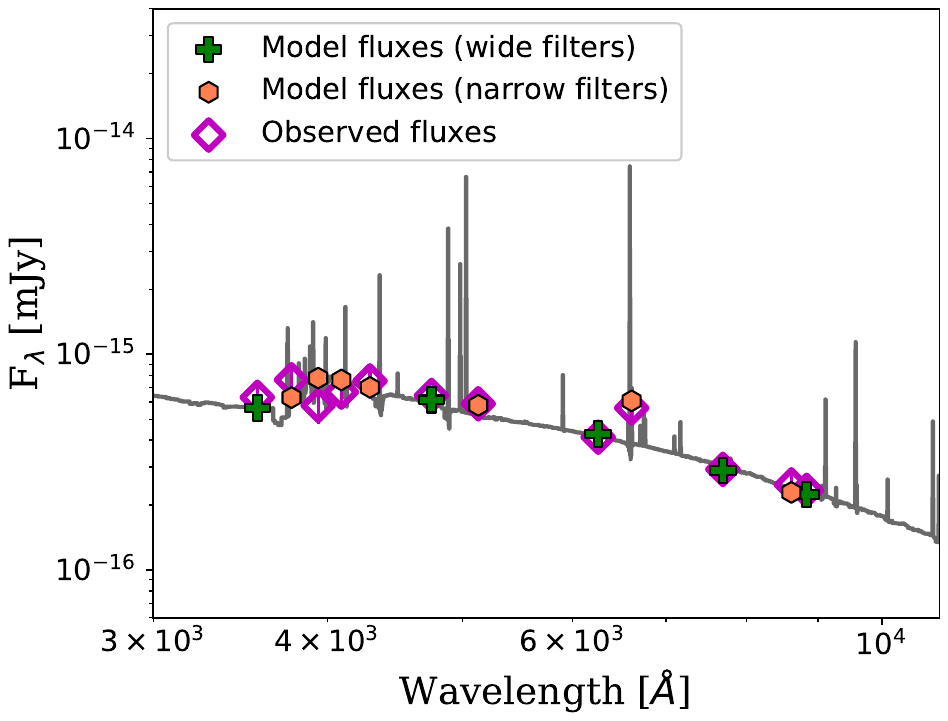}
\caption{Example of SED fitting with CIGALE. Best-fit result for ID 1 ($\chi_r^2 = 0.7$, top) and ID 4 ($\chi_r^2 = 0.8$, bottom). The best-fit stellar metallicity is 1/30th and 1/20th the solar value, respectively. Diamonds represent the observed fluxes, while crosses and hexagons indicate the model fluxes for the wide-band and narrow-band filters, respectively.}
\label{fig:SED}
\end{figure}

In Fig. \ref{fig:SED} we display examples of the SED fitting for two of our candidates.


\begin{figure*}
\includegraphics[scale=0.21,clip]{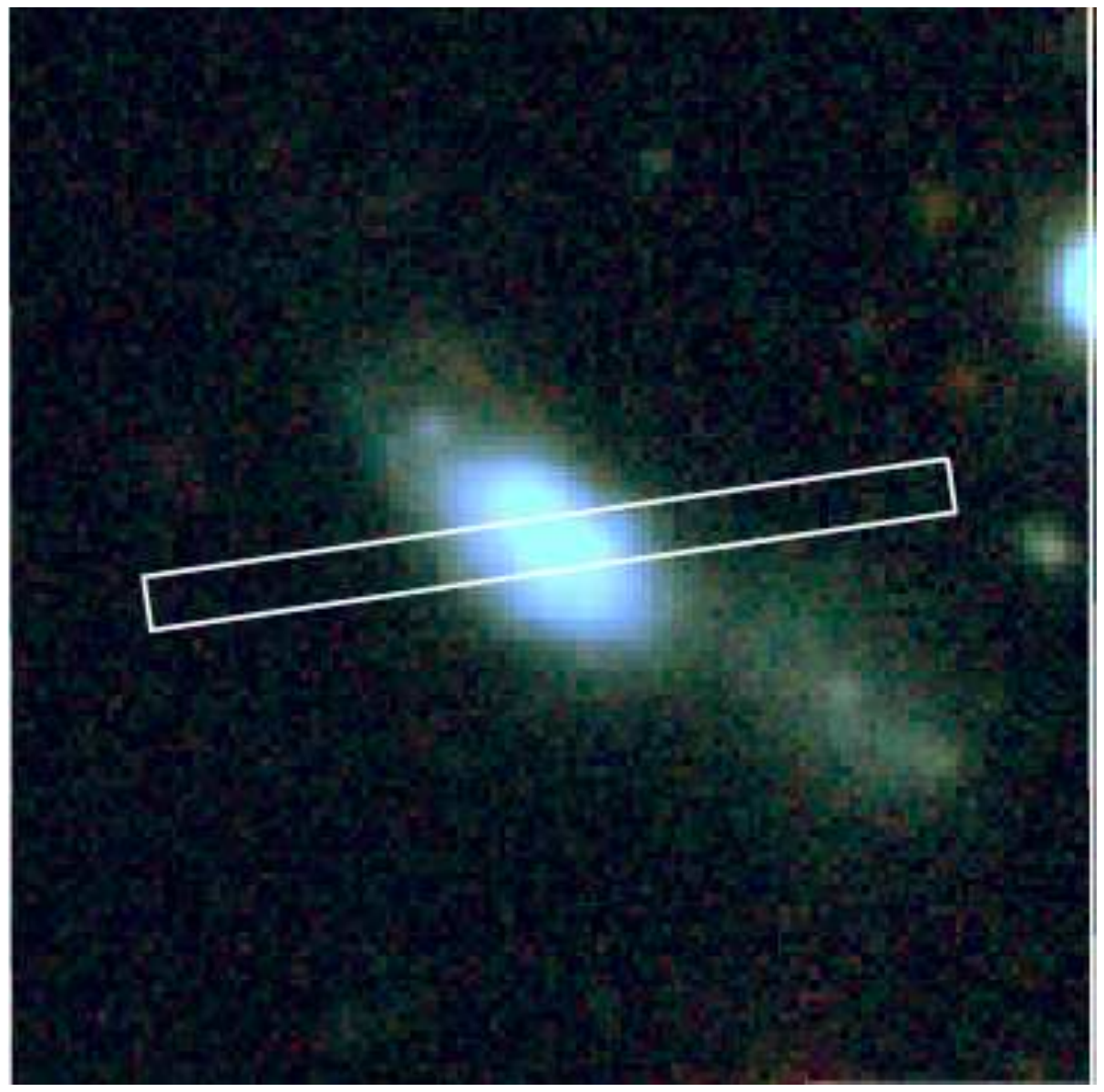}
\includegraphics[scale=0.21,clip]{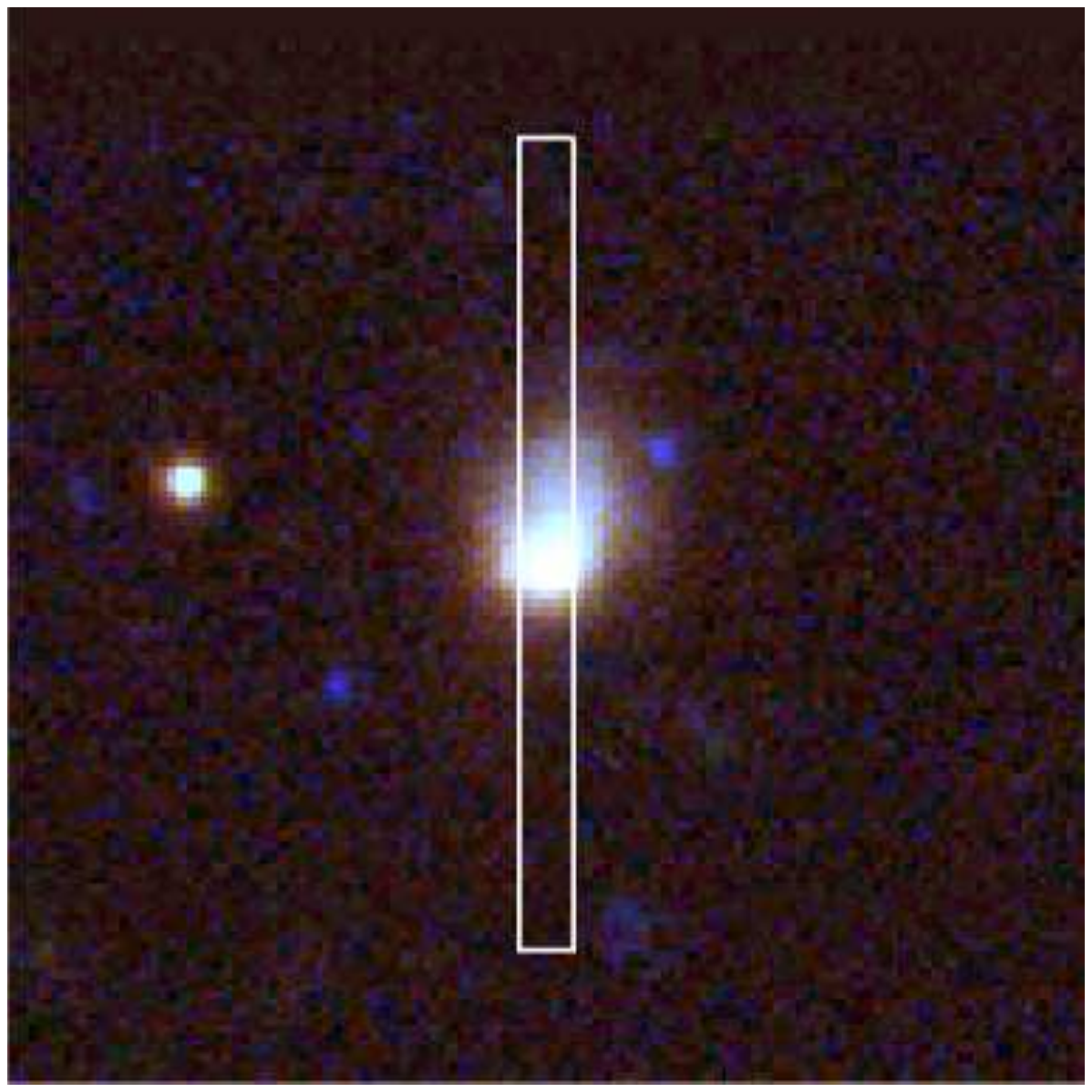}
\includegraphics[scale=0.21,clip]{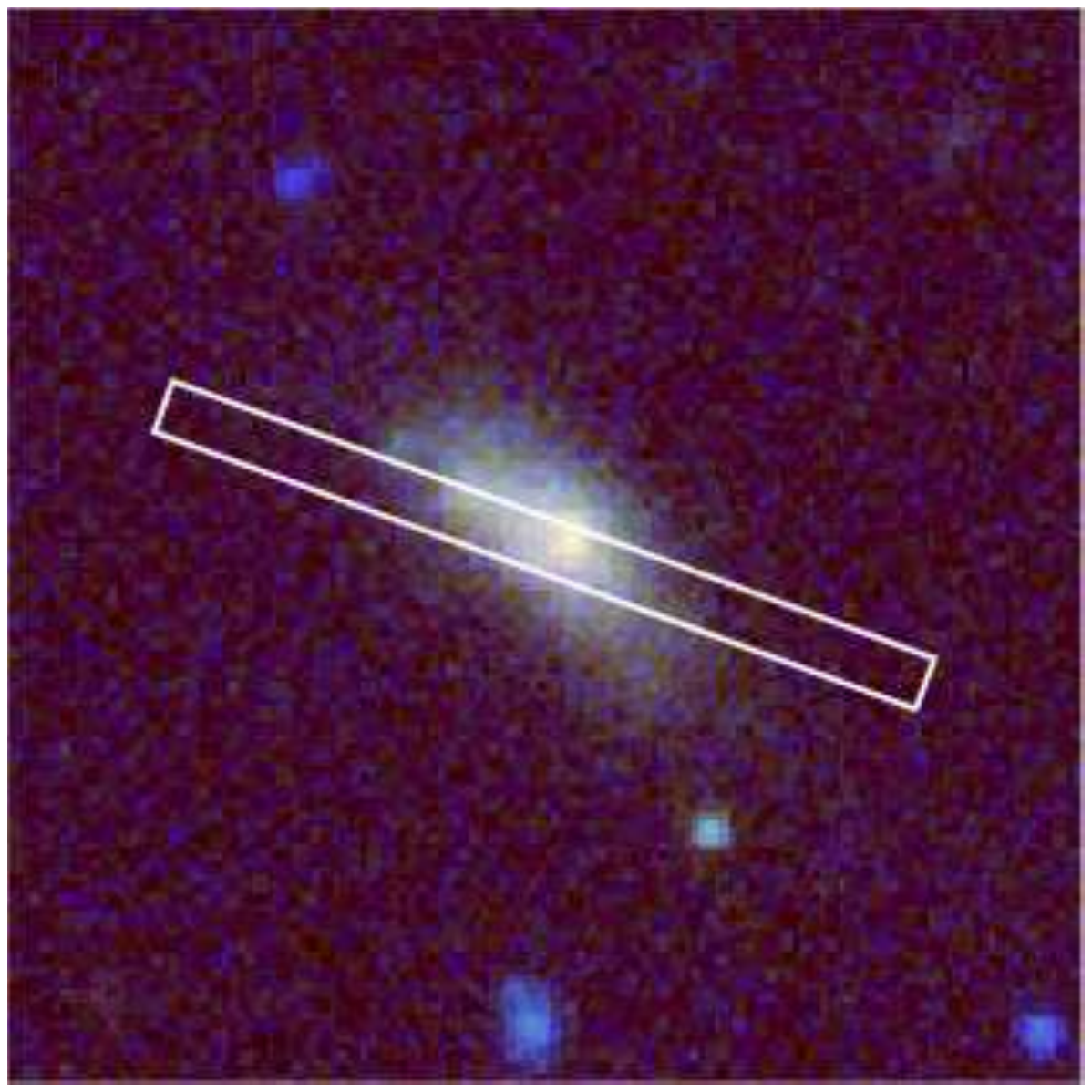}
\includegraphics[scale=0.21,clip]{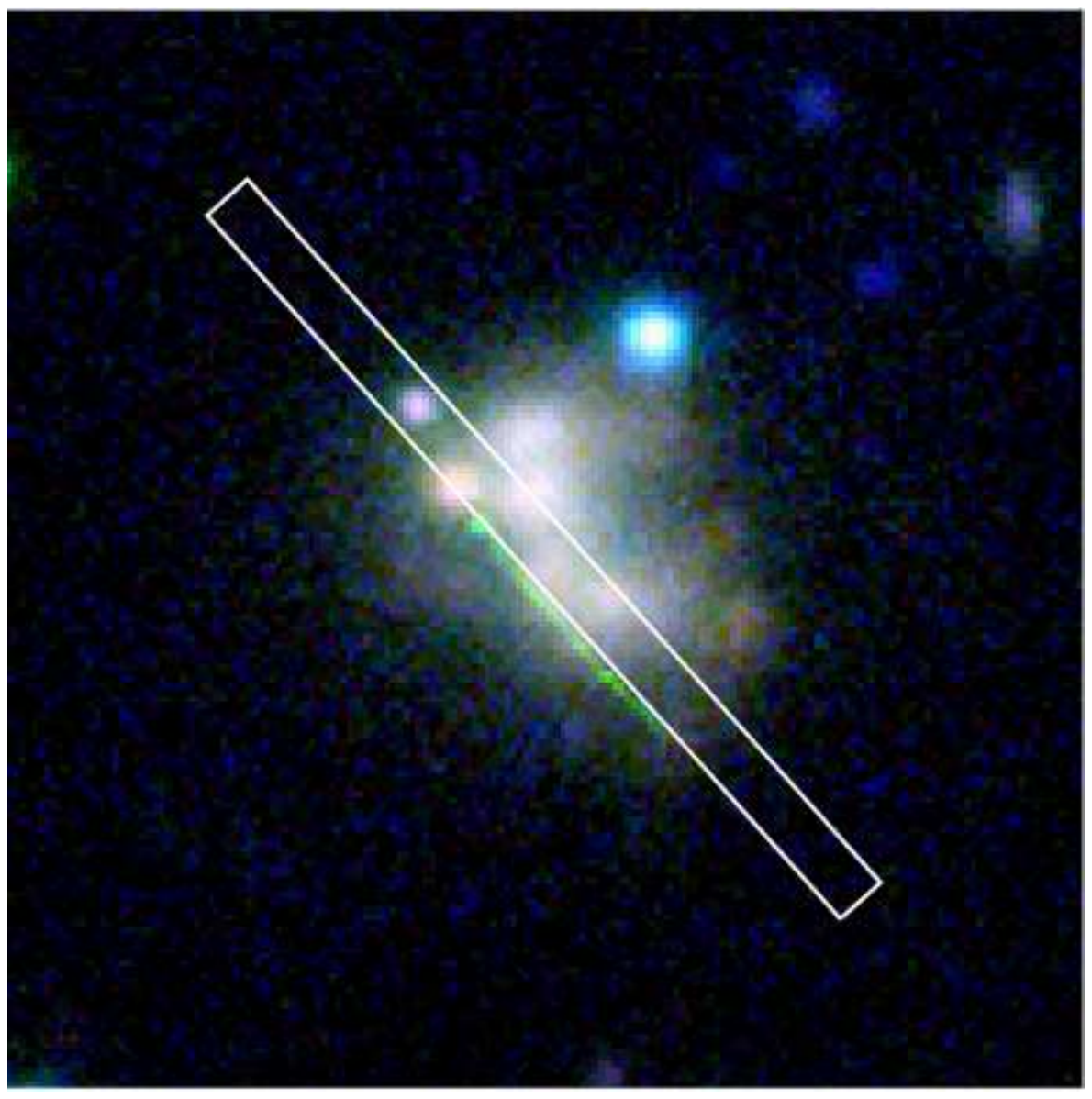}
\includegraphics[bb=7 7 320 320,scale=0.4,clip]{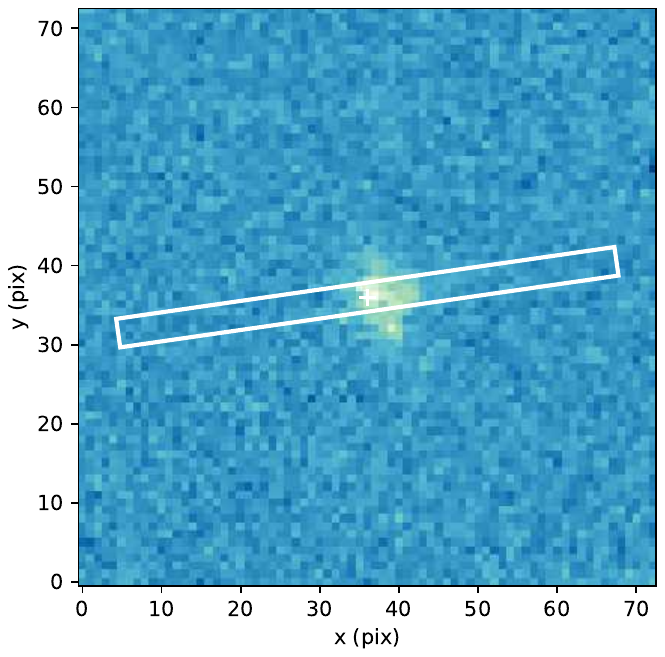}
\includegraphics[bb=7 7 320 320,scale=0.4,clip]{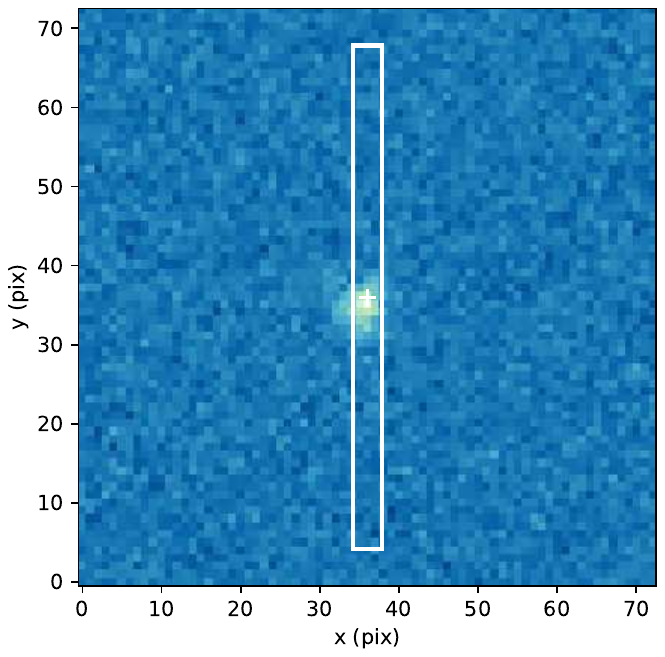}
\includegraphics[bb=7 7 320 320,scale=0.4,clip]{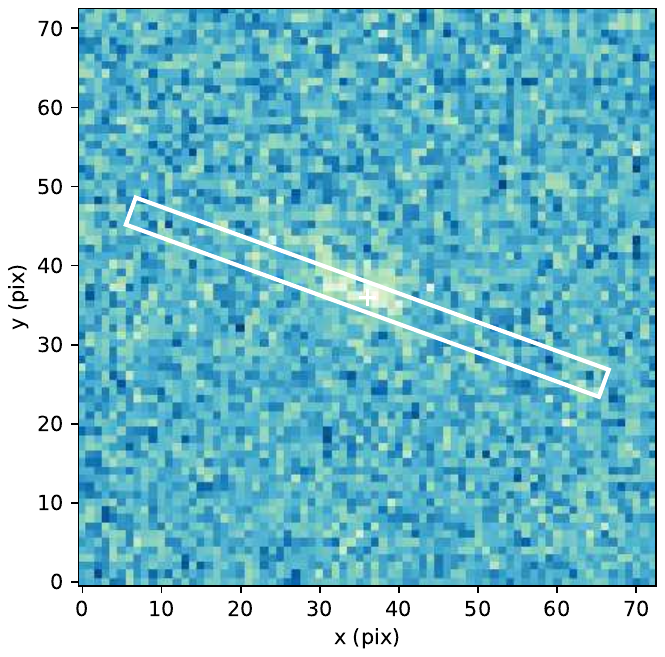}
\includegraphics[bb=7 7 320 320,scale=0.4,clip]{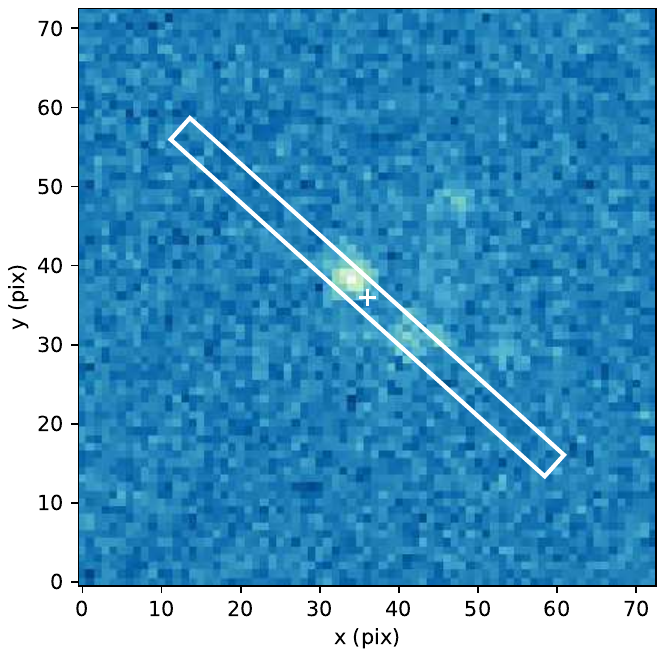}
\includegraphics[bb=55 0 375 330, scale=0.395,clip]{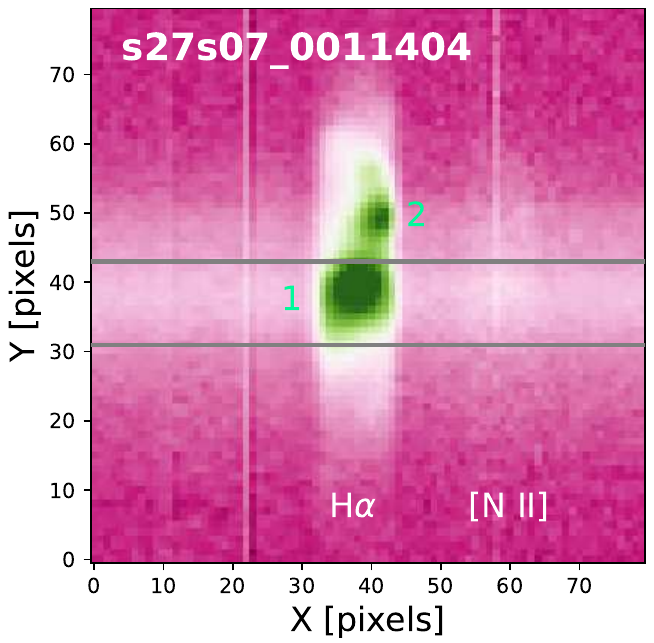}
\includegraphics[bb=55 0 375 330, scale=0.395,clip]{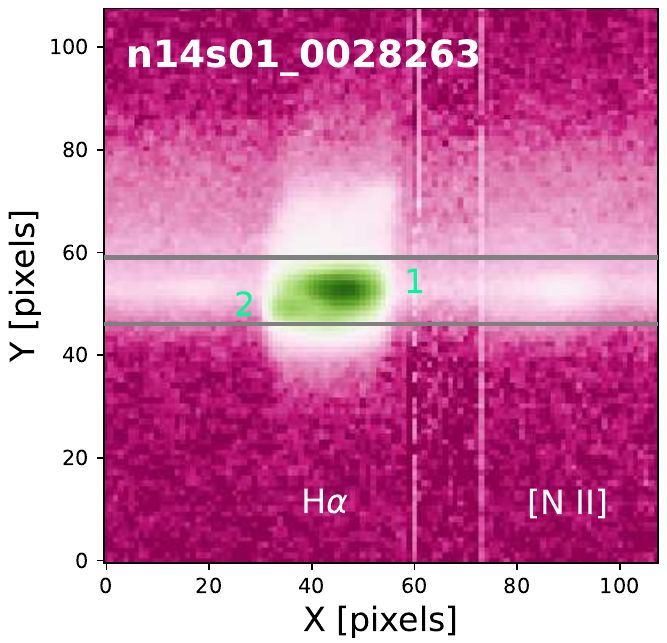}
\includegraphics[bb=55 0 375 330, scale=0.395,clip]{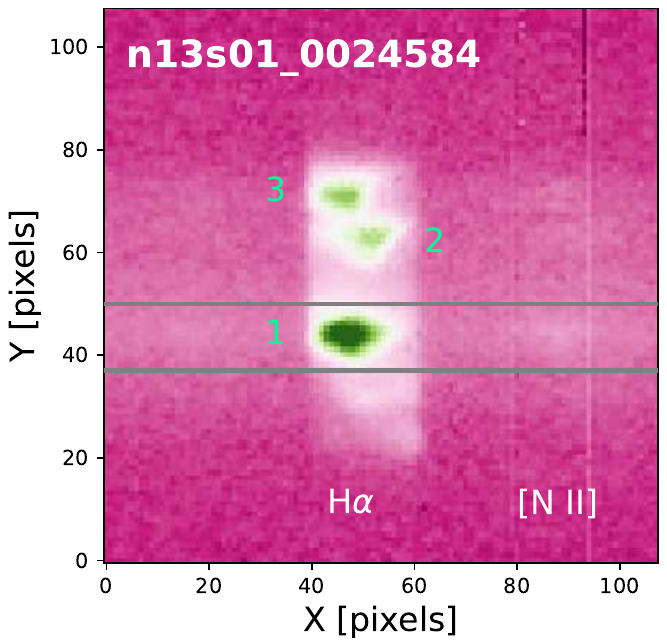}
\includegraphics[bb=55 0 375 330, scale=0.395,clip]{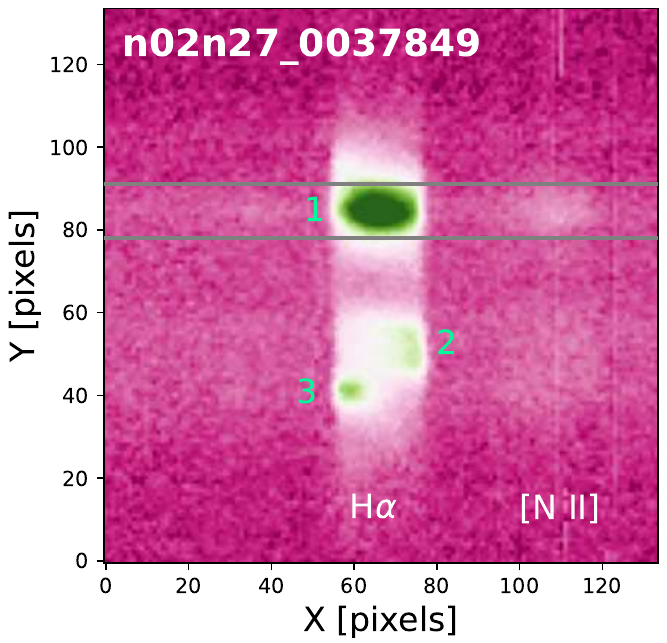}
\caption{Top panels: Legacy images of the four dwarf galaxies observed with GMOS with the position of the slit overlaid. Central panels: S-PLUS $J0660$ continuum-subtracted images of the four targets, illustrating the location of the slit over the main star-forming regions. Bottom panels: 2D long-slit spectra of the four galaxies with the observed wavelength on the horizontal axis and the spatial position on the vertical axis. The image shows the H$\alpha$ and the [N {\sc ii}] doublet emission from the different star forming knots included in the slits. The gray lines delimit the aperture chosen for the extraction of the 1D-spectra of Fig. \ref{fig:specs}. }
\label{fig:halpha_struct}
\end{figure*}

\subsection{Comparison samples}\label{sec:2.3}

Throughout this work, we compare our targets with metal-poor dwarf galaxies selected using similar criteria in other photometric surveys such as the SDSS \citepalias{2018ApJ...863..134H}, and the DES \citepalias{2023ApJ...951..138L}.
The only difference between the selection method between \citetalias{2023ApJ...951..138L} and \citetalias{2018ApJ...863..134H} is that  $u-$band photometry is not available over the DES area, therefore, they modified the  $u-g$ color-cut criterion  into a $NUV - g$ condition based on the GALEX NUV ﬁlter \citep{2005ApJ...619L...1M}. Most targets from these photometric selections have been followed up spectroscopically. For comparison, we include only galaxies with oxygen abundances derived with the direct method.
Stellar masses for both samples have been recalculated using Eq. \ref{eq:mstar}, assuming a Kroupa IMF. 

\section{Gemini-S/GMOS observations and data reduction}\label{sec:3}

The observations were carried out via two Fast Turnaround runs in 2022 and 2023 with the GMOS instrument \citep{2004PASP..116..425H} in the long slit mode  on the Gemini-S telescope  
(Program IDs GS-2022B-FT-210, GS-2023A-FT-204). We used the B600 grism  with a 2$^{\prime\prime}$ slit width.
The wavelength range varies between 4300 and 7500 \AA\ with a central wavelength $\lambda_c =$ 5900 \AA. This choice allowed us to include all the major emission lines needed to determine the oxygen abundance with the direct method, including the
singly ionized ionic population of oxygen traced by the
[O {\sc ii}]$\lambda\lambda$7320,7330 lines.
The spectral  resolution estimated as the full-width-at-half-maximum (FWHM) of the emission lines of comparison lamps  at 5900 $\AA$ is
9 $\AA$, corresponding to $\sigma_{\rm inst}$ = 195 km s$^{-1}$ ($\sigma$ = FWHM/2.355).
The total integration time ranged between 1350 and 7200 seconds.
Data were binned by 2 in the spatial direction and binned by 2 in the spectral direction only for the first galaxy observed in the 2022 observing run (ID 1). 
The central position of the long slit was defined according to the 
peak of H$\alpha$ emission in the S-PLUS continuum-subtracted J0660 images of each galaxy. The position angle of the slit for the first target (ID 1) was set at the parallactic angle. For the remaining targets, however,  the slit  was oriented to cover most of the galactic disk and the H {\sc ii} regions detected in the J0660 narrow-band images (Fig. \ref{fig:halpha_struct}, top and central panels). 
With a slit width of 2$^{\prime\prime}$ generously exceeding the characteristic 
seeing, and all observations conducted under favorable low-airmass conditions (average value $<$ 1.4), our setup was inherently robust against the minimal effects of atmospheric differential refraction, guaranteeing the photometric integrity of the resulting spectra \citep{1982PASP...94..715F}. Spectra of CuAr arc lamps were obtained for wavelength calibrations, along with  bias frames
and dome ﬂats to correct for the detector
bias level and pixel-to-pixel variations, respectively.
The spectrophotometric standard LTT\,2415 was observed with the same experimental set-up and it was used to calibrate the targets  in relative flux.

\begin{figure*}
\includegraphics[width=1\textwidth,height=0.5\textwidth]{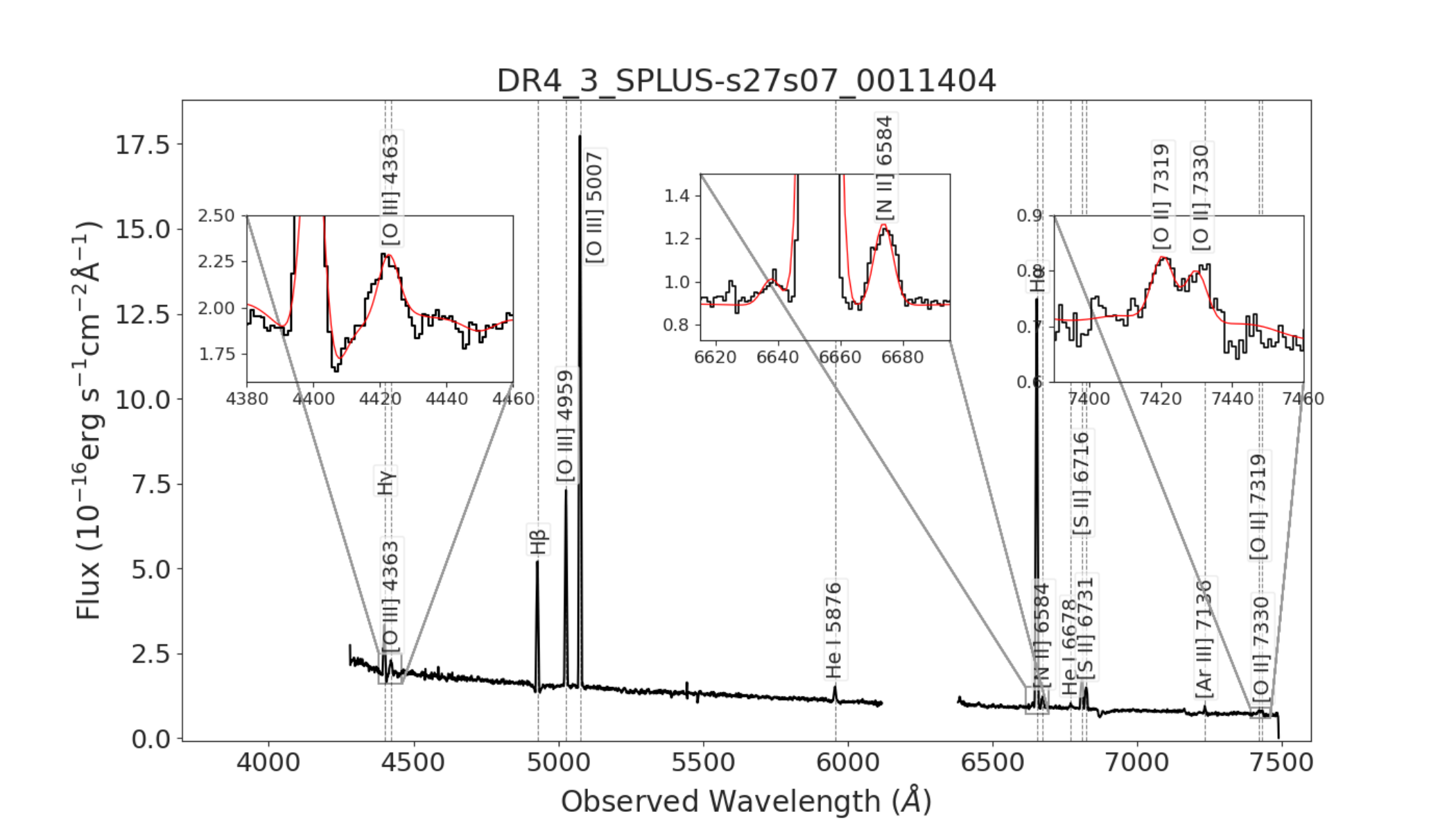} 
\includegraphics[width=1\textwidth,height=0.5\textwidth]{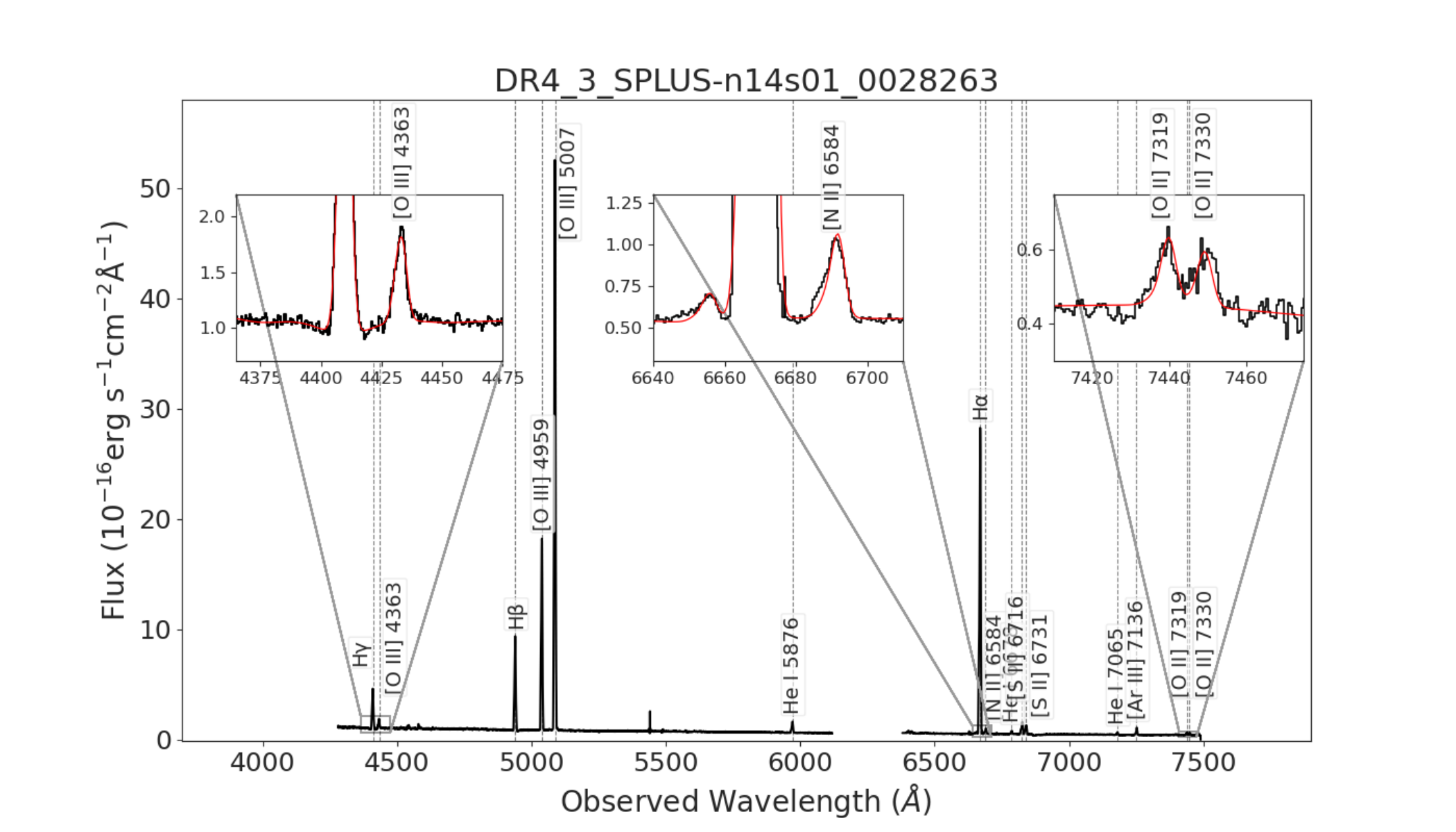} 
\caption{Gemini-S/GMOS spectra of DR4\_3\_SPLUS-s27s07\_0011404 (ID 1, top) and DR4\_3\_SPLUS-n14s01\_0028263 (ID 2, bottom) obtained with the B600 grating. The [O {\sc iii}]$\lambda$4363 and the [N {\sc ii}], [O {\sc ii}] doublets are highlighted in the insets. The red lines show the best-fit model of the spectrum.}
\label{fig:specs}
\end{figure*}

\begin{figure*}
\ContinuedFloat
\includegraphics[width=1\textwidth,height=0.5\textwidth] {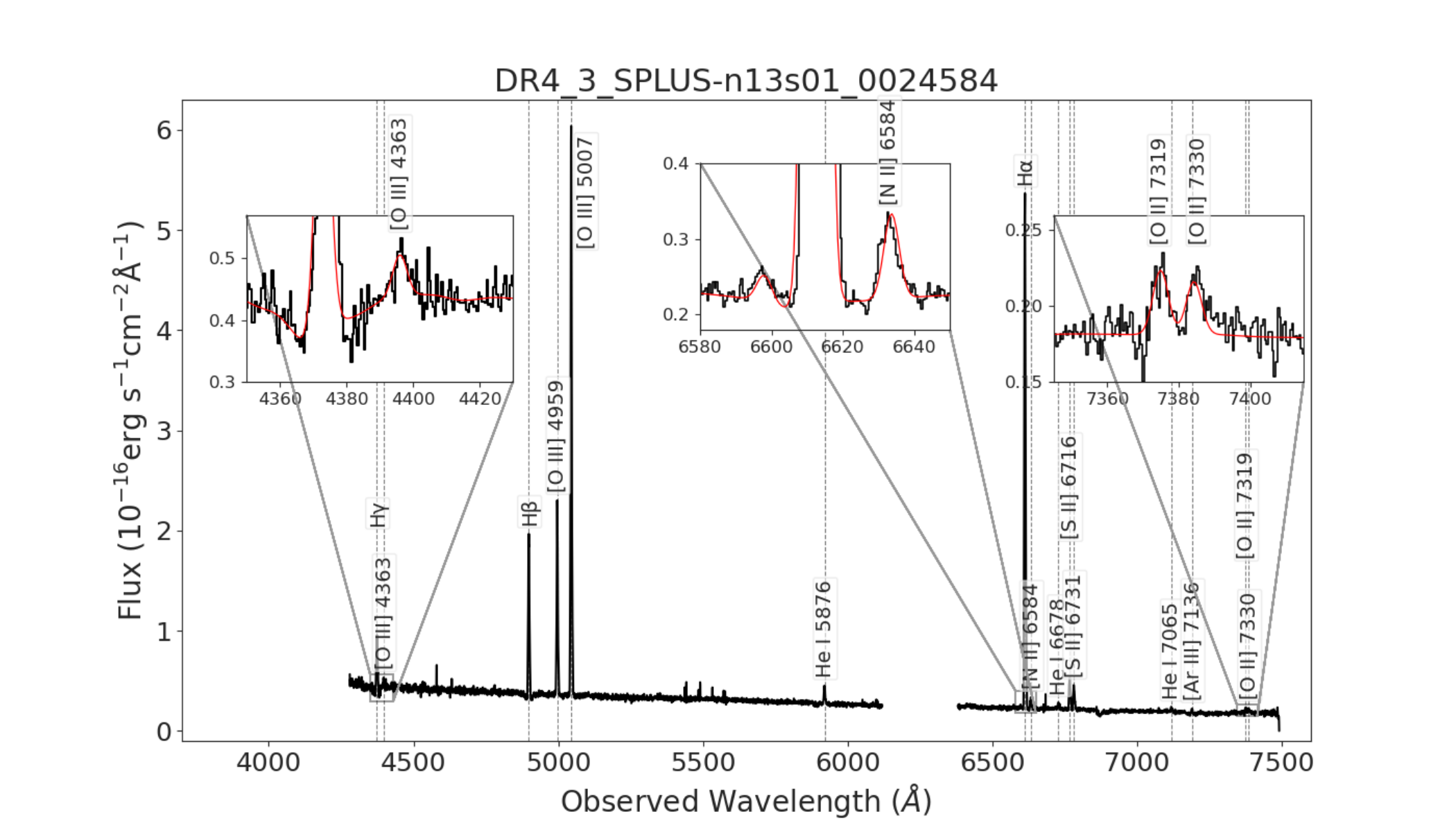} 
\includegraphics[width=1\textwidth,height=0.5\textwidth]{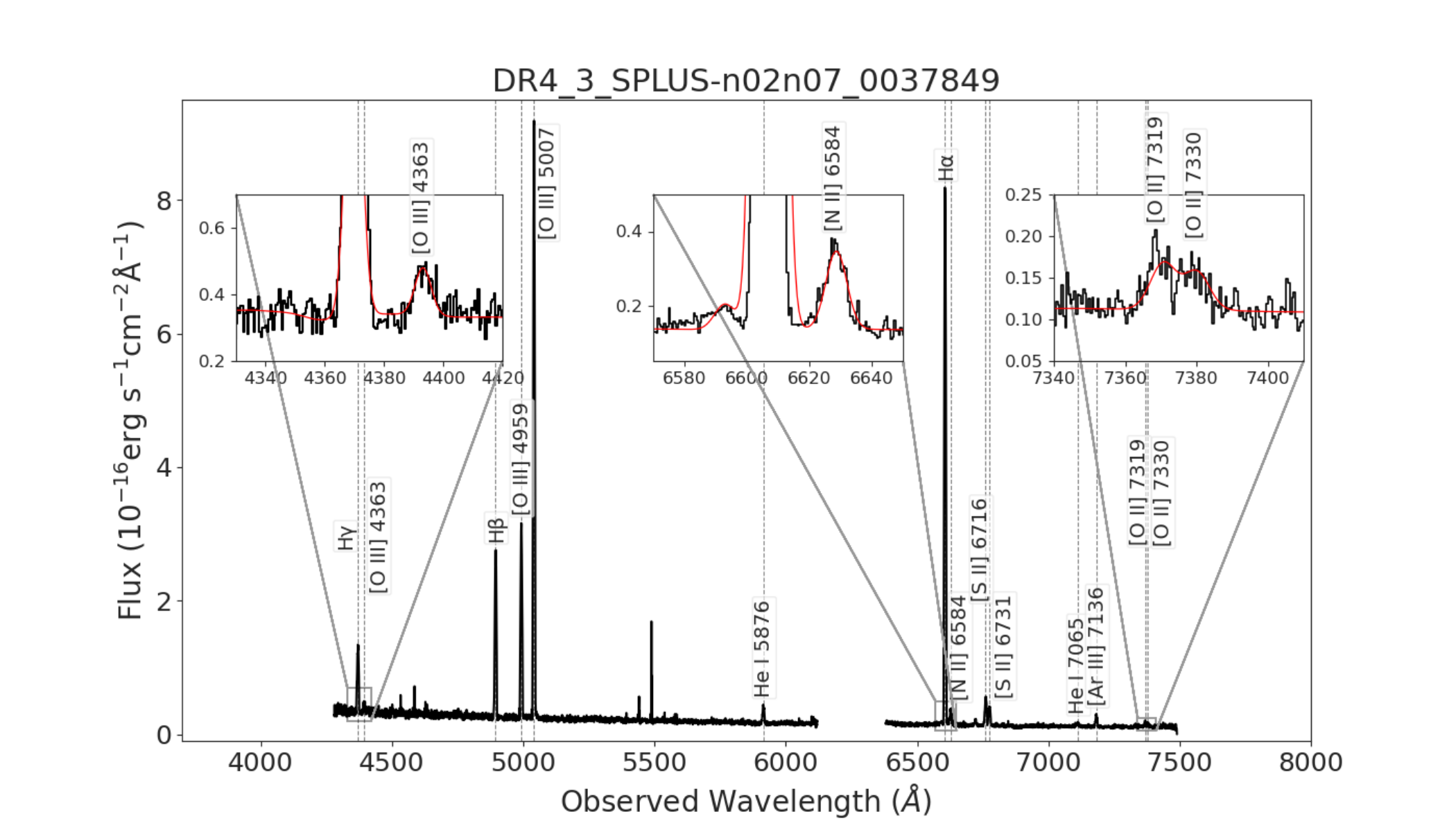} 
\caption{continued. Gemini-S/GMOS spectra of DR4\_3\_SPLUS-n13s01\_0024584 (ID 3, top) and DR4\_3\_SPLUS-n02n27\_0037849 (ID 4, bottom) obtained with the B600 grating.} 
\label{fig:specs}
\end{figure*}

For the two GMOS runs we selected  the targets that were best observable in semesters 2022B, 2023A and that were characterized by the lowest best-fit metallicities in our SED fitting analysis (see \S  \ref{subsec:SED}). 
The data were reduced with the Data Reduction for
Astronomy from Gemini Observatory North and South (\texttt{DRAGONS}) package v3.1.0 \citep{2023RNAAS...7..214L} using the standard recipe for GMOS
long-slit reductions, that includes bias correction, flat-fielding, sky subtraction, wavelength calibration,  flux calibration, and 1D spectra extraction.

Our aim is to derive the oxygen abundances using the direct method, which utilizes the flux ratio of auroral
[O {\sc iii}]$\lambda$4363 to strong lines such as [O {\sc iii}]$\lambda$$\lambda$4959,5007, to measure the electron temperature of the gas, $T_{\rm e}$.
Combining hydrogen recombination lines and strong oxygen
emission features of [O {\sc ii}] and [O {\sc iii}] ions, the total
oxygen abundance can be derived \citep{2006A&A...448..955I,2006agna.book.....O}.


\section{Results}\label{sec:4}

\begin{table*}
\small
\begin{center}
\caption{Observed line fluxes, $F_{\lambda}/F({\rm H}\beta)$, and  extinction corrected line intensities, $I_{\lambda}/I({\rm H}\beta)$, normalized to $F({\rm H}\beta)$=100 for the four galaxies observed with GMOS.} 
\begin{tabular}{lrrrrrrrrrrr}
\hline \hline
\noalign{\smallskip}
 \multicolumn{1}{c}{ } &
\multicolumn{2}{c}{1} &
\multicolumn{2}{c}{2} &
\multicolumn{2}{c}{3} &
\multicolumn{2}{c}{4} \\
\hline
\noalign{\smallskip}
  \multicolumn{1}{c}{$\lambda_{rest}$ [$\AA$]} &
  \multicolumn{1}{c}{$F_{\lambda}/F({\rm H}\beta)$} &
  \multicolumn{1}{c}{$I_{\lambda}/I({\rm H}\beta)$} &
  \multicolumn{1}{c}{$F_{\lambda}/F({\rm H}\beta)$} &
  \multicolumn{1}{c}{$I_{\lambda}/I({\rm H}\beta)$} &
  \multicolumn{1}{c}{$F_{\lambda}/F({\rm H}\beta)$} &
  \multicolumn{1}{c}{$I_{\lambda}/I({\rm H}\beta)$} &
  \multicolumn{1}{c}{$F_{\lambda}/F({\rm H}\beta)$} &
  \multicolumn{1}{c}{$I_{\lambda}/I({\rm H}\beta)$} \\
    \multicolumn{1}{c}{} &
  \multicolumn{1}{c}{} &
  \multicolumn{1}{c}{} &
  \multicolumn{1}{c}{} &
  \multicolumn{1}{c}{} &
  \multicolumn{1}{c}{} &
  \multicolumn{1}{c}{} &
  \multicolumn{1}{c}{} &
  \multicolumn{1}{c}{} \\
\hline
\noalign{\smallskip}
$\lambda$\,4340 H$\gamma$      &   46.86 $\pm$ 0.55    &   47.42  $\pm$   0.64  &  47.47 $\pm$ 0.40   &  48.26 $\pm$   0.32  &   39.36 $\pm$ 0.42   &   41.43  $\pm$   0.34  &   43.73 $\pm$ 0.29   &  47.19 $\pm$  0.27     \\
$\lambda\,$4363 [O {\sc iii}]  &    8.61 $\pm$ 0.50    &    8.71  $\pm$   0.73  &  10.73 $\pm$ 0.29   &  10.90 $\pm$   0.26  &    4.91 $\pm$ 0.39   &    5.14  $\pm$   0.34  &    5.55 $\pm$ 0.28   &   5.94 $\pm$  0.28     \\
$\lambda$\,4471 He {\sc i}     &    3.06 $\pm$ 0.49    &    3.09  $\pm$   0.55  &   3.05 $\pm$ 0.25   &   3.08 $\pm$   0.22  &--$\:\:\:\:\:\:\:\:\;$&--$\:\:\:\:\:\:\:\:\;$   &--$\:\:\:\:\:\:\:\:\;$&--$\:\:\:\:\:\:\:\:\;$  \\
$\lambda$\,4861 H$\beta$       &  100.00 $\pm$ 0.68    &  100.00  $\pm$   0.52  & 100.00 $\pm$ 0.52   & 100.00 $\pm$   0.10  &  100.00 $\pm$ 0.53   &  100.00  $\pm$   0.10  &  100.00 $\pm$ 0.37   & 100.00 $\pm$  1.00     \\
$\lambda$\,4959 [O {\sc iii}]  &  122.79 $\pm$ 0.61    &  123.00  $\pm$   1.00  & 192.49 $\pm$ 0.71   & 192.00 $\pm$   1.00  &  107.61 $\pm$ 0.42   &  107.00  $\pm$   1.00  &  120.54 $\pm$ 0.32   & 119.00 $\pm$  1.00     \\
$\lambda$\,5007 [O {\sc iii}]  &  365.93 $\pm$ 1.80    &  365.00  $\pm$   1.00  & 573.59 $\pm$ 2.13   & 571.00 $\pm$   1.00  &  320.84 $\pm$ 1.24   &  317.00  $\pm$   1.00  &  359.25 $\pm$ 0.97   & 353.00 $\pm$  1.00     \\
$\lambda$\,5876 He {\sc i}     &    7.87 $\pm$ 0.43    &    7.76  $\pm$   0.34  &   9.27 $\pm$ 0.16   &   9.07 $\pm$   0.14  &    9.48 $\pm$ 0.34   &    9.00  $\pm$   0.28  &    9.62 $\pm$ 0.26   &   9.03 $\pm$  0.23     \\
$\lambda$\,6300 [O {\sc i}]    &--$\:\:\:\:\:\:\:\:\;$ &--$\:\:\:\:\:\:\:\:\;$  &   2.24 $\pm$ 0.13   &   2.18 $\pm$   0.11  &--$\:\:\:\:\:\:\:\:\;$&--$\:\:\:\:\:\:\:\:\;$   &--$\:\:\:\:\:\:\:\:\;$& --$\:\:\:\:\:\:\:\:\;$ \\
$\lambda$\,6363 [O {\sc i}]    &--$\:\:\:\:\:\:\:\:\;$ &--$\:\:\:\:\:\:\:\:\;$  &   0.75 $\pm$ 0.04   &   0.72 $\pm$   0.04  &--$\:\:\:\:\:\:\:\:\;$&--$\:\:\:\:\:\:\:\:\;$   &--$\:\:\:\:\:\:\:\:\;$& --$\:\:\:\:\:\:\:\:\;$ \\
$\lambda$\,6563 H$\alpha$      &  223.20 $\pm$ 1.14    &  218.00  $\pm$   1.00  & 226.55 $\pm$ 0.88   & 220.00 $\pm$   1.00  &  237.72 $\pm$ 0.94   &  218.00  $\pm$   1.00  &  253.65 $\pm$ 0.70   & 225.00 $\pm$  1.00     \\
$\lambda$\,6584 [N {\sc ii}]   &    9.06 $\pm$ 0.51    &    8.84  $\pm$   0.74  &   5.87 $\pm$ 0.71   &   5.69 $\pm$   0.64  &    7.36 $\pm$ 0.40   &    6.73  $\pm$   0.32  &    9.51 $\pm$ 0.32   &   8.43 $\pm$  0.27     \\
$\lambda$\,6678 He {\sc i}     &    1.85 $\pm$ 0.40    &    1.81  $\pm$   0.33  &   2.17 $\pm$ 0.12   &   2.09 $\pm$   0.11  &    2.71 $\pm$ 0.32   &    2.45  $\pm$   0.25  &    2.71 $\pm$ 0.25   &   2.37 $\pm$  0.21     \\
$\lambda$\,6716 [S {\sc ii}]   &   16.74 $\pm$ 0.41    &   16.35  $\pm$   0.25  &   9.49 $\pm$ 0.09   &   9.18 $\pm$   0.07  &   15.45 $\pm$ 0.32   &   14.05  $\pm$   0.25  &   13.85 $\pm$ 0.25   &  12.15 $\pm$  0.21     \\
$\lambda$\,6731 [S {\sc ii}]   &   11.88 $\pm$ 0.41    &   11.60  $\pm$   0.21  &   6.98 $\pm$ 0.08   &   6.76 $\pm$   0.07  &   11.12 $\pm$ 0.32   &   10.11  $\pm$   0.25  &    9.27 $\pm$ 0.25   &   8.12 $\pm$  0.21     \\
$\lambda$\,7065 He {\sc i}     &--$\:\:\:\:\:\:\:\:\;$ &--$\:\:\:\:\:\:\:\:\;$  &   1.62 $\pm$ 0.11   &   1.55 $\pm$   0.10  &    2.24 $\pm$ 0.31   &    1.99  $\pm$   0.26  &    1.86 $\pm$ 0.25   &   1.58 $\pm$  0.20     \\
$\lambda$\,7136 [Ar {\sc iii}] &    2.85 $\pm$ 0.39    &    2.77  $\pm$   0.22  &   4.92 $\pm$ 0.08   &   4.73 $\pm$   0.07  &    1.48 $\pm$ 0.31   &    1.32  $\pm$   0.24  &    4.52 $\pm$ 0.25   &   3.85 $\pm$  0.20     \\
$\lambda$\,7320 [O {\sc ii}]   &    1.54 $\pm$ 0.23    &    1.49  $\pm$   0.22  &   0.84 $\pm$ 0.15   &   0.81 $\pm$   0.14  &    1.53 $\pm$ 0.19   &    1.34  $\pm$   0.15  &    1.31 $\pm$ 0.14   &   1.10 $\pm$  0.12     \\
$\lambda$\,7330 [O {\sc ii}]   &    1.88 $\pm$ 0.29    &    1.82  $\pm$   0.26  &   1.03 $\pm$ 0.19   &   0.98 $\pm$   0.17  &    1.87 $\pm$ 0.23   &    1.64  $\pm$   0.18  &    1.60 $\pm$ 0.18   &   1.34 $\pm$  0.14     \\

\hline
\\$F({\rm H}\beta)^a$           &    3.68               &                        &  4.17              &                &     0.95             &                       &    2.11              &                       \\
mean $C({\rm H}\beta)$          &  {3$\times 10^{-4}$}  &				         &  0.0               &                &	 0.41             &                       &	   0.18              &                       \\
\noalign{\smallskip}
\hline \hline
\end{tabular}
\label{tab:fluxes}
\end{center}
\begin{minipage}[r]{16.5cm}
 $^a$ in units of $10^{-15}$ erg s$^{-1}$ cm$^{-2}$.
\end{minipage}
\end{table*}


All the galaxies present two or three star-forming knots within the slit as can be seen in the bottom panels of Fig. \ref{fig:halpha_struct}, showing an expanded view of the 2D  spectra around the H$\alpha$ emission line.
The H$\alpha$ emission of our targets is structured in velocity along the slit.

\subsection{Emission line fluxes and intensities}

The apertures for the extraction of the spectra to measure the oxygen abundances were defined based on the extension of the [O {\sc iii}]$\lambda$4363 emitting region, which is usually much more compact than H$\alpha$. [O {\sc iii}]$\lambda$4363 is detected only in correspondence of the brightest H {\sc ii} region of each galaxy. We defined an aperture size of  
$\sim$ 2$^{\prime\prime}$ for all galaxies,
selected along the spatial axis, as indicated by the gray lines in the bottom panels of Fig. \ref{fig:halpha_struct}.

In Fig. \ref{fig:specs} we display  the 1D spectra obtained for the four objects, with the prominent  emission lines marked. 
The brightest lines are H$\alpha$ and H$\beta$ recombination lines, along with the [O {\sc iii}]$\lambda$4959,$\lambda$5007 lines.
As can be seen in the figure, 
even with long exposures on a 8-meter telescope, the detection of the auroral line is challenging in these galaxies. 

\subsection{Continuum subtraction and line extraction with ppXF.}


We used the penalized pixel-Fitting (pPXF) \citep[ppXF, ]{2004PASP..116..138C,2017MNRAS.466..798C} algorithm to model the stellar continuum and extract the emission line fluxes. pPXF was applied to the four GMOS target and to galaxy No. 6  which has a SDSS spectrum.
The stellar continuum was modeled to account for absorption features present in some of the targets, reducing their impact on the measurement of the emission line fluxes (e.g. Balmer absorption). The fit was performed using single stellar population (SSP) templates from the E-MILES library \citep{2016MNRAS.463.3409V,2023MNRAS.526.3273C}. pPXF constructs an optimal combination of these templates and fits them to the observed spectrum using Gauss–Hermite expansions for the line-of-sight velocity distribution.
Multiplicative polynomials of degree 8–10 (depending on the galaxy) were employed to match the overall spectral shape of the data. The emission lines were simultaneously fitted together with the continuum and the line-of-sight velocity was derived by fitting the emission lines simultaneously, assuming a fixed velocity width.
The instrumental resolution was taken into account when setting the initial velocity dispersion.
The uncertainties on the emission-line fluxes were taken directly from the formal errors provided by pPXF, which are derived from the covariance matrix of the fit and account for both the noise in the spectrum and the coupling between fitted parameters.
The complete list of the emission lines that were fit in each galaxy is given in Table \ref{tab:fluxes}. 


\subsection{Extinction correction}

The measured emission-line fluxes were first corrected for galactic extinction using the \citet{1989ApJ...345..245C} law and then corrected for internal extinction with the Nebular Empirical Analysis Tool \citep[\texttt{NEAT},][]{2012MNRAS.422.3516W}, determined from the ratios of hydrogen Balmer lines in an iterative procedure. 
The extinction coefficient $c$H($\beta$) is initially estimated using the observed Balmer line ratios (H$\alpha$/H$\beta$ and H$\gamma$/H$\beta$) compared to their theoretical values at the electron temperature $T_e$ = 10000 K and density $N_e$ = 1000 cm$^{-3}$. A flux-weighted average of the individual ratios is adopted as the first estimate of $c$H($\beta$), which is then used to deredden the line fluxes. Electron temperatures and densities are derived from the dereddened spectrum (as described in Sect. 4.5), and a new value of $c$H($\beta$) is recalculated using the corresponding theoretical Balmer ratios. This procedure is iterated until $c$H($\beta$) converges \citep{2012MNRAS.422.3516W}.
The extinction coefficient for our sample was derived using the  Small Magellanic Cloud (SMC) extinction law, \citep{1984A&A...132..389P}\footnote{This extinction curve is the one that is implemented by default in the tool that we used to determine the ionized gas properties (see Sect. \ref{subsec:4.4}). Using this curve instead of the more recent extinction measurements by \citet{2003ApJ...594..279G} does not significantly affect our results}, 
which provides a good representation of a metal-poor interstellar medium. In most of our targets, the observed Balmer decrement (H$\alpha$/H$\beta$) was lower than the theoretical value expected for case B recombination \citep{2009A&A...506L...1A,2017ApJ...847...38Y,2024arXiv240409015S}. With the exception of target No. 3, the extinction coefficients derived from the H$\gamma$/H$\beta$ ratios are small suggesting the presence of low internal extinction (Table \ref{tab:parameters}).

The observed fluxes of the detected emission lines, $F({\lambda}$), normalized to $F$({\rm H}$\beta$) and multiplied by 100, the extinction coefficients $c({\rm H}\beta)$, and the extinction-corrected intensities $I({\lambda})$ 
normalized to $I({{\rm H}\beta})$ 
are given in Table \ref{tab:fluxes}.
The corresponding values for target ID 6, derived from the SDSS spectrum, are provided in the Appendix (Table \ref{app:tab1}).



\begin{figure}
\centering
\includegraphics[bb=0 0 500 400, clip, scale=0.70]{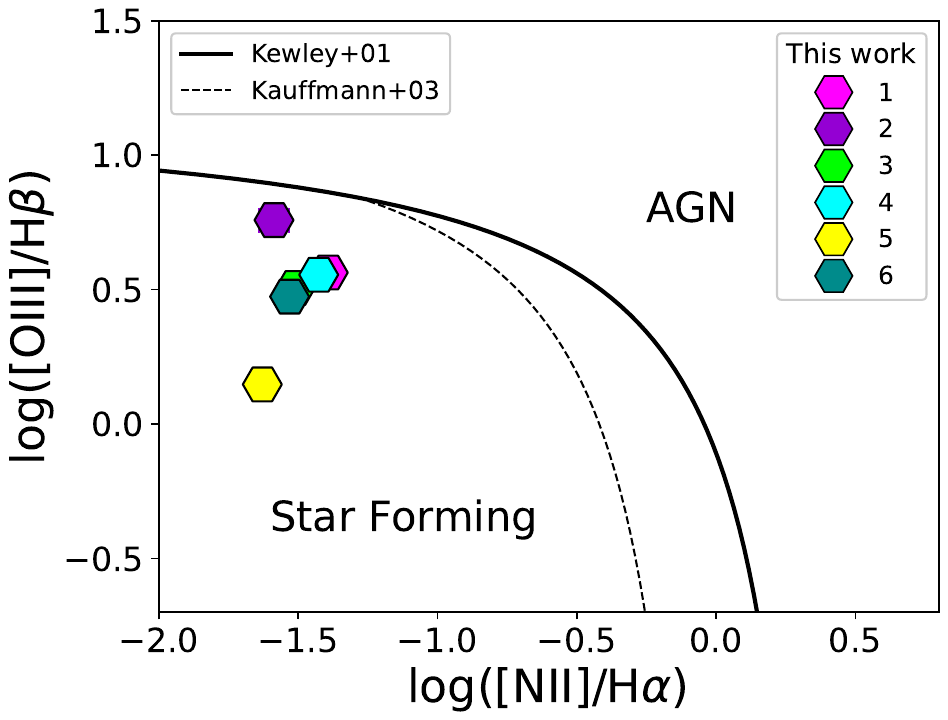}
\caption{BPT diagram of the six targets in this work, colored according to their respective ID number. 
The dashed line \citep{2001ApJ...556..121K} and solid line \citep{2003MNRAS.346.1055K}  indicate the regions where ionization is driven by star formation, AGN activity, or a combination of the two mechanisms.}
\label{fig:BPT}
\end{figure}

\subsection{BPT diagram}

Before deriving the oxygen abundances, we used
the Baldwin-Phillips-Terlevich (BPT) diagrams \citep{1981PASP...93....5B,1987ApJS...63..295V}
to investigate the
source of ionization in these galaxies and determine whether it is driven by star formation or an active galactic nucleus (AGN). AGN in dwarf galaxies are rare to be found, nonetheless an increasing number of studies are discovering intermediate-mass black holes in low-mass systems \citep{2013ApJ...775..116R,2015ApJ...813...82R,2024ApJ...974...14S,2024arXiv241100091P}. In Fig. \ref{fig:BPT}
we show the widely used emission-line diagnostic diagram  displaying [N {\sc ii}]/H$\alpha$ versus [O {\sc iii}]/H$\beta$. The
region above the solid line \citep[defined by ][]{2003MNRAS.346.1055K} indicates that an AGN is the main mechanism of ionization, the region below the dashed line \citep{2001ApJ...556..121K} corresponds to a "pure" star formation ionization mechanism, while the region in between the two curves
contain contributions from both AGN and star formation (composite galaxies). The figure shows that the six dwarfs under analysis in this work are typical star-forming galaxies with low  [N {\sc ii}]/H$\alpha$ ratios. Similar results are found if we use the [S {\sc ii}]/H$\alpha$ versus [O {\sc iii}]/H$\beta$ (not shown).

\subsection{Oxygen and nitrogen abundances}\label{subsec:4.4}




We determined the physical conditions of the star forming regions with the direct method 
\citep{1992MNRAS.255..325P,2005ApJ...631..231P,2006agna.book.....O}. Calculations of both the density and temperature of the electron gas were carried out using \texttt{NEAT} \citep{2012MNRAS.422.3516W}.

\begin{table*}[t]
\begin{center}
\caption{Electron temperatures, densities, ionization correction factors, and chemical abundances of the H {\sc ii} regions where O{\sc iii}$\lambda$4363 emission is detected.} 
\footnotesize
\begin{tabular}{lrrrrrr}
\hline \hline
\noalign{\smallskip}
 \multicolumn{1}{c}{ } &
\multicolumn{1}{c}{1} &
\multicolumn{1}{c}{2} &
\multicolumn{1}{c}{3} &
\multicolumn{1}{c}{4} &
\multicolumn{1}{c}{5} &
\multicolumn{1}{c}{6} \\
\noalign{\smallskip}
 \hline
\noalign{\smallskip}
N$_e$~[cm$^{-3}$]         &			$  100$	                & $  100$                     & $  100$		                & $  100$		             & 920$\pm$200             & $ 100$                \\[0.02cm]
T$_e$~[K]                 &     16600  $\pm$ 700            & ${15000}\pm  200 $          & ${14900} \pm 400$           & ${14200}^{+ 300}_{-300}$   & 18400$\pm$900           & 15600$\pm$1000            \\[0.05cm]
O$^+$/H$\times10^{-6}$    & 5.83$^{+1.28}_{-1.05}$          & 4.88$^{+0.66}_{-0.64}$      & 6.15$^{+1.00}_{-0.86}$      & 8.42$^{+1.05}_{-0.94}$     & 9.38$^{+1.32}_{-1.16}$  & 8.51$^{+2.77}_{-1.90}$ \\
O$^{++}$/H$\times10^{-5}$ & 3.05$^{+0.31}_{-0.28}$          & 6.13$^{+0.17}_{-0.17}$      & 3.33$^{+0.26}_{-0.24}$      & 4.34$^{+0.23}_{-0.22}$     & 0.94$^{+0.11}_{-0.09}$  & 2.79$^{+0.49}_{-0.36}$ \\
O/H$\times10^{-5}$        &  ${3.63}^{+ 0.42}_{-0.38}$      & ${ 6.62}^{+0.60}_{-0.60}$   & ${ 3.94}^{+ 0.54}_{ -0.51}$ & ${ 5.19}^{+0.62}_{-0.60}$  & 1.88$^{+0.24}_{-0.21}$  & 3.65$^{+0.75}_{-0.54}$   \\[0.05cm]
ICF(O)                    &                 1               &            1                &          1                  &         1                  &      1                  &       1                   \\[0.05cm]                 
N$^+$/H$\times10^{-7}$    & 6.16$^{+0.73}_{-0.65}$          & 4.82$^{+0.55}_{-0.54}$      & 4.68$^{+0.36}_{-0.34}$      & 7.93$^{+0.41}_{-0.39}$     & 3.71$^{+0.38}_{-0.35}$  & 6.40$^{+0.89}_{-0.70}$ \\
N/H$\times10^{-6}$        &   $3.83^{+0.51}_{-0.45}$        & $6.55^{+1.20}_{-1.00}$      & $3.00^{+0.27}_{-0.25}$      & $4.89^{+0.37}_{-0.34}$     & 0.74$\pm$0.07           & 2.75$^{+0.26}_{-0.23}$   \\[0.05cm]
ICF(N)                    &        6.2 $\pm$ 0.7            & 13.7 $\pm$ 1.6              &         6.4 $\pm$ 0.5       & ${6.2}\pm 0.5 $            &   2.00$\pm$0.05         &      4.3$\pm$0.5       \\[0.05cm]                 
12+log(O/H)               &    7.56 $\pm$   0.06	        & 7.82$\pm {0.04}$     	      & 7.59$\pm{0.06}$        	    & 7.71$\pm$ 0.05       	  & 7.28$\pm$0.05           & 7.56$\pm$ 0.08   \\[0.05cm]
log(N/O)                  &	 -0.98 $\pm$ 0.07               & -1.00 $\pm {0.09}$          & -1.12 $\pm{0.07}$      	    & -1.02$\pm 0.07$	          & -1.40$\pm$0.07          &  -1.12$\pm0.09$ \\
\noalign{\smallskip}
\hline \hline
\end{tabular}
\label{tab:parameters}
\end{center}
\end{table*}

The 
 auroral line [O{\sc iii}]$\lambda$4363 is detected in all the galaxies (Fig. \ref{fig:specs}), including the two additional objects with previous spectroscopic observations (ID 5 and 6).
Temperatures and densities were derived using an iterative process from the relevant diagnostic lines with \texttt{NEAT}.
We used the ratio
of the 
[S {\sc ii}]$\lambda$6717$,\lambda$6731 lines to estimate the electron densities, $N_{\rm e}$. For all GMOS targets  the ratio is close to or slightly above 1.4 for all galaxies, indicating that the star-forming regions lie in the low-density regime \citep{2006agna.book.....O}. We therefore ran \texttt{NEAT} assuming a constant electron density value of $N_e =$ 100 cm$^{-3}$. This assumption has a negligible effect on the derived oxygen abundances, since the direct-method abundances are primarily sensitive to $T_{\rm e}$  \citep{2006agna.book.....O}. 
The electron temperature $T_{\rm e}$([O {\sc iii}]) was calculated from the [O {\sc iii}]$\lambda$4363/($\lambda$4959 + $\lambda$5007) emission-line intensity ratio and it was used to derive the abundances of O$^{++}$ and O$^{+}$.
$T_{\rm e}$ ranges between $\sim$14\,000 K and $\sim$20000 K. Total oxygen abundances ${\rm O/H} = ({\rm O^{+}/H} + {\rm O^{++}/H})$, were obtained from the 
intensities of the [O {\sc ii}]$\lambda$7320$,\lambda$7330 and  [O {\sc iii}] lines.
%
%
The contribution of oxygen ionization states not emitting in the optical region is negligible for the total oxygen abundance \citep{2006A&A...448..955I,2013ApJ...765..140A}. \texttt{NEAT} computes the oxygen ionization correction factor, ICF(O), following \citet{1994MNRAS.271..257K}: ICF(O) = (O/H)/{[(He$^+$/He$^{++}$]/He$^+$)}$^{2/3}$. In practice, this factor is very close to unity, since both O$^+$ and O$^{++}$ are observed in our spectra.

For nitrogen, Table~\ref{tab:fluxes} and Fig.~\ref{fig:specs} show that only [N II]$\lambda$6584 is detected. \texttt{NEAT} therefore estimates the ionic abundance (N$^+$/H$^+$) from this line, and converts it into the total nitrogen abundance using
ICF(N)=(O/H)/(O$^+$/H$^+$) \citep{1994MNRAS.271..257K}.
For both N$^+$ and O$^+$ ionic abundances, we adopt T$_e$[O III]. The more appropriate low-ionization temperature (derived from N$^+$, O$^+$, or S$^+$ lines) cannot be used, since the auroral lines of these species are not detected (see Table~\ref{tab:fluxes}).
Lastly, the N/O abundance ratio is derived directly from N$^+$/O$^+$, in order to avoid relying on the ICF required to obtain the total nitrogen abundance. As noted above, ICF(N) is necessarily larger than ICF(O), that is essentially equal to unity.
Uncertainties are propagated from line flux measurements to the derived abundances with a Monte Carlo method implemented by \texttt{NEAT}. 
Table~\ref{tab:parameters} presents the total abundances, together with the ionic abundances and the ICFs used to derive the total values.


For target No. 5, \citet{2009A&A...505...63G} reported an oxygen abundance of 12 + log(O/H) = 7.35 $\pm$ 0.04. For consistency with our analysis, we used the extinction corrected emission line intensities from their work and recalculated the oxygen and nitrogen abundances with \texttt{NEAT} finding 12 + log(O/H) = 7.28 $\pm$ 0.05 (Table \ref{app:tab1}). 
Galaxy No. 6  was observed by the SDSS; we extracted the emission line fluxes as described above and derived the oxygen abundance 
using the same [O {\sc iii}] and [O {\sc ii}] lines that were observed in the GMOS spectra (see Table \ref{app:tab1}).
We find oxygen abundances in the range $7.28 <$ 12 + log(O/H) $< 7.82$ (i.e. between 4\% and 13\% of the solar value), confirming that the galaxies are metal poor. Electron temperatures, electron 
densities, oxygen and nitrogen abundances, ICF factors are displayed in Table \ref{tab:parameters}.

\subsection{The nitrogen-to-oxygen ratio}

\begin{figure}
\centering
\includegraphics[bb=0 0 500 400, clip, scale=0.70]{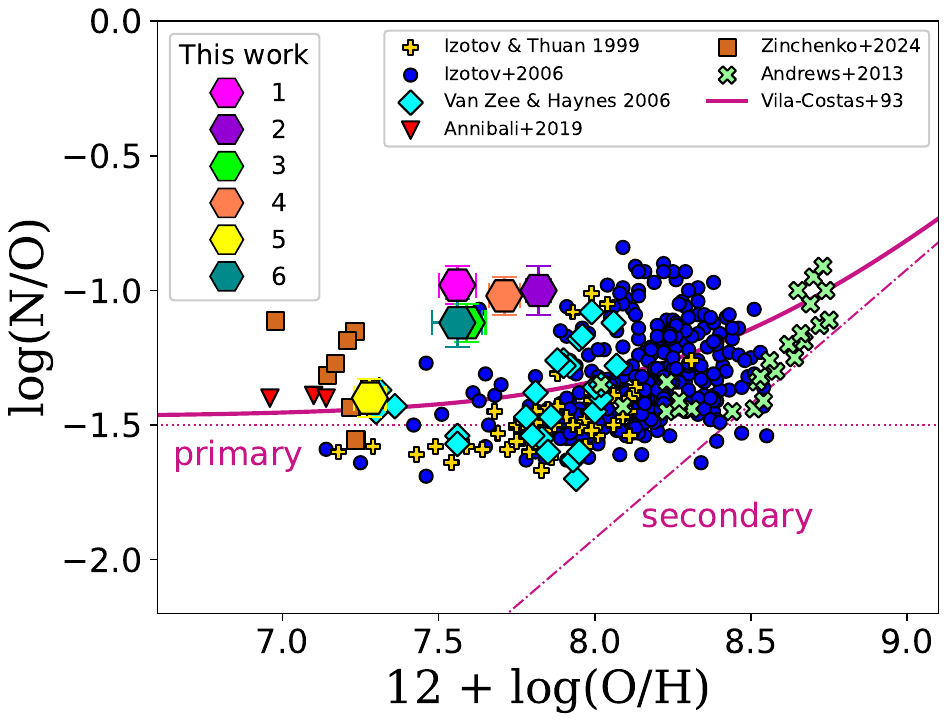} 
\caption{N/O versus oxygen abundance. The S-PLUS galaxies observed with GMOS (hexagons, colored according to their respective ID number) are compared to other samples with available measurements of the oxygen abundances derived with the direct method: plus \citep{1999ApJ...511..639I}, circles \citep{2006A&A...448..955I}, diamonds \citep{2006ApJ...636..214V}, 
crosses \citep{2013ApJ...765..140A}, 
downward triangles \citep{2019MNRAS.482.3892A}, squares \citep{2024A&A...690A..28Z}.  We also display the model of \citet{1993MNRAS.265..199V} showing the primary (dot-dashed line) and secondary (dashed line) nitrogen formation components and their combination (solid line).} 
\label{fig:NO}
\end{figure}

The nitrogen-to-oxygen ratio, N/O, is considered  an indicator of the chemical evolutionary age of a galaxy.
While oxygen is mostly formed by primary nucleosynthesis in massive stars ($>\,10$ M$_{\odot}$) and injected into the ISM  by core-collapse supernovae
\citep{2013ARA&A..51..457N},
nitrogen production is more complex and it can have both a primary and secondary origin.
Secondary nitrogen synthesis can occur in stars with masses around or above 1.3 M$_{\odot}$ as part of the CNO cycle of oxygen and carbon during hydrogen burning. However stellar evolution models predict that it is mostly produced in intermediate-mass stars (4 - 8 M$_{\odot}$) and released in the ISM during the asymptotic giant branch phase \citep[AGB,][]{1981A&A....94..175R,1995ApJS..101..181W,2002A&A...381L..25M}.
At low metallicities, nitrogen is produced from carbon and
oxygen created via the helium burning process; in
this case, nitrogen behaves like a primary element, its abundance being proportional to that of the other primary heavy elements
resulting in a ﬁxed ratio of N/O, independently of the oxygen abundance.
\citet{1999ApJ...511..639I} identified that the average value of the N/O ratio of extremely metal-poor galaxies (12 + log(O/H) $\lesssim$ 7.6) is approximately constant with a very small dispersion $< \log$(N/O) $> = -1.6 \pm 0.02$, the so-called N/O plateau (Fig. \ref{fig:NO}). This was interpreted as the evidence of nitrogen being produced as a primary element by massive stars at low O/H.
At high metallicities (12 + log(O/H) $\gtrsim 8.3$), the ISM metal abundance becomes high
enough that the production of secondary nitrogen and N/O increases linearly with oxygen abundance \citep{1980FCPh....5..287T,1985ESOC...21..155P,1993MNRAS.265..199V,2012MNRAS.421.1624P, 2013ApJ...765..140A, 2016MNRAS.458.3466V}.

%

More recent studies found that 
galaxies with low oxygen abundances show an excess N/O \citep{2006ApJ...636..214V,2013AJ....146....3S,2024ApJ...962...50W,2024A&A...690A..28Z}, indicating that other factors 
can affect the ratio at low O/H.
For example, strong winds carrying nitrogen-enriched gas from
post-main sequence hot and massive Wolf-Rayet stars can raise N/O
\citep{1989MNRAS.239..297W,2001A&A...373..555M,2007ApJ...656..168L,2014A&A...561A..64T}.
The increase in the dispersion could also be explained if the galaxy is going through a post-starburst phase: after
the production of carbon and primary nitrogen by massive
stars during the last major event of star formation, there is a delayed additional
contribution of primary nitrogen produced by intermediate-mass stars
in the post-starburst phase, unbalanced by the oxygen release of core-collapse supernovae \citep{1990ApJ...363..142G,2006ApJ...636..214V,2019MNRAS.482.3892A}.
Another possibility is the rapid infall of metal-poor gas from the intergalactic medium that mixes with the ambient more metal-rich ISM \citep{2005A&A...434..531K}.  This process dilutes the oxygen abundance (by increasing the hydrogen content in O/H),  without significanlty altering the N/O ratio. As a result, the mixed gas will be characterized by an apparent excess in the N/O -- O/H plane \citep{2016ApJ...819..110S,2021ApJ...908..183L}.

Although probably small, another source of dispersion in the N/O versus O/H relation is the fact that oxygen can also
be produced in low-metallicity galaxies through the third dredge-up in AGB stars, mainly at 12+log(O/H) $\leq$ 7.7. This has been observed in 
the ISM of Local Group irregular dwarf galaxies  \citep{2009MNRAS.398..280M}. However, it is not possible to quantify this effect in galaxies where only H~{\sc ii} regions are detected, as in our study.


In Fig. \ref{fig:NO}, we show the N/O ratios of our targets compared to samples of star-forming galaxies from the literature with different ranges of metallicity \citep{1999ApJ...511..639I,2006A&A...448..955I,2006ApJ...636..214V,2013ApJ...765..140A,2019MNRAS.482.3892A,2024A&A...690A..28Z} and the model of \citet{1993MNRAS.265..199V}. This figure  shows the N/O plateau identified by \citet{1999ApJ...511..639I} in extremely metal-poor galaxies (plus symbols), although more recent studies, also displayed in the figure, present a larger scatter at low metallicities raising the question about the actual existence of a plateau in this regime \citep{2006ApJ...636..214V,2024A&A...690A..28Z}. %

\begin{table*}
\begin{center}
\caption{Comparison between H$\alpha$ fluxes from spectroscopic (GMOS) and photometric (three-filter method) measurements and star formation rates obtained with GMOS observations and SED fitting.}
\small
\begin{tabular}{cccccc}
\hline \hline
\noalign{\smallskip}
ID & $\log$($F_{{\rm H}\alpha}^{\rm spec}$) & $\log$($F_{{\rm H}\alpha}^{\rm phot}$) & $\log$(SFR)$^{\rm spec}$  & aperture & $\log$(SFR)$^{\rm SED}$\\
 & [erg s$^{-1}$ cm$^{-2}$]  & [erg s$^{-1}$ cm$^{-2}$] & [M$_{\odot}$ yr$^{-1}$]   & [$^{\prime\prime}$] & [M$_{\odot}$ yr$^{-1}$] \\
\noalign{\smallskip}
\hline \hline
1 & -13.753$\pm$ 0.002 & -13.66 & -1.04 $\pm$ 0.02 &  7.9  & -1.32 $\pm$ 0.02\\
2 & -13.728$\pm$ 0.001 & -13.85 & -1.09 $\pm$ 0.02 &  7.8  & -1.46 $\pm$ 0.02\\
3 & -14.211$\pm$ 0.002 & -14.11 & -2.12 $\pm$ 0.03 &  10.5 & -2.29 $\pm$ 0.02\\
4 & -13.891$\pm$ 0.001 & -13.53 & -1.99 $\pm$ 0.03 &  13.7 & -2.05 $\pm$ 0.02\\
5 & $\:\,$-12.936$\pm$ 0.007$^a$ & -13.30 & -1.99 $\pm$ 0.01 &  --   & -2.50 $ \pm$ 0.02\\
6 & -13.814$\pm$ 0.004 & -13.50 & -2.24 $\pm$ 0.07 &  3   & -2.38 $\pm$ 0.02\\
\noalign{\smallskip}
 \hline\hline
\end{tabular}
\label{tab:FHa}
\end{center}
\footnotesize{$^a$ \citet{2009A&A...505...63G}}
\end{table*}

We measured the nitrogen abundances and derived the N/O ratios for the six galaxies of our sample. 
We find   $-1.40 <$ $\log$(N/O) $< - 0.98$: five out of six dwarfs 
show a higher than expected N/O.
Particularly, ID 1 is the galaxy with the largest N/O (log(N/O) = -0.98$\pm 0.07$), which would be more typical of a galaxy with
a ten times larger metallicity.
We did not detect the He {\sc ii} $\lambda$4686 line, which would suggest the presence of Wolf-Rayet stars in any of the dwarfs. However, a detailed analysis of the N/O ratio and its dependence on oxygen abundance is beyond the scope of this paper. Further observations are required to detect additional nitrogen lines, which would allow a more reliable determination of the oxygen abundances in these galaxies.





\begin{figure}
\includegraphics[bb=0 0 460 460, clip, scale=0.2695]{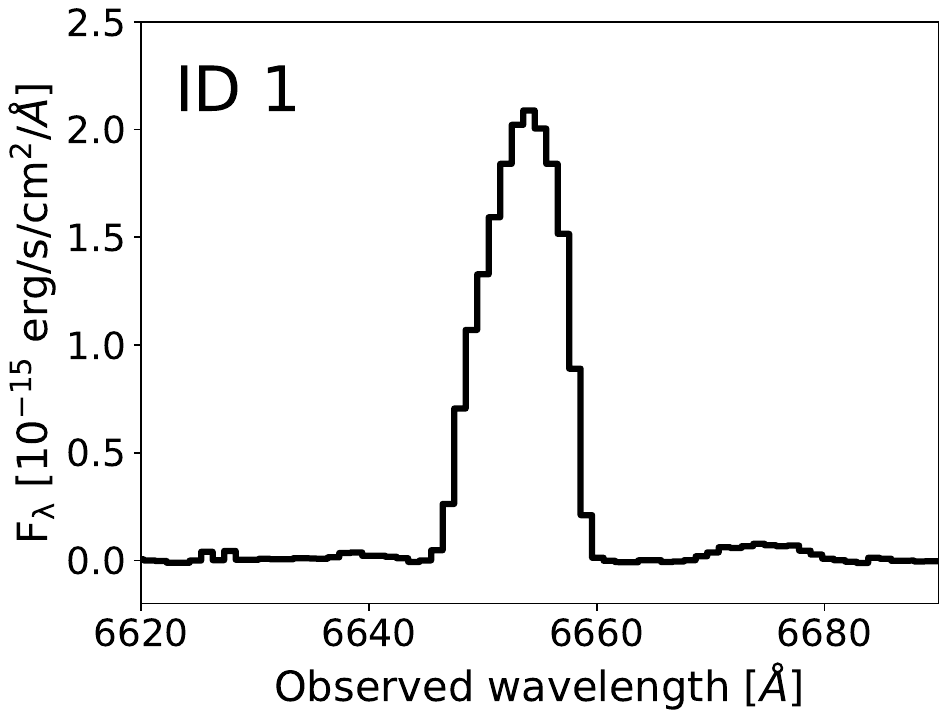}
\includegraphics[bb=0 0 460 460, clip,scale=0.2695]{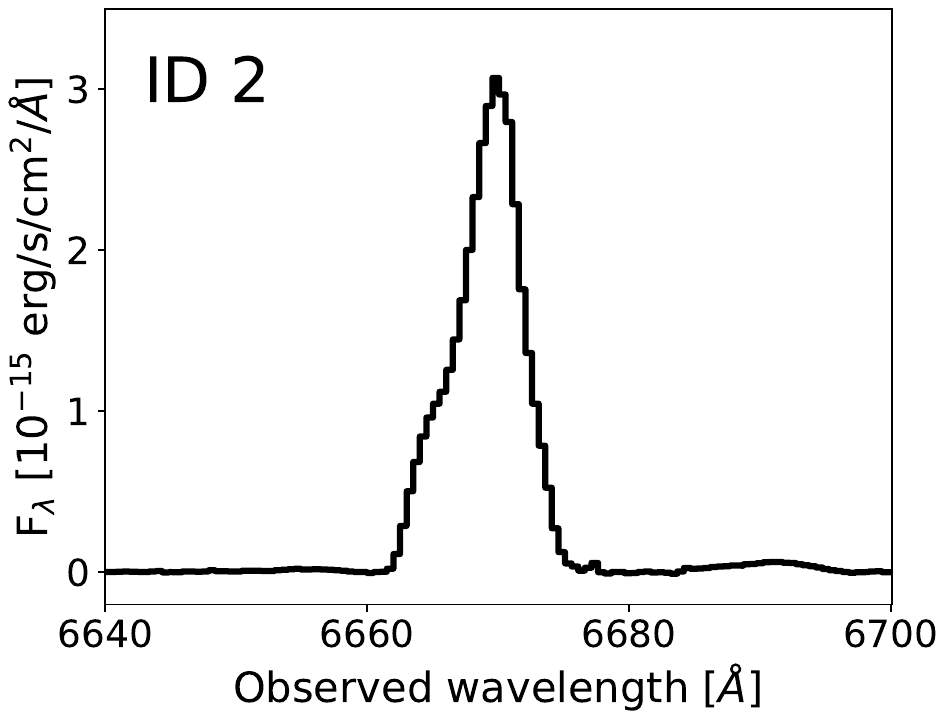}
\includegraphics[bb=0 0 460 460, clip,scale=0.2695]{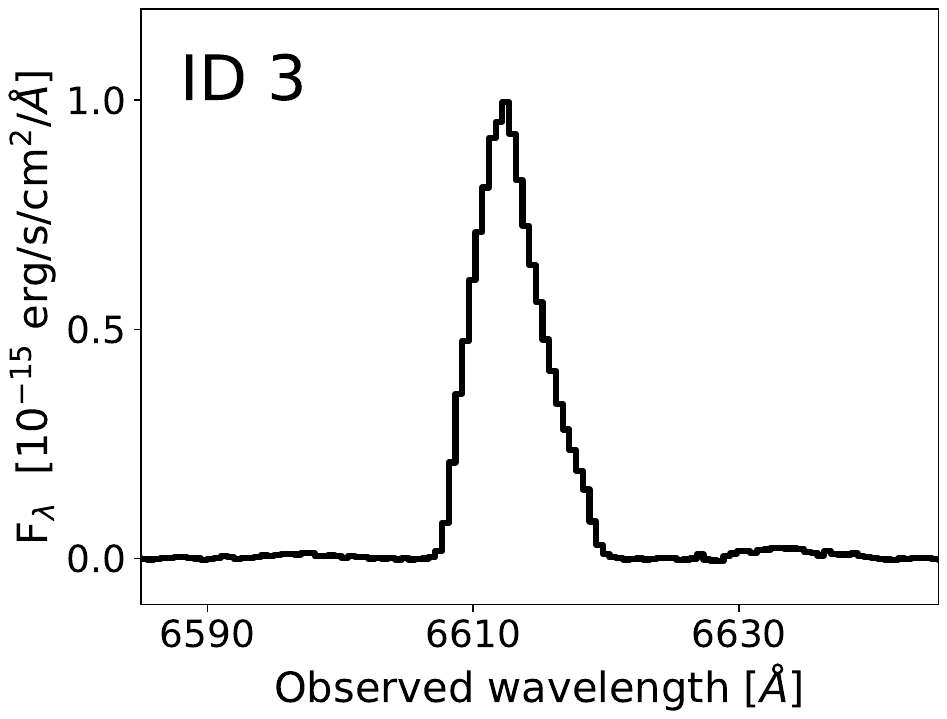}
\includegraphics[bb=0 0 460 460, clip,scale=0.2695]{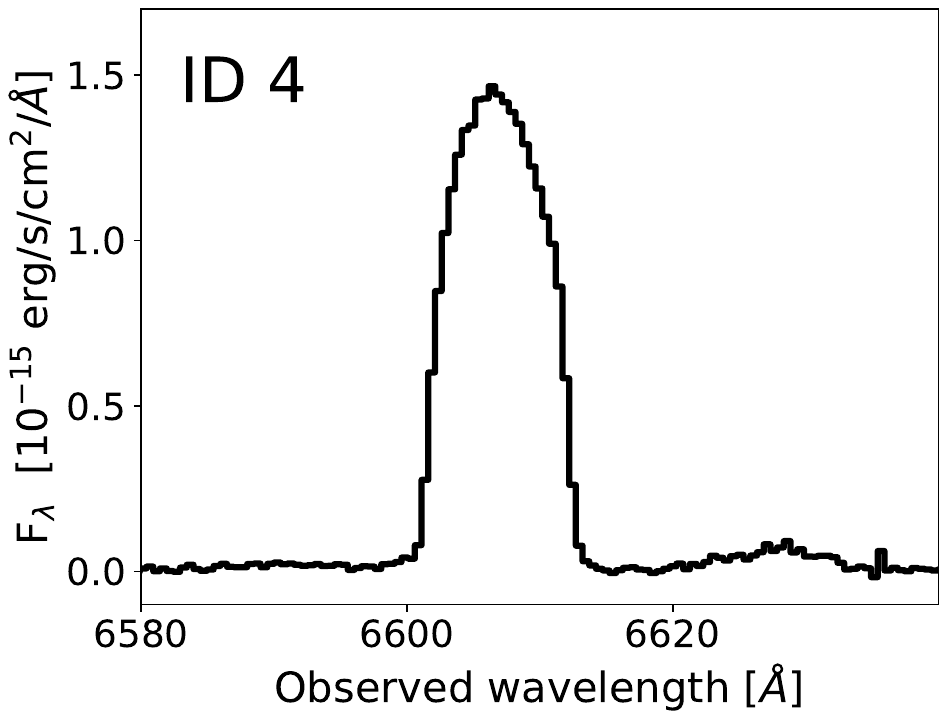}
\caption{H$\alpha$ line profiles of the four targets observed with GMOS, summed over the extent of the emission from all the star-forming knots. 
}
\label{fig:halpha_prof}
\end{figure}


\subsection{H$\alpha$ line profile fitting and star formation rates}\label{subsec:4.5}


To derive the star formation rates of our
targets, we extracted 1D spectra using larger apertures than those employed to estimate the oxygen abundances, in order to include the entire H$\alpha$ emission
detected in the GMOS spectra (see Fig. \ref{fig:halpha_struct}). The aperture sizes vary between 7$\farcs$8 and 13$\farcs$7 (Tab. \ref{tab:FHa}).
Figure \ref{fig:halpha_prof} shows the region of the spectra centered around the 
H$\alpha$ line, 
obtained for these larger apertures.
H$\alpha$ fluxes, displayed in Table \ref{tab:FHa}, were obtained 
by integrating the line profiles after subtracting the continuum  with \texttt{PySpecKit} \citep{2022AJ....163..291G}.


\begin{figure}
\centering
\includegraphics[bb= 0 0 500 400, clip, scale=0.70]{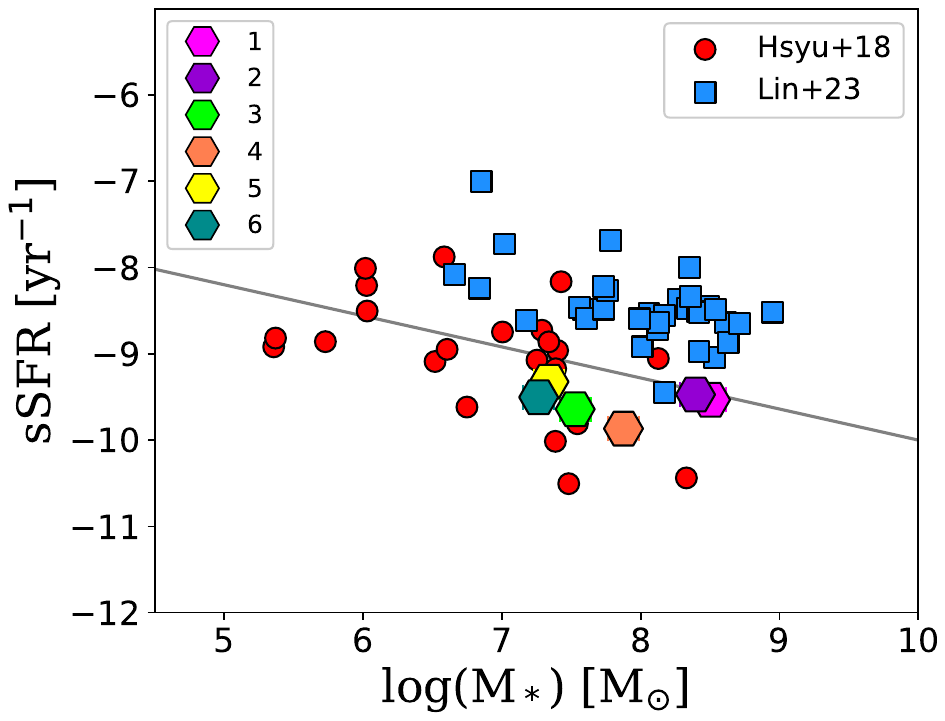} 
\caption{Specific star formation rate versus stellar mass. The S-PLUS galaxies (star symbols) are compared to metal-poor dwarfs selected in other photometric surveys such as the SDSS (H18, circles) and DES (L23, squares).}
\label{fig:ssfr}
\end{figure}



We calculated the H$\alpha$ luminosity for each galaxy using the
H$\alpha$ line fluxes 
listed in Table \ref{tab:FHa}, the extinction coefficients given in Table \ref{tab:fluxes}, and the redshifts provided in Table \ref{tab:properties}. H$\alpha$ luminosities are in the range
$10^{39} < \log$(LH$\alpha$) $< 10^{40}$ erg s$^{-1}$. The H$\alpha$ luminosities were converted into SFR using the calibration of \citet{2012ARA&A..50..531K} for a \citet{2001MNRAS.322..231K} IMF:

\begin{equation}
    \log (SFR) \, [\rm{M}_{\odot} \rm{yr}^{-1}] = \log \it{L}(\rm{H}\alpha) - 41.27,
\end{equation}
where $L(\rm{H}\alpha)$ is in units of erg s$^{-1}$.

Although metal-poor galaxies can have starburst-level SFRs \citep{2019MNRAS.483.5491I,2024MNRAS.527..281I},
our sample shows normal star formation activities 
with SFRs ranging from 0.01 to 0.1~M$_{\odot}$~yr$^{-1}$ and specific star formation rates (sSFRs) around 10$^{-10}$~yr$^{-1}$. 
Figure \ref{fig:ssfr} compares sSFRs and stellar masses of our sample with those estimated by H18 and L23. 
Stellar masses have been derived using the same methodology for all three samples, i.e., from the r - i colour described in Eq. 1. While the H18 stellar masses were already estimated in this way, we have recalculated the stellar masses of the L23 sample using the SDSS photometry provided in this work.
These two samples cover a wider range of these parameters. Particularly, the selection criteria of H23 are such that their sample is more biased towards higher redshift, more active metal-poor galaxies. The
solid line shows the star formation sequence defined by \citet{2007ApJS..173..315S} in the SDSS as a reference for the overall trend between the two parameters.

As can be seen in Fig. \ref{fig:ssfr},
lower mass galaxies have higher sSFRs, which are consistent with the “downsizing” scenario \citep{1996AJ....112..839C} predicting that lower mass galaxies are more gas-rich and capable of sustaining significant star formation activity at present epoch.
The S-PLUS targets have sSFRs that are compatible with their stellar masses, according to the relation of  \citet{2007ApJS..173..315S}, but overall they appear less active compare to the other samples of metal-poor dwarfs.
The SFRs of our targets may be underestimated since the slits cover only a small fraction of the galactic discs. For comparison we derive the H$\alpha$ fluxes from the J0660 narrow-band filter following the three-filter method of \citet{2015A&A...580A..47V}.
The method uses the $r$ and $i$ bands to trace the linear continuum and the
J0660 filter to identify the line emission. It has already been implemented in similar data from the S-PLUS and J-PLUS surveys \citep{2015A&A...580A..47V,2021MNRAS.500.1323L,2024MNRAS.532..270G,2024MNRAS.530.3787S}. The method provides a H$\alpha$ + [N {\sc ii}] flux estimate, but from the spectra we can see that the contribution of [N {\sc ii}] is always negligible compared to the the H$\alpha$ flux, thus, we do not apply any correction for [N {\sc ii}] contamination as suggested in \citet{2015A&A...580A..47V}. The fluxes obtained in this way are within at most a factor of 2 compared to those obtained from the GMOS spectra, implying that we are not missing a large fraction of the H$\alpha$ emission outside the slits. Comparison with the output of the SED fitting procedure discussed in Sect. \ref{subsec:SED} shows similar results.
The SFRs derived from the GMOS spectra agree within a factor of 3 with  those estimated from SED fitting (Table \ref{tab:FHa}).


\begin{figure}
\centering
\includegraphics[bb= 0 0 500 400, clip, scale=0.49]{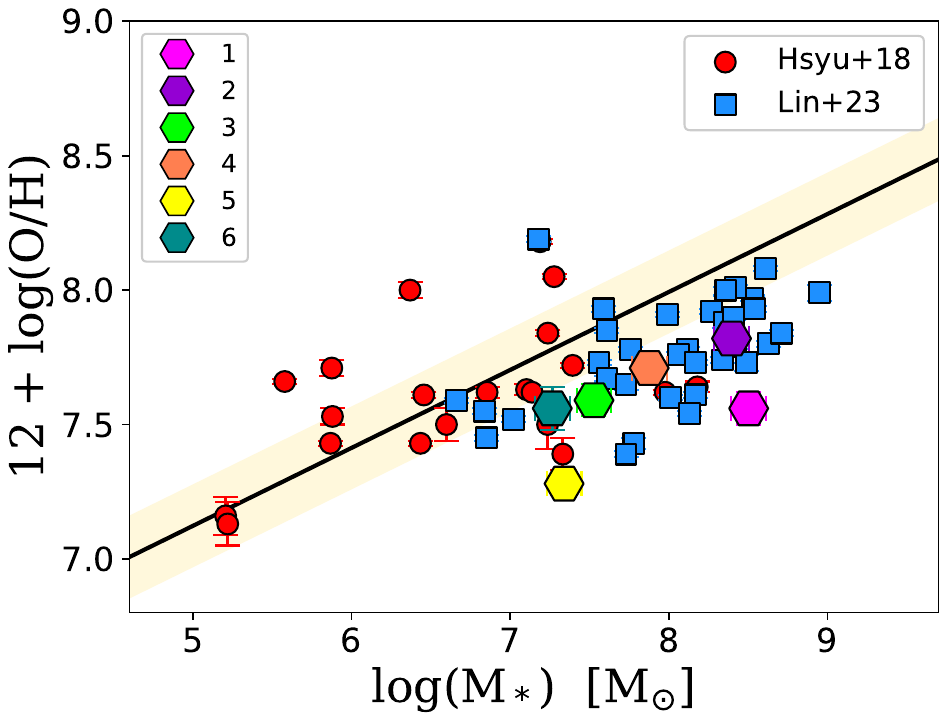}
\includegraphics[bb= 0 0 500 400, clip, scale=0.49]{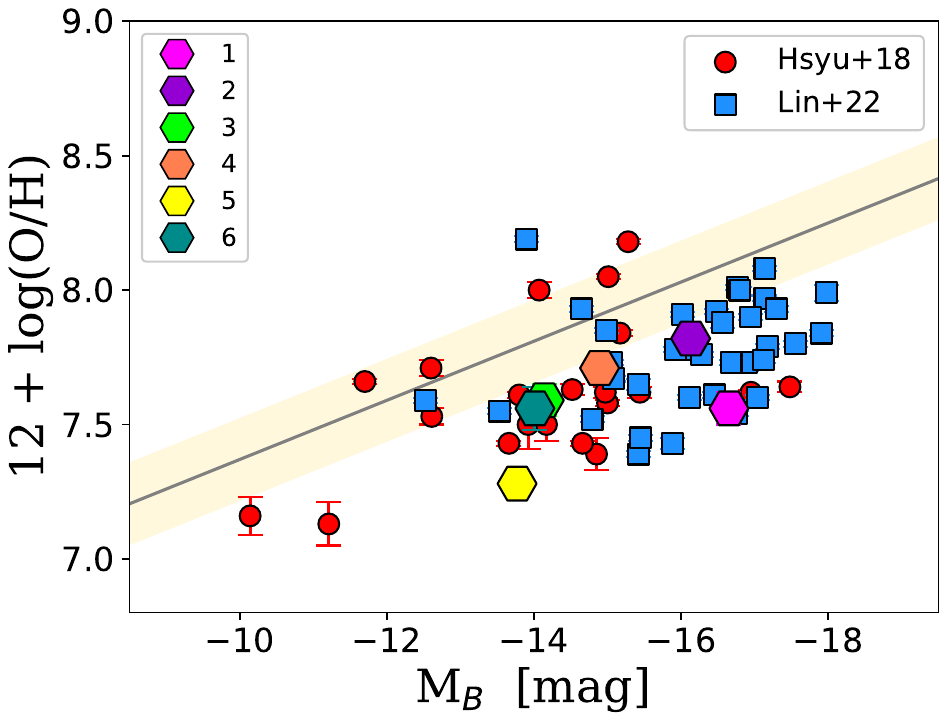}
\caption{Left: Mass-metallicity relation for our sample of metal-poor dwarfs (hexagon symbols) compared to galaxies selected in different photometric surveys - SDSS (circles) and DES (squares) using the same color cuts. The galaxies are compared to the relation found in the Local Volume Legacy (LVL) survey \citep[black line]{2012ApJ...754...98B}, rescaled to a Kroupa IMF. The galaxies selected from the S-PLUS survey (hexagons) show an offset larger than 1$\sigma$ (shaded region) from the main relation of the LVL galaxies. ID 1 and 5 are the most extreme outlier displaying the largest offset from the MZR. Stellar masses and oxygen abundances have been derived using the same methodology for all three samples.  Right: Luminosity-metallicity relation for the three samples compared to the LVL relation (black line).}
\label{fig:MZR}
\end{figure}

\subsection{Mass-metallicity and luminosity-metallicity relations}\label{subsec:4.6}

The relation between the stellar mass
of galaxies and the metallicity of their interstellar medium, known as the mass–metallicity relation \citep[MZR,][]{1979A&A....80..155L}, is the result of the interplay between the processes that drive and regulate the baryon cycle within galaxies: star formation, stellar feedback, and gas accretion \citep{2011MNRAS.416.1354D,2013ApJ...772..119L}. The relation holds for a broad range of stellar masses, from $\sim 10^7$ M$_{\odot}$ to $\sim 10^{12}$ M$_{\odot}$.  It is steeper at low $M_*$, then it flattens at a characteristic stellar mass
reaching an asymptotic saturation metallicity \citep{2004ApJ...613..898T,2020MNRAS.491..944C}.
The scatter in the MZR has been observed to correlate with
different galaxy properties.
\citet{2008ApJ...672L.107E} reported that
galaxies with higher sSFRs or larger half-light radii present systematically lower gas-phase metallicities for a given stellar mass.
\citet{2010MNRAS.408.2115M} and \citet{2010A&A...521L..53L}
showed that the MZR has a secondary dependence on the star formation rate, finding  
an anticorrelation between SFR and metallicity at fixed stellar mass (i.e. galaxies with higher SFRs tend to have lower metallicities).
They proposed that local galaxies are distributed in a three-dimensional space defined by mass, metallicity and SFR (the fundamental metallicity relation).
Nonetheless, other studies based on integral field unit observations did not find any significant dependence of the observed MZR on the SFR \citep{2017MNRAS.469.2121S,2017ApJ...844...80B}, and no consensus has been reached so far.

\begin{figure}
\centering
\includegraphics[bb= 0 0 500 400, clip, scale=0.70]{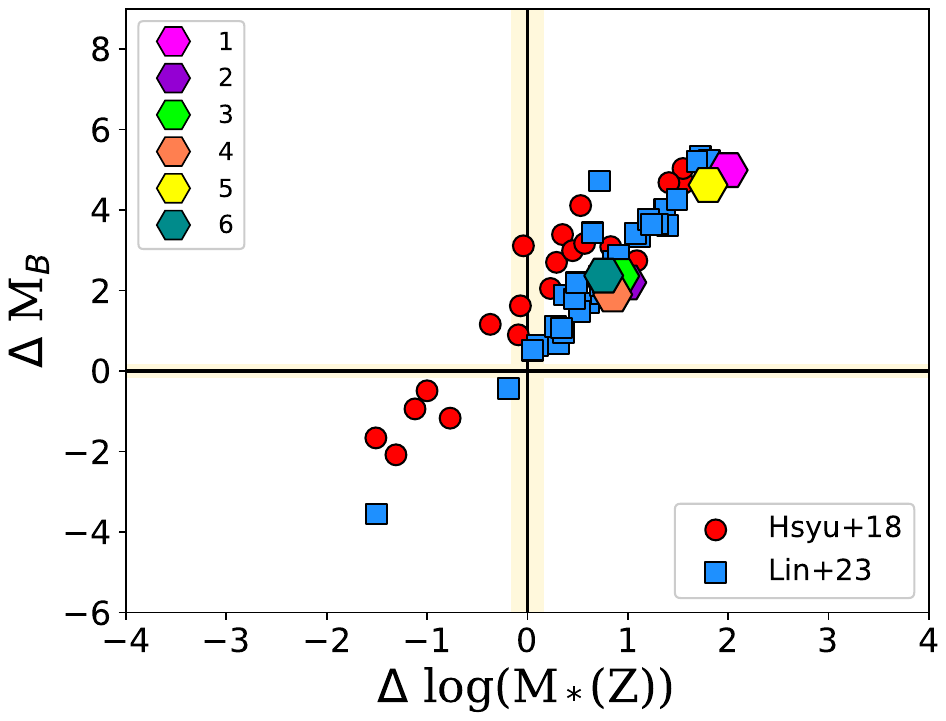}
\caption{Offsets from the mass-metallicity (x axis) and from the luminosity metallicity relations for the sample of dwarf galaxies selected in different photometric surveys (SDSS, DES, S-PLUS). The offsets are calculated from the relation found in the Local Volume Legacy \citep{2012ApJ...754...98B}. The yellow shaded area shows the 1$\sigma$ scatter of the $L-Z$ and $M-Z$ relation in the LVL. Three out of six galaxies selected from S-PLUS (hexagons) show the largest offsets in both relations compared with the galaxies in the other surveys. }
\label{fig:offsets}
\end{figure}

The behaviour of the MZR at low mass has been explained in different ways: 
either low-mass galaxies have been less efficient at producing metals \citep{2007ApJ...655L..17B}, or gas outflows due to supernova feedback
are much more efficient in ejecting metals out of low-mass systems, resulting in
a lower effective yield with decreasing stellar mass
\citep{1986ApJ...303...39D,2017MNRAS.472.3354D}.
\citet{2014MNRAS.444.1705G} studied the MZR at the low-mass regime using planetary nebulae as tracers of the oxygen abundance. They found that both dwarf irregular and dwarf spheroidal galaxies in the Local Group follow the same MZR, and that their MZR is also consistent with the global MZR of star-forming galaxies  \citep[see also Figs. 6 and 7 in][]{2019IAUS..344..161G}.

Having obtained the oxygen abundances, in Fig. \ref{fig:MZR} we investigate the MZR for these galaxies  
and infer whether they follow the general trend observed in galaxies of the Local Volume Legacy (LVL) survey \citep{2012ApJ...754...98B}.
Our sample of galaxies (hexagon symbols), as well as other objects from the comparison samples, appear to be offset from the main relation defined by the LVL (black line in Fig. \ref{fig:MZR}).
The two most metal-poor galaxies of our sample, ID 1 and 5, are the most extreme outliers.
Particularly, the estimated oxygen abundance of ID 1 ($\sim$ 1/20th the solar value) would be typical of a galaxy with a stellar mass of  1.6 $\times$10$^6$ M$_{\odot}$, almost 200 times smaller than the value we measured (Tab. \ref{tab:properties}, Fig. \ref{fig:MZR}).

The presence of metal-poor outliers at a given stellar mass is in general attributed to
external events. This can happen if star formation occurs in
pockets of pristine gas accreted from the intergalactic medium \citep{2016IAUS..308..390S,2017MNRAS.472.3354D,2022ApJ...938...96J,2023MNRAS.522.2089D}, or if it is triggered by a merger/interaction of a close galaxy pairs \citep{2008AJ....135.1877E,2012A&A...537A..72L,2020MNRAS.498.1939G}.
In these scenarios, the metallicity of the ISM will be diluted and the galaxy would appear as an outlier in the MZR relation as well as in the luminosity-metallicity relation (LZR). Such an effect would be even stronger in the LZR \citep{2010MNRAS.406.1238E,2020ApJ...891..181M} due to the high luminosity of young massive stars and the contribution of strong emission lines, making the star-forming dwarfs brighter in the $B$ band when compared to galaxies of similar mass with a lower star formation activity. In the right panel of Fig. \ref{fig:MZR} we display the luminosity metallicity relation of all the samples. Following \citetalias{2023ApJ...951..138L} we
derive the luminosity of the galaxies in the $B$-band using the empirical relation of \citet{2014MNRAS.445..890C} based on $g$ and $i$ magnitude:

\begin{equation}
B = i + (1.27 \pm 0.03) \times (g - i) + (0.16  \pm 0.01) .
\end{equation}

The galaxies are compared to the LVL survey and, as expected, the offset from the main relation is even more prominent in the luminosity-metallicity relation (right panel of Fig. \ref{fig:MZR}). 
Outflows due to supernova feedback ejecting enriched gas from the
star-forming regions could generate metal losses in low-mass systems \citep{2015ApJ...815L..17M,2019ApJ...886...74M}. However, given the low star formation rates
observed, it is less likely that strong outflows due to stellar
feedback may be causing the observed metal deficiency \citep{2020MNRAS.498.1939G}.

To better illustrate the discrepancy in the MZR and the LZR we
show in Fig. \ref{fig:offsets} the luminosity and
stellar mass offsets of the three samples from Fig. \ref{fig:MZR}. We deﬁne the offsets ($\Delta M_*$, $\Delta M_B$) as
the horizontal distance of each object from the MZR and LZR deﬁned by the relation obtained for the LVL galaxies\footnote{We calculate $B$ magnitudes using the empirical color transformation of \cite{2014MNRAS.445..890C}: $B$ = $i$  + 1.27 $(g  - i)$ + 0.16}. $\Delta M_*$ ($\Delta M_B$) represents the logarithmic difference between the observed stellar mass ($B$ absolute magnitude) of the galaxy and the expected stellar mass ($B$ absolute magnitude) for a given metallicity, if it followed the LVL mass (luminosity)-metallicity relation (see \citetalias{2023ApJ...951..138L} for details).
Several galaxies from the \citetalias{2023ApJ...951..138L} and \citetalias{2018ApJ...863..134H} samples exhibit small or negative scatter from the MZR, 
however, almost half of the galaxies of both samples are located on the upper right side of the plot, with the S-PLUS galaxies being among the most extreme outliers in both relations.
DR4\_3\_SPLUS-s27s07\_0011404 (ID 1), the most metal-poor galaxy among the targets observed with GMOS (see Table~\ref{tab:parameters}), is the most extreme outlier, lying at the top-right corner of Fig. \ref{fig:offsets}.
This galaxy and target No. 5, seem to have a fainter nearby companion at a projected distances of 3 and 2~kpc, respectively (see Fig.~\ref{fig:img}). 
However, while in the first case the two systems appear to be interacting, in the second case  it is not clear from the Legacy image whether the faint galaxy is actually a companion or just a background object.

A detailed study of the environment of the sample of dwarf galaxies selected in this work is beyond the scope of this paper. Preliminary analysis, using the density catalogue of galaxies in the S-PLUS survey based on the $k$-th nearest neighbor method (Lopes-Silva et al., in prep.), indicate that these systems mostly reside in low-density environments.
This will, however, be discussed in more detail in a future work (Gonçalves et al., in prep.).

\begin{figure}
\includegraphics[bb= 0 0 460 400, clip, scale=0.55]{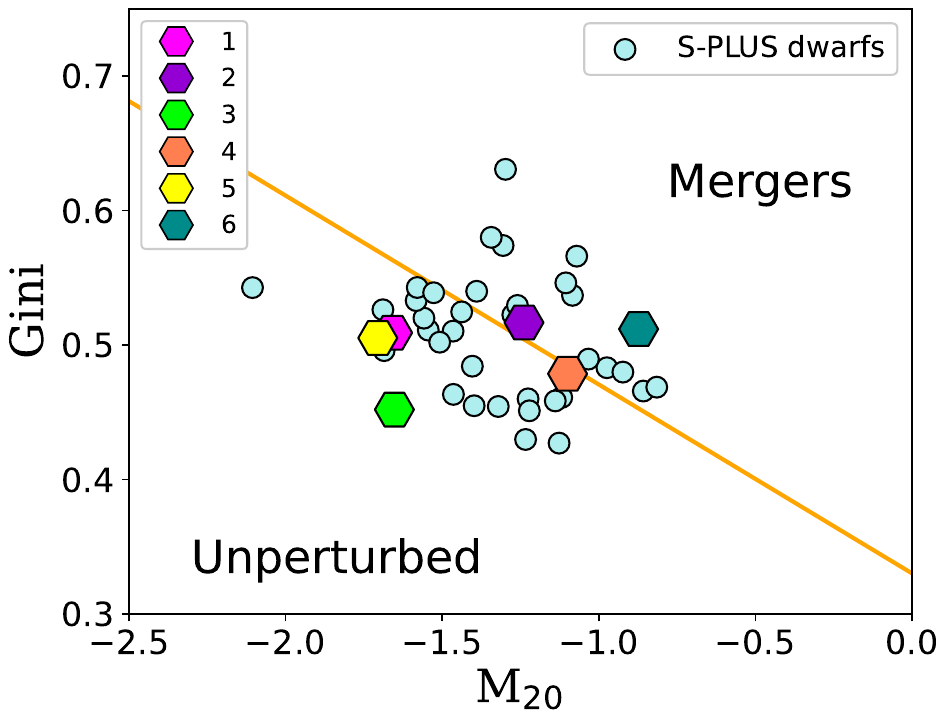}
\includegraphics[bb= 0 0 460 400, clip, scale=0.55]{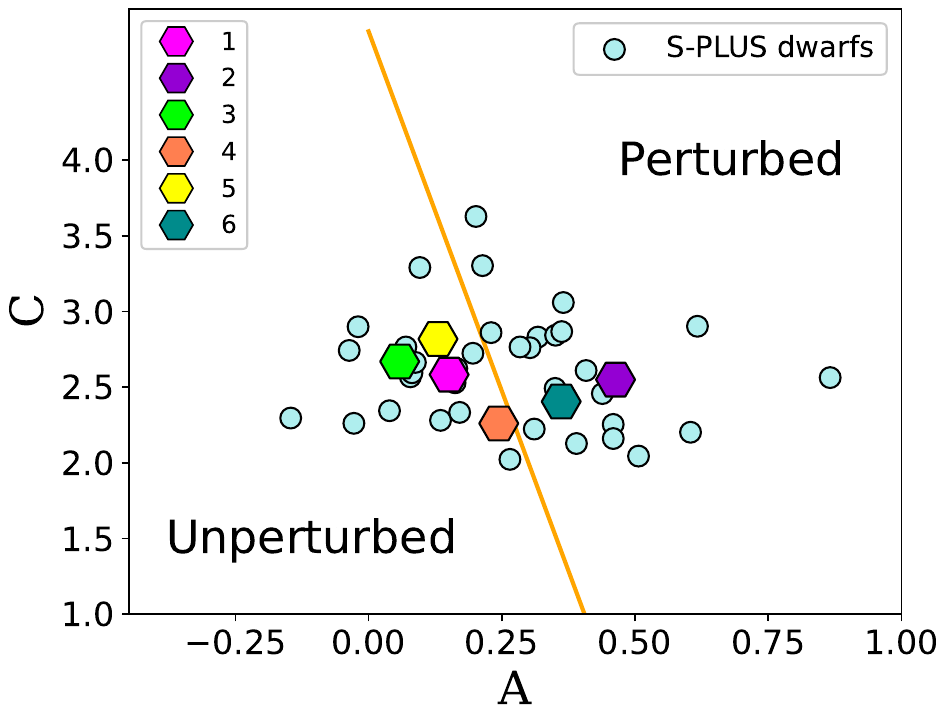}
\caption{Gini-$M_{20}$ diagram (top panel) and  $C$ - $A$ diagram (bottom panel) for our S-PLUS sample including all 47 metal-poor dwarf galaxy candidates (circles). Hexagon symbols highlight the galaxies followed-up with GMOS. High Gini and $M_{20}$ values imply that a galaxy is in a merger phase. Similarly, high concentration and asymmetry values indicate perturbed morphologies.
The yellow lines show the separation between mergers (or perturbed) and non-interacting galaxies as established by \citet{2008MNRAS.391.1137L} and \citet{2024MNRAS.528.1125K} for the Gini-$M_{20}$ and $C$-$A$ diagrams, respectively.  Galaxy 2 and 6 lie above the threshold values that separate major-merger (and perturbed) candidates from non-interacting galaxies in both diagnostic diagrams. 
}
\label{fig:morphology}
\end{figure}

\subsection{Optical morphologies}

As a further test, we looked for evidence of perturbations in the optical morphologies that may give hints on galaxy-galaxy interactions or mergers.
We analyzed the $g-$band images of all 47 galaxies of our sample (see Sect. \ref{sec:2.1}) which were taken from the DESI Legacy Imaging Survey. 
Following \citet{2024MNRAS.528.1125K} 
we built the diagnostic diagrams involving combinations of
non-parametric morphological  parameters such as the
Gini coefficient ($G$), the second moment of the brightest pixels of a
galaxy containing 20 per cent of the total flux \citep[$M_{20}$;][]{2004AJ....128..163L}, and the concentration–asymmetry–smoothness system \citep[CAS; see][for details]{2003ApJS..147....1C},  which were obtained  using the {\sc astromorphlib} software developed by \citet{hernandez22}. These diagnostics have been
extensively applied to quantify galaxy morphologies and to
identify systems with signatures of recent or ongoing mergers.
Particularly, the position of a galaxy in the
$G  - M_{20}$
plane allows us to separate major mergers from non-interacting galaxies. 
Mergers occupy the region defined by the following relation,
$S(G, M_{20})$ = 0.14 $M_{20} + G - 0.33 > 0$ \citep[][solid line in the top panel of Fig. \ref{fig:morphology}]{2008MNRAS.391.1137L}.
Recently, \citet{2024MNRAS.528.1125K} investigated the correlation between CAS morphological parameters to identify galaxies with perturbed morphologies that could be tracing either tidal interactions or ram pressure stripping effects. They identified the $C - A$ diagram as the best diagnostic to separate perturbed  from unperturbed galaxies and the separation between these two regions occur at $C$ = $-$9.5 $A +$ 4.85 (solid line in the bottom panel of Fig. \ref{fig:morphology}).

\begin{figure}
\centering
\includegraphics[scale=0.235]{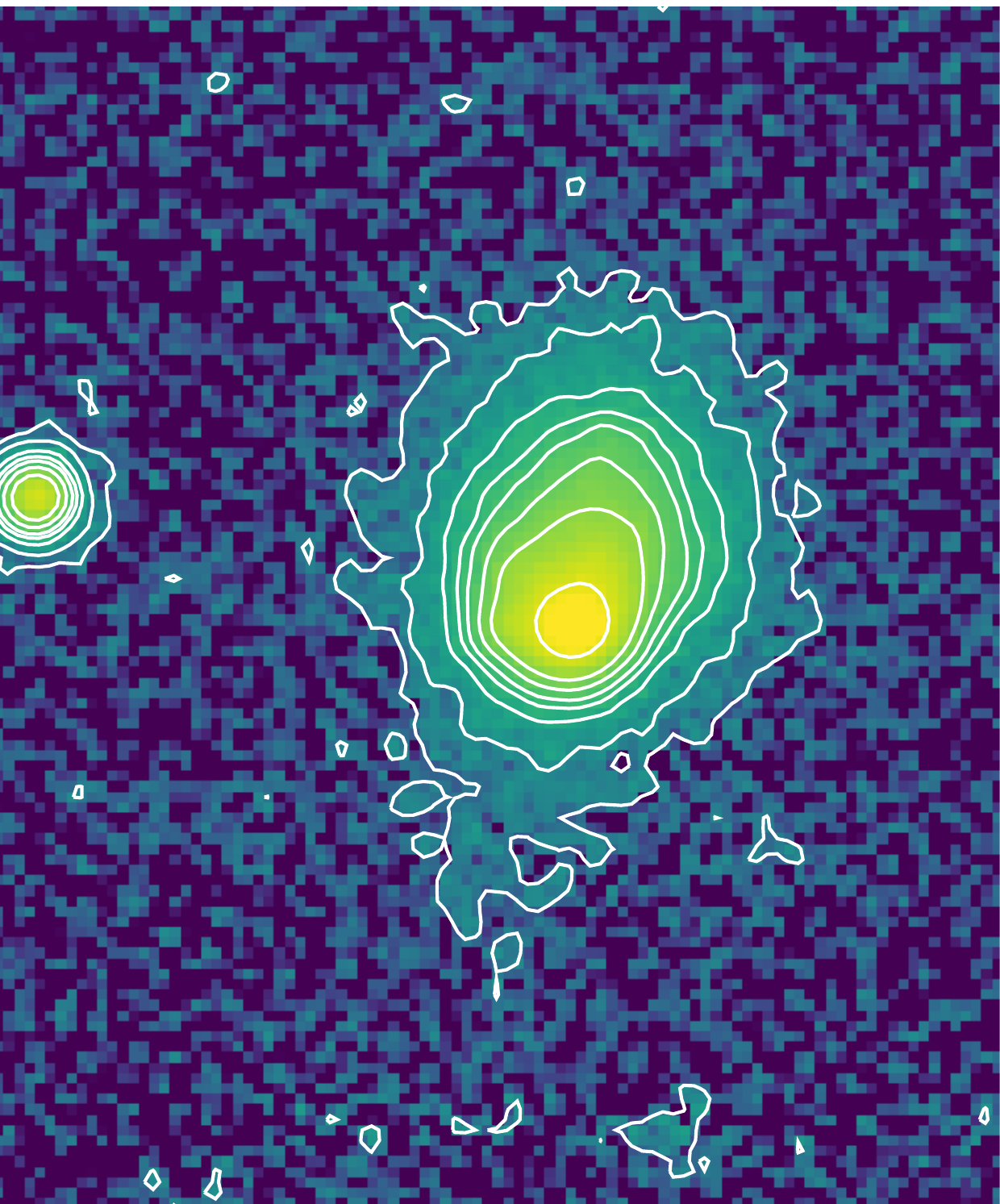}
\includegraphics[scale=0.27]{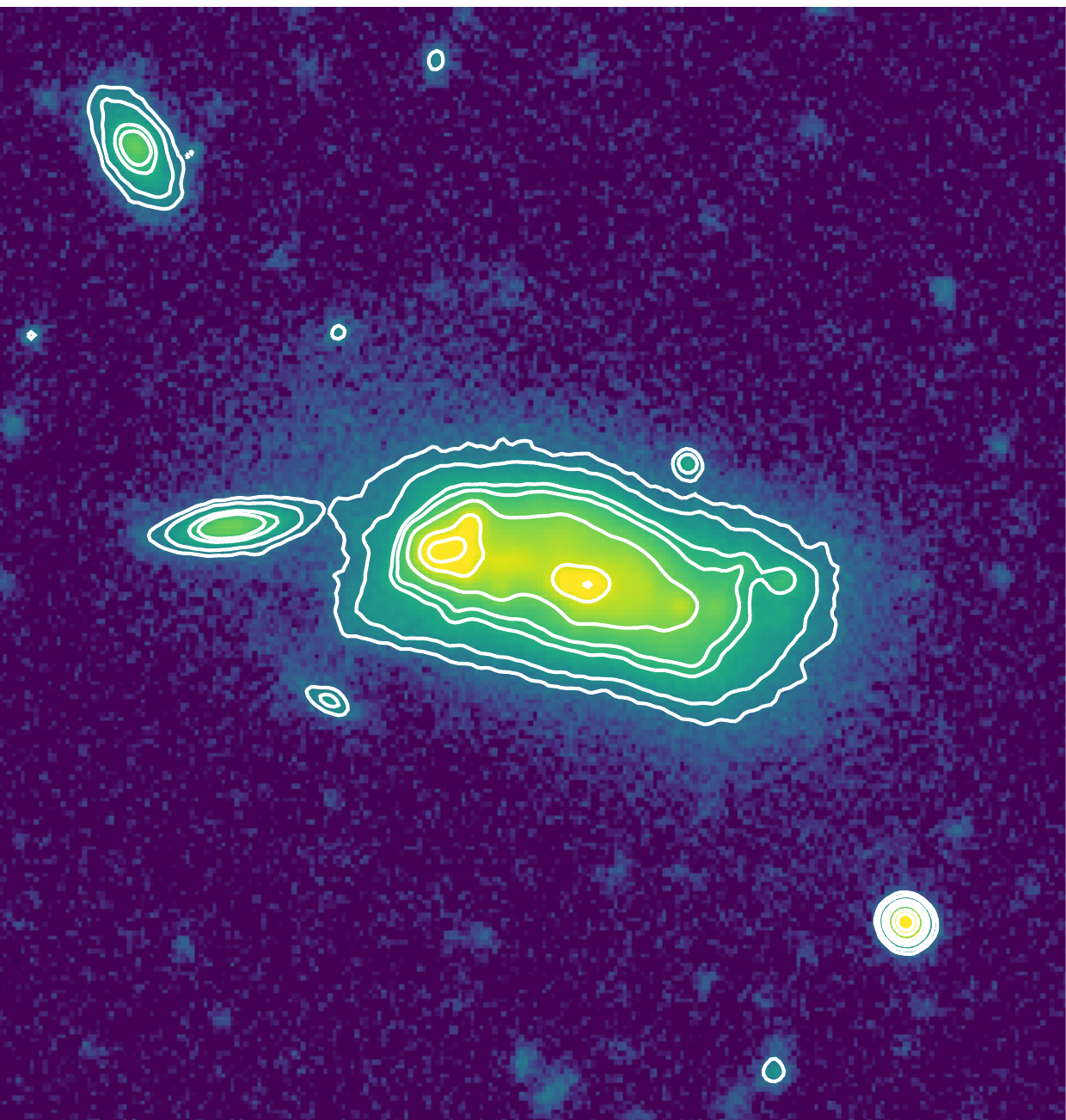}
\caption{Legacy $g$-band images of two galaxies in our sample, ID 2 (left panel) and ID 6 (right panel), with $C$, $A$, $G$, and $M_{20}$ indices indicating that the galaxies are in an interaction or merging phase.
}
\label{fig:im_morph}
\end{figure}

\begin{figure}
\centering
\includegraphics[scale=0.52]{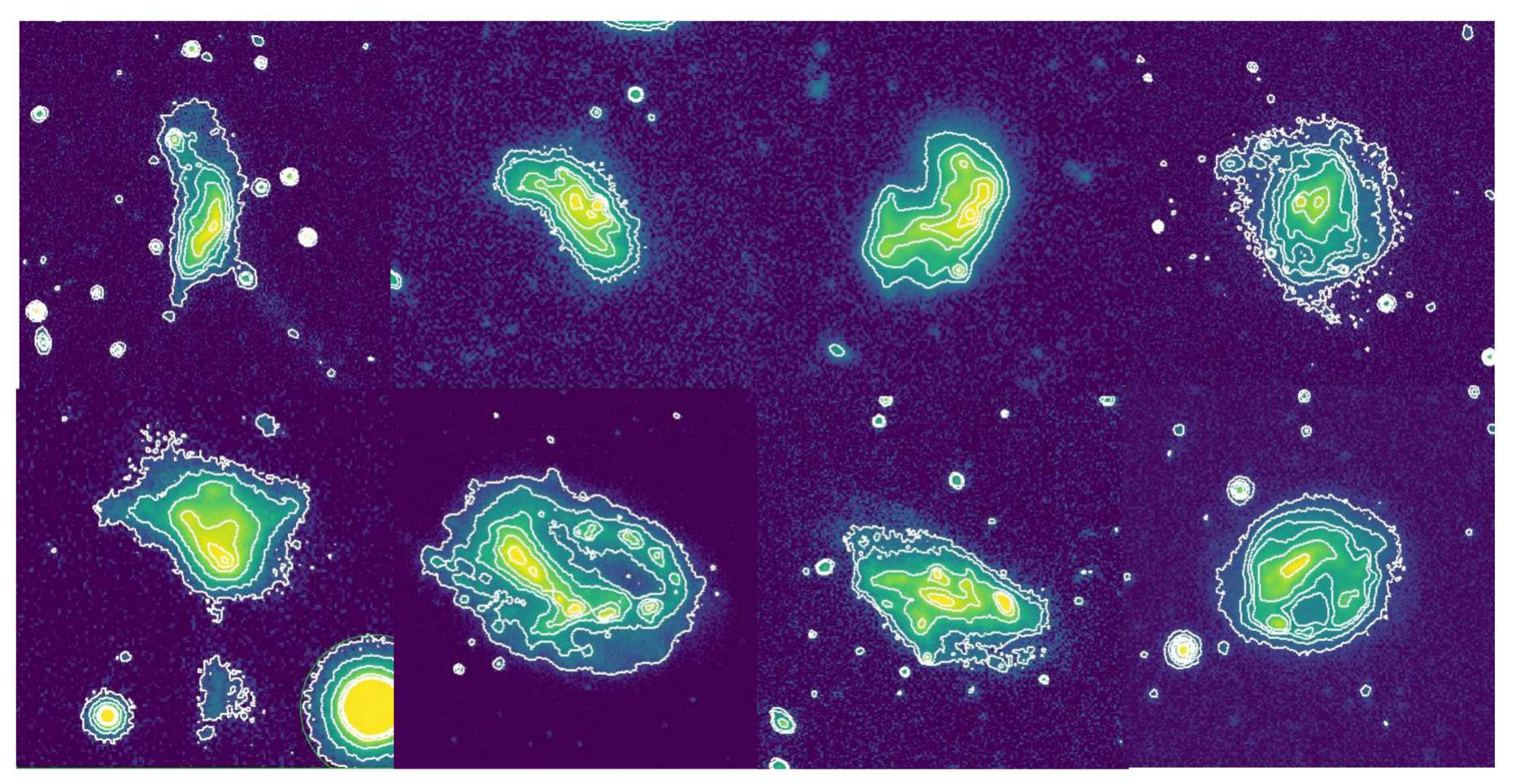}
\caption{Legacy $g$-band images of other galaxies  with $C$, $A$, $G$, and $M_{20}$ indices indicating that the galaxies are in an interaction or merging phase.
}
\label{fig_app:im_morph}
\end{figure}

Half of the dwarf galaxies selected with S-PLUS appear to lie in the merger/interacting regions defined by the Gini-$M_{20}$ and the  $C - A$ diagrams, including two targets of this work. ID 2 exhibits an extended, low surface brightness tail, while  ID 6 shows a clumpy morphology with star-forming knots distributed throughout its disc, and a high Asymmetry ($A$ = 0.36).  Notably, its Legacy image does not reveal features such as tidal tails or stellar streams (Fig. \ref{fig:im_morph}).
Visual inspection of the Legacy images for other galaxies with high asymmetry and Gini-$M_{20}$ values reveals elongated and disturbed morphologies, including tails, rings, and possible companions, all indicative of recent or ongoing interactions
(Fig. \ref{fig_app:im_morph}).

This analysis may suggest that the low-metallicity of these systems is related to 
a dwarf-dwarf merger or interaction.
However, this result would not rule out that the remaining galaxies might be experiencing
minor merger events. \citet{2012MNRAS.419.2703H} report that in late minor
mergers in which the less massive galaxies have been almost entirely
dissolved, the application of these diagnostics is less effective.
Another possibility within the dwarf-dwarf galaxy interaction scenario, is that the stellar discs have not yet been significantly disturbed  by the interaction, while the external gas component of the less massive companion has been already stripped  and accreted by the most massive galaxy \citep{2011ApJ...735...71W}, as it occurs in the case of IZw18  \citep{2012A&A...537A..72L}. Lastly, if the star formation process is triggered by accretion from the intergalactic medium, the stellar disc will not be perturbed, but the galaxies will appear as metal-poor outliers in the mass-metallicity relation \citep{2016IAUS..308..390S,2022MNRAS.512.6164Z}.





\section{Summary and conclusions}\label{sec:5}

We selected a sample of 47 star-forming dwarfs from the S-PLUS survey that are metal-poor galaxy candidates based on their $ugriz$ colors, and SED fitting using the whole set of 12 filters  provided by S-PLUS.
We started a spectroscopic follow-up campaign with the Gemini-S telescope to confirm their low metal abundances and to verify the photometric selection method that we applied. We observed four galaxies and derived O/H using the direct method. Oxygen abundances for two additional systems were derived from previous studies and spectroscopic measurements reported in the literature. We obtained the following results:

1. All the objects with spectroscopic follow-up were found to be metal-poor galaxies, with $7.28 <$ 12 + log(O/H) $< 7.82$, i.e., between  4\% and 13\% of the solar abundance, confirming the effectiveness of our method based on color cuts and multi-band SED fitting. 

2. We estimated nitrogen abundances and nitrogen-to-oxygen ratios. Five out of six galaxies 
show excess N/O for a given O/H. Particularly, ID 1, the most metal-deficient target of the sample,
exhibits a N/O more typical of a galaxy with ten times higher metallicity.

3. We derived the MZR relation and compared our targets with other samples of dwarf galaxies selected in the SDSS and DES surveys. The S-PLUS galaxies are outliers of the MZR, implying that they are metal-deficient for their stellar masses. For example, the oxygen abundance of ID 1 (approximately 1/20th of the solar value) is typical of a galaxy with a stellar mass nearly 200 times smaller than what we measured.

4. We analyzed the optical morphology of all the 47 metal-poor dwarf galaxy candidates.
Half of the sample appears to have perturbed morphologies that might be related to a recent major merger event, according to their location in the Gini-$M_{20}$ and $C - A$ diagnostic diagrams.
Among the six galaxies with a spectroscopic follow-up, only two -- ID 2 and 6 -- appear to have disturbed optical morphologies according to these diagnostics.
Two other systems -- ID 1 and 5 -- present faint companions in the optical images within a few kpc project distances.

5. The analysis of the H$\alpha$ emission line profile of target ID 4 highlights the presence of a star-forming knot with a velocity difference of $\Delta$v $\sim$ 370 km s$^{-1}$ from the closest H {\sc ii} region at approximately
120 pc. Such a velocity difference is significantly larger than the typical rotation velocity gradient of a dwarf galaxy. This might imply that the knot is kinematically distinct from the rest of the galaxy; however, further analysis is required to confirm this possibility.

We suggest that an external event may have occurred in these galaxies, triggering star formation and lowering the oxygen abundance. Such an event might be  associated, at least in a few cases, with an interaction/merger with a dwarf galaxy companion. 
Given the low star formation rates observed, it is less likely that strong outflows due to stellar feedback may be causing the observed metal deficiency.
Integral field unit observations of the ionized gas, as well as  high resolution maps of the atomic gas, will be required to confirm the 
dwarf-dwarf galaxy interaction scenario.

In future work, we plan to extend the spectroscopic analysis to the other galaxies of the sample in order to assemble a statistically significant set of metal-poor galaxies in the southern hemisphere and to investigate more thoroughly the processes that affect and regulate the metallicity of low-mass systems.

\begin{acknowledgements}
MG acknowledges support from FAPERJ grant E-26/211.370/2021. DRG acknowledges FAPERJ (E-26/211.370/2021; E-26/211.527/2023) and CNPq (313016/2020-8; 403011/2022-1; 315307/2023-4) grants. ACK thanks  FAPESP for the support grant 2020/16416-5 and the Conselho Nacional de Desenvolvimento Científico e Tecnológico (CNPq). CEFL is supported by ANID’s Millennium Institute of Astrophysics (MAS) AIM23-0001, by ANID/FONDECYT Regular grant 123163, and by DIUDA 88231R11. LAGS acknowledges financial support from CONICET.
PKH gratefully acknowledges the Fundação de Amparo à Pesquisa do Estado de São Paulo (FAPESP) for the support grant 2023/14272-4. MEDR acknowledges support from {\it Agencia Nacional de Promoci\'on de la Investigaci\'on, el Desarrollo Tecnol\'ogico y la Innovaci\'on} (Agencia I+D+i, PICT-2021-GRF-TI-00290, Argentina). CMdO thanks FAPESP for funding through grant 2019/26492-3 and CNPq through grant 309209/2019-6.
CL-D acknowledges a grant from the ESO Comité Mixto ORP037/2022, and the support from the Agencia Nacional de Investigación y Desarrollo (ANID) through Fondecyt project 3250511.
ARL acknowledges the grant 2025/09544-0 from São Paulo Research Foundation (FAPESP).
ARL and AVSC acknowledge financial support from Consejo Nacional de Investigaciones Científicas y Técnicas (CONICET) (PIP 1504), Agencia I+D+i and Universidad Nacional de La Plata (Argentina).
ACS acknowledges support from FAPERGS (grants 23/2551-0001832-2 and 24/2551-0001548-5), CNPq (grants 314301/2021-6, 312940/2025-4, 445231/2024-6, and 404233/2024-4), and CAPES (grant 88887.004427/2024-00).

Based on observations obtained at the international Gemini Observatory, a program of NSF NOIRLab, which is managed by the Association of Universities for Research in Astronomy (AURA) under a cooperative agreement with the U.S. National Science Foundation on behalf of the Gemini Observatory partnership: the U.S. National Science Foundation (United States), National Research Council (Canada), Agencia Nacional de Investigaci\'{o}n y Desarrollo (Chile), Ministerio de Ciencia, Tecnolog\'{i}a e Innovaci\'{o}n (Argentina), Minist\'{e}rio da Ci\^{e}ncia, Tecnologia, Inova\c{c}\~{o}es e Comunica\c{c}\~{o}es (Brazil), and Korea Astronomy and Space Science Institute (Republic of Korea).

The S-PLUS project, including the T80-South robotic telescope and the S-PLUS scientific survey, was founded as a partnership between the Fundação de Amparo à Pesquisa do Estado de São Paulo (FAPESP), the Observatório Nacional (ON), the Federal University of Sergipe (UFS), and the Federal University of Santa Catarina (UFSC), with important financial and practical contributions from other collaborating institutes in Brazil, Chile (Universidad de La Serena), and Spain (Centro de Estudios de Física del Cosmos de Aragón, CEFCA). We further acknowledge financial support from the São Paulo Research Foundation (FAPESP) grant 2019/263492-3, the Brazilian National Research Council (CNPq), the Coordination for the Improvement of Higher Education Personnel (CAPES), the Carlos Chagas Filho Rio de Janeiro State Research Foundation (FAPERJ), and the Brazilian Innovation Agency (FINEP).

The S-PLUS collaboration members are grateful for the contributions from CTIO staff in helping in the construction, commissioning, and maintenance of the T80-South telescope and camera. We are also indebted to Rene Laporte, INPE, and Keith Taylor for their essential contributions to the project. From CEFCA, we particularly would like to thank Antonio Marín-Franch for his invaluable contributions in the early phases of the project, David Cristóbal-Hornillos and his team for their help with the installation of the data reduction package JYPE version 0.9.9, César Íñiguez for providing 2D measurements of the filter transmissions, and all other staff members for their support with various aspects of the project.

The Legacy Surveys consist of three individual and complementary projects: the Dark Energy Camera Legacy Survey (DECaLS; Proposal ID 2014B-0404; PIs: David Schlegel and Arjun Dey), the Beijing-Arizona Sky Survey (BASS; NOAO Prop. ID 2015A-0801; PIs: Zhou Xu and Xiaohui Fan), and the Mayall z-band Legacy Survey (MzLS; Prop. ID 2016A-0453; PI: Arjun Dey). DECaLS, BASS and MzLS together include data obtained, respectively, at the Blanco telescope, Cerro Tololo Inter-American Observatory, NSF’s NOIRLab; the Bok telescope, Steward Observatory, University of Arizona; and the Mayall telescope, Kitt Peak National Observatory, NOIRLab. Pipeline processing and analyses of the data were supported by NOIRLab and the Lawrence Berkeley National Laboratory (LBNL). The Legacy Surveys project is honored to be permitted to conduct astronomical research on Iolkam Du’ag (Kitt Peak), a mountain with particular significance to the Tohono O’odham Nation.

\end{acknowledgements}

\begin{appendix}

\section{Emission line fluxes of DR4\_3\_SPLUS-n03s22\_0021537}

\begin{table}[!h]
\small
\begin{center}
\caption{Observed line fluxes, $F_{\lambda}/F({\rm H}\beta)$, and  extinction corrected line intensities, $I_{\lambda}/I({\rm H}\beta)$, normalized to $F({\rm H}\beta)$=100 for DR4\_3\_SPLUS-n03s22\_0021537 (ID 6).} 
\begin{tabular}{lrrrrrrrrrrr}
\hline \hline
\noalign{\smallskip}
\noalign{\smallskip}
  \multicolumn{1}{c}{$\lambda_{rest}$ [$\AA$]} &
  \multicolumn{1}{c}{$F_{\lambda}/F({\rm H}\beta)$} &
  \multicolumn{1}{c}{$I_{\lambda}/I({\rm H}\beta)$} \\
    \multicolumn{1}{c}{} &
  \multicolumn{1}{c}{} &
  \multicolumn{1}{c}{} \\
\hline
\noalign{\smallskip}
$\lambda$\,4340 H$\gamma$      &  50.27 $\pm$ 1.53 &  55.40 $\pm$   1.40  \\  
$\lambda$\,4363 [O {\sc iii}]  &   5.58 $\pm$ 0.78 &   6.14 $\pm$   0.82  \\
$\lambda$\,4861 H$\beta$       &   2.99 $\pm$ 0.67 &   3.22 $\pm$   0.67  \\
$\lambda$\,4959 [O {\sc iii}]  & 100.00 $\pm$ 1.94 & 100.00 $\pm$   1.00  \\
$\lambda$\,5007 [O {\sc iii}]  &  99.94 $\pm$ 1.57 &  98.37 $\pm$   0.70  \\
$\lambda$\,5876 He {\sc i}     & 297.89 $\pm$ 4.67 & 291.00 $\pm$   2.00  \\
$\lambda$\,6300 [O {\sc i}]    &  11.07 $\pm$ 0.65 &  10.23 $\pm$   0.56  \\
$\lambda$\,6312 [S {\sc iii}]  &   3.31 $\pm$ 0.40 &   2.92 $\pm$   0.34  \\
$\lambda$\,6363 [O {\sc i}]    &   1.10 $\pm$ 0.13 &   0.96 $\pm$   0.11  \\
$\lambda$\,6563 H$\alpha$      & 324.63 $\pm$ 5.07 & 280.00 $\pm$   2.00  \\
$\lambda$\,6584 [N {\sc ii}]   &   9.57 $\pm$ 0.50 &   8.22 $\pm$   0.39  \\
$\lambda$\,6678 He {\sc i}     &   2.72 $\pm$ 0.34 &   2.30 $\pm$   0.28  \\
$\lambda$\,6716 [S {\sc ii}]   &  19.83 $\pm$ 0.54 &  16.82 $\pm$   0.37  \\
$\lambda$\,6731 [S {\sc ii}]   &  13.91 $\pm$ 0.47 &  11.79 $\pm$   0.35  \\
$\lambda$\,7065 He {\sc i}     &   1.94 $\pm$ 0.32 &   1.59 $\pm$   0.25  \\
$\lambda$\,7136 [Ar {\sc iii}] &   5.73 $\pm$ 0.38 &   4.68 $\pm$   0.29  \\
$\lambda$\,7320 [O {\sc ii}]   &   2.12 $\pm$ 0.26 &   1.70 $\pm$   0.19  \\
$\lambda$\,7330 [O {\sc ii}]   &   2.59 $\pm$ 0.31 &   2.06 $\pm$   0.24  \\

\hline
\\$F({\rm H}\beta)$           &    4.82               &                        \\
mean $C({\rm H}\beta)$          & ${0.24}$       &				         \\
\noalign{\smallskip}
\hline \hline
\end{tabular}
\label{app:tab1}
\end{center}
\begin{minipage}[r]{16.5cm}
\end{minipage}
\end{table}


\end{appendix}

\bibliography{SPLUS_Gemini.bib}{}

\begin{thebibliography}{}
\expandafter\ifx\csname natexlab\endcsname\relax\def\natexlab#1{#1}\fi
\providecommand{\url}[1]{\href{#1}{#1}}
\providecommand{\dodoi}[1]{doi:~\href{http://doi.org/#1}{\nolinkurl{#1}}}
\providecommand{\doeprint}[1]{\href{http://ascl.net/#1}{\nolinkurl{http://ascl.net/#1}}}
\providecommand{\doarXiv}[1]{\href{https://arxiv.org/abs/#1}{\nolinkurl{https://arxiv.org/abs/#1}}}

\bibitem[{{Almeida-Fernandes} {et~al.}(2022){Almeida-Fernandes}, {SamPedro}, {Herpich}, {Molino}, {Barbosa}, {Buzzo}, {Overzier}, {de Lima}, {Nakazono}, {Oliveira Schwarz}, {Perottoni}, {Bolutavicius}, {Guti{\'e}rrez-Soto}, {Santos-Silva}, {Vitorelli}, {Werle}, {Whitten}, {Costa Duarte}, {Bom}, {Coelho}, {Sodr{\'e}}, {Placco}, {Teixeira}, {Alonso-Garc{\'\i}a}, {Barbosa}, {Beers}, {Bonatto}, {Chies-Santos}, {Hartmann}, {Lopes de Oliveira}, {Navarete}, {Kanaan}, {Ribeiro}, {Schoenell}, \& {Mendes de Oliveira}}]{2022MNRAS.511.4590A}
{Almeida-Fernandes}, F., {SamPedro}, L., {Herpich}, F.~R., {et~al.} 2022, \mnras, 511, 4590, \dodoi{10.1093/mnras/stac284}

\bibitem[{{Andrews} \& {Martini}(2013)}]{2013ApJ...765..140A}
{Andrews}, B.~H., \& {Martini}, P. 2013, \apj, 765, 140, \dodoi{10.1088/0004-637X/765/2/140}

\bibitem[{{Annibali} {et~al.}(2019){Annibali}, {La Torre}, {Tosi}, {Nipoti}, {Cusano}, {Aloisi}, {Bellazzini}, {Ciotti}, {Marchetti}, {Mignoli}, {Romano}, \& {Sacchi}}]{2019MNRAS.482.3892A}
{Annibali}, F., {La Torre}, V., {Tosi}, M., {et~al.} 2019, \mnras, 482, 3892, \dodoi{10.1093/mnras/sty2911}

\bibitem[{{Asplund} {et~al.}(2021){Asplund}, {Amarsi}, \& {Grevesse}}]{2021A&A...653A.141A}
{Asplund}, M., {Amarsi}, A.~M., \& {Grevesse}, N. 2021, \aap, 653, A141, \dodoi{10.1051/0004-6361/202140445}

\bibitem[{{Atek} {et~al.}(2009){Atek}, {Kunth}, {Schaerer}, {Hayes}, {Deharveng}, {{\"O}stlin}, \& {Mas-Hesse}}]{2009A&A...506L...1A}
{Atek}, H., {Kunth}, D., {Schaerer}, D., {et~al.} 2009, \aap, 506, L1, \dodoi{10.1051/0004-6361/200912787}

\bibitem[{{Baldwin} {et~al.}(1981){Baldwin}, {Phillips}, \& {Terlevich}}]{1981PASP...93....5B}
{Baldwin}, J.~A., {Phillips}, M.~M., \& {Terlevich}, R. 1981, \pasp, 93, 5, \dodoi{10.1086/130766}

\bibitem[{{Barrera-Ballesteros} {et~al.}(2017){Barrera-Ballesteros}, {S{\'a}nchez}, {Heckman}, {Blanc}, \& {MaNGA Team}}]{2017ApJ...844...80B}
{Barrera-Ballesteros}, J.~K., {S{\'a}nchez}, S.~F., {Heckman}, T., {Blanc}, G.~A., \& {MaNGA Team}. 2017, \apj, 844, 80, \dodoi{10.3847/1538-4357/aa7aa9}

\bibitem[{{Bell} {et~al.}(2003){Bell}, {McIntosh}, {Katz}, \& {Weinberg}}]{2003ApJS..149..289B}
{Bell}, E.~F., {McIntosh}, D.~H., {Katz}, N., \& {Weinberg}, M.~D. 2003, \apjs, 149, 289, \dodoi{10.1086/378847}

\bibitem[{{Benitez} {et~al.}(2014){Benitez}, {Dupke}, {Moles}, {Sodre}, {Cenarro}, {Marin-Franch}, {Taylor}, {Cristobal}, {Fernandez-Soto}, {Mendes de Oliveira}, {Cepa-Nogue}, {Abramo}, {Alcaniz}, {Overzier}, {Hernandez-Monteagudo}, {Alfaro}, {Kanaan}, {Carvano}, {Reis}, {Martinez Gonzalez}, {Ascaso}, {Ballesteros}, {Xavier}, {Varela}, {Ederoclite}, {Vazquez Ramio}, {Broadhurst}, {Cypriano}, {Angulo}, {Diego}, {Zandivarez}, {Diaz}, {Melchior}, {Umetsu}, {Spinelli}, {Zitrin}, {Coe}, {Yepes}, {Vielva}, {Sahni}, {Marcos-Caballero}, {Kitaura}, {Maroto}, {Masip}, {Tsujikawa}, {Carneiro}, {Gonzalez Nuevo}, {Carvalho}, {Reboucas}, {Carvalho}, {Abdalla}, {Bernui}, {Pigozzo}, {Ferreira}, {Chandrachani Devi}, {Bengaly}, {Campista}, {Amorim}, {Asari}, {Bongiovanni}, {Bonoli}, {Bruzual}, {Cardiel}, {Cava}, {Cid Fernandes}, {Coelho}, {Cortesi}, {Delgado}, {Diaz Garcia}, {Espinosa}, {Galliano}, {Gonzalez-Serrano}, {Falcon-Barroso}, {Fritz}, {Fernandes}, {Gorgas}, {Hoyos}, {Jimenez-Teja}, {Lopez-Aguerri}, {Lopez-San Juan},
  {Mateus}, {Molino}, {Novais}, {OMill}, {Oteo}, {Perez-Gonzalez}, {Poggianti}, {Proctor}, {Ricciardelli}, {Sanchez-Blazquez}, {Storchi-Bergmann}, {Telles}, {Schoennell}, {Trujillo}, {Vazdekis}, {Viironen}, {Daflon}, {Aparicio-Villegas}, {Rocha}, {Ribeiro}, {Borges}, {Martins}, {Marcolino}, {Martinez-Delgado}, {Perez-Torres}, {Siffert}, {Calvao}, {Sako}, {Kessler}, {Alvarez-Candal}, {De Pra}, {Roig}, {Lazzaro}, {Gorosabel}, {Lopes de Oliveira}, {Lima-Neto}, {Irwin}, {Liu}, {Alvarez}, {Balmes}, {Chueca}, {Costa-Duarte}, {da Costa}, {Dantas}, {Diaz}, {Fabregat}, {Ferrari}, {Gavela}, {Gracia}, {Gruel}, {Gutierrez}, {Guzman}, {Hernandez-Fernandez}, {Herranz}, {Hurtado-Gil}, {Jablonsky}, {Laporte}, {Le Tiran}, {Licandro}, {Lima}, {Martin}, {Martinez}, {Montero}, {Penteado}, {Pereira}, {Peris}, {Quilis}, {Sanchez-Portal}, {Soja}, {Solano}, {Torra}, \& {Valdivielso}}]{2014arXiv1403.5237B}
{Benitez}, N., {Dupke}, R., {Moles}, M., {et~al.} 2014, arXiv e-prints, arXiv:1403.5237, \dodoi{10.48550/arXiv.1403.5237}

\bibitem[{{Berg} {et~al.}(2012){Berg}, {Skillman}, {Marble}, {van Zee}, {Engelbracht}, {Lee}, {Kennicutt}, {Calzetti}, {Dale}, \& {Johnson}}]{2012ApJ...754...98B}
{Berg}, D.~A., {Skillman}, E.~D., {Marble}, A.~R., {et~al.} 2012, \apj, 754, 98, \dodoi{10.1088/0004-637X/754/2/98}

\bibitem[{{Boquien} {et~al.}(2019){Boquien}, {Burgarella}, {Roehlly}, {Buat}, {Ciesla}, {Corre}, {Inoue}, \& {Salas}}]{2019A&A...622A.103B}
{Boquien}, M., {Burgarella}, D., {Roehlly}, Y., {et~al.} 2019, \aap, 622, A103, \dodoi{10.1051/0004-6361/201834156}

\bibitem[{{Bouwens} {et~al.}(2016){Bouwens}, {Oesch}, {Labb{\'e}}, {Illingworth}, {Fazio}, {Coe}, {Holwerda}, {Smit}, {Stefanon}, {1van Dokkum}, {Trenti}, {Ashby}, {Huang}, {Spitler}, {Straatman}, {Bradley}, \& {Magee}}]{2016ApJ...830...67B}
{Bouwens}, R.~J., {Oesch}, P.~A., {Labb{\'e}}, I., {et~al.} 2016, \apj, 830, 67, \dodoi{10.3847/0004-637X/830/2/67}

\bibitem[{{Bovill} \& {Ricotti}(2009)}]{2009ApJ...693.1859B}
{Bovill}, M.~S., \& {Ricotti}, M. 2009, \apj, 693, 1859, \dodoi{10.1088/0004-637X/693/2/1859}

\bibitem[{{Bromm} \& {Yoshida}(2011)}]{2011ARA&A..49..373B}
{Bromm}, V., \& {Yoshida}, N. 2011, \araa, 49, 373, \dodoi{10.1146/annurev-astro-081710-102608}

\bibitem[{{Brooks} {et~al.}(2007){Brooks}, {Governato}, {Booth}, {Willman}, {Gardner}, {Wadsley}, {Stinson}, \& {Quinn}}]{2007ApJ...655L..17B}
{Brooks}, A.~M., {Governato}, F., {Booth}, C.~M., {et~al.} 2007, \apjl, 655, L17, \dodoi{10.1086/511765}

\bibitem[{{Cappellari}(2017)}]{2017MNRAS.466..798C}
{Cappellari}, M. 2017, \mnras, 466, 798, \dodoi{10.1093/mnras/stw3020}

\bibitem[{{Cappellari}(2023)}]{2023MNRAS.526.3273C}
---. 2023, \mnras, 526, 3273, \dodoi{10.1093/mnras/stad2597}

\bibitem[{{Cappellari} \& {Emsellem}(2004)}]{2004PASP..116..138C}
{Cappellari}, M., \& {Emsellem}, E. 2004, \pasp, 116, 138, \dodoi{10.1086/381875}

\bibitem[{{Cardelli} {et~al.}(1989){Cardelli}, {Clayton}, \& {Mathis}}]{1989ApJ...345..245C}
{Cardelli}, J.~A., {Clayton}, G.~C., \& {Mathis}, J.~S. 1989, \apj, 345, 245, \dodoi{10.1086/167900}

\bibitem[{{Cenarro} {et~al.}(2019){Cenarro}, {Moles}, {Crist{\'o}bal-Hornillos}, {Mar{\'\i}n-Franch}, {Ederoclite}, {Varela}, {L{\'o}pez-Sanjuan}, {Hern{\'a}ndez-Monteagudo}, {Angulo}, {V{\'a}zquez Rami{\'o}}, {Viironen}, {Bonoli}, {Orsi}, {Hurier}, {San Roman}, {Greisel}, {Vilella-Rojo}, {D{\'\i}az-Garc{\'\i}a}, {Logro{\~n}o-Garc{\'\i}a}, {Gurung-L{\'o}pez}, {Spinoso}, {Izquierdo-Villalba}, {Aguerri}, {Allende Prieto}, {Bonatto}, {Carvano}, {Chies-Santos}, {Daflon}, {Dupke}, {Falc{\'o}n-Barroso}, {Gon{\c{c}}alves}, {Jim{\'e}nez-Teja}, {Molino}, {Placco}, {Solano}, {Whitten}, {Abril}, {Ant{\'o}n}, {Bello}, {Bielsa de Toledo}, {Castillo-Ram{\'\i}rez}, {Chueca}, {Civera}, {D{\'\i}az-Mart{\'\i}n}, {Dom{\'\i}nguez-Mart{\'\i}nez}, {Garzar{\'a}n-Calderaro}, {Hern{\'a}ndez-Fuertes}, {Iglesias-Marzoa}, {I{\~n}iguez}, {Jim{\'e}nez Ruiz}, {Kruuse}, {Lamadrid}, {Lasso-Cabrera}, {L{\'o}pez-Alegre}, {L{\'o}pez-Sainz}, {Ma{\'\i}cas}, {Moreno-Signes}, {Muniesa}, {Rodr{\'\i}guez-Llano}, {Rueda-Teruel}, {Rueda-Teruel},
  {Soriano-Lagu{\'\i}a}, {Tilve}, {Valdivielso}, {Yanes-D{\'\i}az}, {Alcaniz}, {Mendes de Oliveira}, {Sodr{\'e}}, {Coelho}, {Lopes de Oliveira}, {Tamm}, {Xavier}, {Abramo}, {Akras}, {Alfaro}, {Alvarez-Candal}, {Ascaso}, {Beasley}, {Beers}, {Borges Fernandes}, {Bruzual}, {Buzzo}, {Carrasco}, {Cepa}, {Cortesi}, {Costa-Duarte}, {De Pr{\'a}}, {Favole}, {Galarza}, {Galbany}, {Garcia}, {Gonz{\'a}lez Delgado}, {Gonz{\'a}lez-Serrano}, {Guti{\'e}rrez-Soto}, {Hernandez-Jimenez}, {Kanaan}, {Kuncarayakti}, {Landim}, {Laur}, {Licandro}, {Lima Neto}, {Lyman}, {Ma{\'\i}z Apell{\'a}niz}, {Miralda-Escud{\'e}}, {Morate}, {Nogueira-Cavalcante}, {Novais}, {Oncins}, {Oteo}, {Overzier}, {Pereira}, {Rebassa-Mansergas}, {Reis}, {Roig}, {Sako}, {Salvador-Rusi{\~n}ol}, {Sampedro}, {S{\'a}nchez-Bl{\'a}zquez}, {Santos}, {Schmidtobreick}, {Siffert}, {Telles}, \& {Vilchez}}]{2019A&A...622A.176C}
{Cenarro}, A.~J., {Moles}, M., {Crist{\'o}bal-Hornillos}, D., {et~al.} 2019, \aap, 622, A176, \dodoi{10.1051/0004-6361/201833036}

\bibitem[{{Chemerynska} {et~al.}(2024){Chemerynska}, {Atek}, {Dayal}, {Furtak}, {Feldmann}, {Greene}, {Maseda}, {Nanayakkara}, {Oesch}, {Fujimoto}, {Labb{\'e}}, {Bezanson}, {Brammer}, {Cutler}, {Leja}, {Pan}, {Price}, {Wang}, {Weaver}, \& {Whitaker}}]{2024ApJ...976L..15C}
{Chemerynska}, I., {Atek}, H., {Dayal}, P., {et~al.} 2024, \apjl, 976, L15, \dodoi{10.3847/2041-8213/ad8dc9}

\bibitem[{{Conselice}(2003)}]{2003ApJS..147....1C}
{Conselice}, C.~J. 2003, \apjs, 147, 1, \dodoi{10.1086/375001}

\bibitem[{{Cook} {et~al.}(2014){Cook}, {Dale}, {Johnson}, {Van Zee}, {Lee}, {Kennicutt}, {Calzetti}, {Staudaher}, \& {Engelbracht}}]{2014MNRAS.445..890C}
{Cook}, D.~O., {Dale}, D.~A., {Johnson}, B.~D., {et~al.} 2014, \mnras, 445, 890, \dodoi{10.1093/mnras/stu1581}

\bibitem[{{Cowie} {et~al.}(1996){Cowie}, {Songaila}, {Hu}, \& {Cohen}}]{1996AJ....112..839C}
{Cowie}, L.~L., {Songaila}, A., {Hu}, E.~M., \& {Cohen}, J.~G. 1996, \aj, 112, 839, \dodoi{10.1086/118058}

\bibitem[{{Curti} {et~al.}(2020){Curti}, {Mannucci}, {Cresci}, \& {Maiolino}}]{2020MNRAS.491..944C}
{Curti}, M., {Mannucci}, F., {Cresci}, G., \& {Maiolino}, R. 2020, \mnras, 491, 944, \dodoi{10.1093/mnras/stz2910}

\bibitem[{{Dav{\'e}} {et~al.}(2011){Dav{\'e}}, {Finlator}, \& {Oppenheimer}}]{2011MNRAS.416.1354D}
{Dav{\'e}}, R., {Finlator}, K., \& {Oppenheimer}, B.~D. 2011, \mnras, 416, 1354, \dodoi{10.1111/j.1365-2966.2011.19132.x}

\bibitem[{{De Rossi} {et~al.}(2017){De Rossi}, {Bower}, {Font}, {Schaye}, \& {Theuns}}]{2017MNRAS.472.3354D}
{De Rossi}, M.~E., {Bower}, R.~G., {Font}, A.~S., {Schaye}, J., \& {Theuns}, T. 2017, \mnras, 472, 3354, \dodoi{10.1093/mnras/stx2158}

\bibitem[{{Dekel} \& {Silk}(1986)}]{1986ApJ...303...39D}
{Dekel}, A., \& {Silk}, J. 1986, \apj, 303, 39, \dodoi{10.1086/164050}

\bibitem[{{del Valle-Espinosa} {et~al.}(2023){del Valle-Espinosa}, {S{\'a}nchez-Janssen}, {Amor{\'\i}n}, {Fern{\'a}ndez}, {S{\'a}nchez Almeida}, {Garc{\'\i}a Lorenzo}, \& {Papaderos}}]{2023MNRAS.522.2089D}
{del Valle-Espinosa}, M.~G., {S{\'a}nchez-Janssen}, R., {Amor{\'\i}n}, R., {et~al.} 2023, \mnras, 522, 2089, \dodoi{10.1093/mnras/stad1087}

\bibitem[{{Dey} {et~al.}(2019){Dey}, {Schlegel}, {Lang}, {Blum}, {Burleigh}, {Fan}, {Findlay}, {Finkbeiner}, {Herrera}, {Juneau}, {Landriau}, {Levi}, {McGreer}, {Meisner}, {Myers}, {Moustakas}, {Nugent}, {Patej}, {Schlafly}, {Walker}, {Valdes}, {Weaver}, {Y{\`e}che}, {Zou}, {Zhou}, {Abareshi}, {Abbott}, {Abolfathi}, {Aguilera}, {Alam}, {Allen}, {Alvarez}, {Annis}, {Ansarinejad}, {Aubert}, {Beechert}, {Bell}, {BenZvi}, {Beutler}, {Bielby}, {Bolton}, {Brice{\~n}o}, {Buckley-Geer}, {Butler}, {Calamida}, {Carlberg}, {Carter}, {Casas}, {Castander}, {Choi}, {Comparat}, {Cukanovaite}, {Delubac}, {DeVries}, {Dey}, {Dhungana}, {Dickinson}, {Ding}, {Donaldson}, {Duan}, {Duckworth}, {Eftekharzadeh}, {Eisenstein}, {Etourneau}, {Fagrelius}, {Farihi}, {Fitzpatrick}, {Font-Ribera}, {Fulmer}, {G{\"a}nsicke}, {Gaztanaga}, {George}, {Gerdes}, {Gontcho}, {Gorgoni}, {Green}, {Guy}, {Harmer}, {Hernandez}, {Honscheid}, {Huang}, {James}, {Jannuzi}, {Jiang}, {Joyce}, {Karcher}, {Karkar}, {Kehoe}, {Kneib}, {Kueter-Young}, {Lan},
  {Lauer}, {Le Guillou}, {Le Van Suu}, {Lee}, {Lesser}, {Perreault Levasseur}, {Li}, {Mann}, {Marshall}, {Mart{\'\i}nez-V{\'a}zquez}, {Martini}, {du Mas des Bourboux}, {McManus}, {Meier}, {M{\'e}nard}, {Metcalfe}, {Mu{\~n}oz-Guti{\'e}rrez}, {Najita}, {Napier}, {Narayan}, {Newman}, {Nie}, {Nord}, {Norman}, {Olsen}, {Paat}, {Palanque-Delabrouille}, {Peng}, {Poppett}, {Poremba}, {Prakash}, {Rabinowitz}, {Raichoor}, {Rezaie}, {Robertson}, {Roe}, {Ross}, {Ross}, {Rudnick}, {Safonova}, {Saha}, {S{\'a}nchez}, {Savary}, {Schweiker}, {Scott}, {Seo}, {Shan}, {Silva}, {Slepian}, {Soto}, {Sprayberry}, {Staten}, {Stillman}, {Stupak}, {Summers}, {Sien Tie}, {Tirado}, {Vargas-Maga{\~n}a}, {Vivas}, {Wechsler}, {Williams}, {Yang}, {Yang}, {Yapici}, {Zaritsky}, {Zenteno}, {Zhang}, {Zhang}, {Zhou}, \& {Zhou}}]{2019AJ....157..168D}
{Dey}, A., {Schlegel}, D.~J., {Lang}, D., {et~al.} 2019, \aj, 157, 168, \dodoi{10.3847/1538-3881/ab089d}

\bibitem[{{D{\'\i}az-Garc{\'\i}a} {et~al.}(2015){D{\'\i}az-Garc{\'\i}a}, {Cenarro}, {L{\'o}pez-Sanjuan}, {Ferreras}, {Varela}, {Viironen}, {Crist{\'o}bal-Hornillos}, {Moles}, {Mar{\'\i}n-Franch}, {Arnalte-Mur}, {Ascaso}, {Cervi{\~n}o}, {Gonz{\'a}lez Delgado}, {M{\'a}rquez}, {Masegosa}, {Molino}, {Povi{\'c}}, {Alfaro}, {Aparicio-Villegas}, {Ben{\'\i}tez}, {Broadhurst}, {Cabrera-Ca{\~n}o}, {Castander}, {Cepa}, {Fern{\'a}ndez-Soto}, {Husillos}, {Infante}, {Aguerri}, {Mart{\'\i}nez}, {del Olmo}, {Perea}, {Prada}, {Quintana}, \& {Gruel}}]{2015A&A...582A..14D}
{D{\'\i}az-Garc{\'\i}a}, L.~A., {Cenarro}, A.~J., {L{\'o}pez-Sanjuan}, C., {et~al.} 2015, \aap, 582, A14, \dodoi{10.1051/0004-6361/201425582}

\bibitem[{{Ekta} \& {Chengalur}(2010)}]{2010MNRAS.406.1238E}
{Ekta}, B., \& {Chengalur}, J.~N. 2010, \mnras, 406, 1238, \dodoi{10.1111/j.1365-2966.2010.16756.x}

\bibitem[{{Ellison} {et~al.}(2008{\natexlab{a}}){Ellison}, {Patton}, {Simard}, \& {McConnachie}}]{2008ApJ...672L.107E}
{Ellison}, S.~L., {Patton}, D.~R., {Simard}, L., \& {McConnachie}, A.~W. 2008{\natexlab{a}}, \apjl, 672, L107, \dodoi{10.1086/527296}

\bibitem[{{Ellison} {et~al.}(2008{\natexlab{b}}){Ellison}, {Patton}, {Simard}, \& {McConnachie}}]{2008AJ....135.1877E}
---. 2008{\natexlab{b}}, \aj, 135, 1877, \dodoi{10.1088/0004-6256/135/5/1877}

\bibitem[{{Filippenko}(1982)}]{1982PASP...94..715F}
{Filippenko}, A.~V. 1982, \pasp, 94, 715, \dodoi{10.1086/131052}

\bibitem[{{Garnett}(1990)}]{1990ApJ...363..142G}
{Garnett}, D.~R. 1990, \apj, 363, 142, \dodoi{10.1086/169324}

\bibitem[{{Ginsburg} {et~al.}(2022){Ginsburg}, {Sokolov}, {de Val-Borro}, {Rosolowsky}, {Pineda}, {Sip{\H{o}}cz}, \& {Henshaw}}]{2022AJ....163..291G}
{Ginsburg}, A., {Sokolov}, V., {de Val-Borro}, M., {et~al.} 2022, \aj, 163, 291, \dodoi{10.3847/1538-3881/ac695a}

\bibitem[{{Gon{\c{c}}alves}(2019)}]{2019IAUS..344..161G}
{Gon{\c{c}}alves}, D.~R. 2019, in Dwarf Galaxies: From the Deep Universe to the Present, ed. K.~B.~W. {McQuinn} \& S.~{Stierwalt}, Vol. 344, 161--177, \dodoi{10.1017/S1743921318007408}

\bibitem[{{Gon{\c{c}}alves} {et~al.}(2014){Gon{\c{c}}alves}, {Magrini}, {Teodorescu}, \& {Carneiro}}]{2014MNRAS.444.1705G}
{Gon{\c{c}}alves}, D.~R., {Magrini}, L., {Teodorescu}, A.~M., \& {Carneiro}, C.~M. 2014, \mnras, 444, 1705, \dodoi{10.1093/mnras/stu1464}

\bibitem[{{Gondhalekar} {et~al.}(2024){Gondhalekar}, {Chies-Santos}, {de Souza}, {Queiroz}, {Lopes}, {Ferrari}, {Azevedo}, {Monteiro-Pereira}, {Overzier}, {Smith Castelli}, {Jaff{\'e}}, {Haack}, {Rahna}, {Shen}, {Mu}, {Lima-Dias}, {Barbosa}, {Oliveira Schwarz}, {Riffel}, {Jimenez-Teja}, {Grossi}, {Mendes de Oliveira}, {Schoenell}, {Ribeiro}, \& {Kanaan}}]{2024MNRAS.532..270G}
{Gondhalekar}, Y., {Chies-Santos}, A.~L., {de Souza}, R.~S., {et~al.} 2024, \mnras, 532, 270, \dodoi{10.1093/mnras/stae1410}

\bibitem[{{Gonz{\'a}lez Delgado} {et~al.}(2021){Gonz{\'a}lez Delgado}, {D{\'\i}az-Garc{\'\i}a}, {de Amorim}, {Bruzual}, {Cid Fernandes}, {P{\'e}rez}, {Bonoli}, {Cenarro}, {Coelho}, {Cortesi}, {Garc{\'\i}a-Benito}, {L{\'o}pez Fern{\'a}ndez}, {Mart{\'\i}nez-Solaeche}, {Rodr{\'\i}guez-Mart{\'\i}n}, {Magris}, {Mej{\'\i}a-Narvaez}, {Brito-Silva}, {Abramo}, {Diego}, {Dupke}, {Hern{\'a}n-Caballero}, {Hern{\'a}ndez-Monteagudo}, {L{\'o}pez-Sanjuan}, {Mar{\'\i}n-Franch}, {Marra}, {Moles}, {Montero-Dorta}, {Queiroz}, {Sodr{\'e}}, {Varela}, {V{\'a}zquez Rami{\'o}}, {V{\'\i}lchez}, {Baqui}, {Ben{\'\i}tez}, {Crist{\'o}bal-Hornillos}, {Ederoclite}, {Mendes de Oliveira}, {Civera}, {Muniesa}, {Taylor}, {Tempel}, \& {J-PAS Collaboration}}]{2021A&A...649A..79G}
{Gonz{\'a}lez Delgado}, R.~M., {D{\'\i}az-Garc{\'\i}a}, L.~A., {de Amorim}, A., {et~al.} 2021, \aap, 649, A79, \dodoi{10.1051/0004-6361/202039849}

\bibitem[{{Gordon} {et~al.}(2003){Gordon}, {Clayton}, {Misselt}, {Landolt}, \& {Wolff}}]{2003ApJ...594..279G}
{Gordon}, K.~D., {Clayton}, G.~C., {Misselt}, K.~A., {Landolt}, A.~U., \& {Wolff}, M.~J. 2003, \apj, 594, 279, \dodoi{10.1086/376774}

\bibitem[{{Grossi}(2019)}]{2019IAUS..344..319G}
{Grossi}, M. 2019, in Dwarf Galaxies: From the Deep Universe to the Present, ed. K.~B.~W. {McQuinn} \& S.~{Stierwalt}, Vol. 344, 319--330, \dodoi{10.1017/S1743921318007159}

\bibitem[{{Grossi} {et~al.}(2020){Grossi}, {Garc{\'\i}a-Benito}, {Cortesi}, {Gon{\c{c}}alves}, {Gon{\c{c}}alves}, {Lopes}, {Men{\'e}ndez-Delmestre}, \& {Telles}}]{2020MNRAS.498.1939G}
{Grossi}, M., {Garc{\'\i}a-Benito}, R., {Cortesi}, A., {et~al.} 2020, \mnras, 498, 1939, \dodoi{10.1093/mnras/staa2382}

\bibitem[{{Guseva} {et~al.}(2017){Guseva}, {Izotov}, {Fricke}, \& {Henkel}}]{2017A&A...599A..65G}
{Guseva}, N.~G., {Izotov}, Y.~I., {Fricke}, K.~J., \& {Henkel}, C. 2017, \aap, 599, A65, \dodoi{10.1051/0004-6361/201629181}

\bibitem[{{Guseva} {et~al.}(2009){Guseva}, {Papaderos}, {Meyer}, {Izotov}, \& {Fricke}}]{2009A&A...505...63G}
{Guseva}, N.~G., {Papaderos}, P., {Meyer}, H.~T., {Izotov}, Y.~I., \& {Fricke}, K.~J. 2009, \aap, 505, 63, \dodoi{10.1051/0004-6361/200912414}

\bibitem[{{Guti{\'e}rrez-Soto} {et~al.}(2020){Guti{\'e}rrez-Soto}, {Gon{\c{c}}alves}, {Akras}, {Cortesi}, {L{\'o}pez-Sanjuan}, {Guerrero}, {Daflon}, {Borges Fernandes}, {Mendes de Oliveira}, {Ederoclite}, {Sodr{\'e}}, {Pereira}, {Kanaan}, {Werle}, {V{\'a}zquez Rami{\'o}}, {Alcaniz}, {Angulo}, {Cenarro}, {Crist{\'o}bal-Hornillos}, {Dupke}, {Hern{\'a}ndez-Monteagudo}, {Mar{\'\i}n-Franch}, {Moles}, {Varela}, {Ribeiro}, {Schoenell}, {Alvarez-Candal}, {Galbany}, {Jim{\'e}nez-Esteban}, {Logro{\~n}o-Garc{\'\i}a}, \& {Sobral}}]{2020A&A...633A.123G}
{Guti{\'e}rrez-Soto}, L.~A., {Gon{\c{c}}alves}, D.~R., {Akras}, S., {et~al.} 2020, \aap, 633, A123, \dodoi{10.1051/0004-6361/201935700}

\bibitem[{Hernandez-Jimenez \& Krabbe(2022)}]{hernandez22}
Hernandez-Jimenez, J.~A., \& Krabbe, A.~C. 2022, {Astromorphlib: Python scripts to analyze the morphology of isolated and interacting galaxies}, 0.2,  Zenodo, \dodoi{10.5281/zenodo.6940848}

\bibitem[{{Hirschauer} {et~al.}(2022){Hirschauer}, {Salzer}, {Haurberg}, {Gronwall}, \& {Janowiecki}}]{2022ApJ...925..131H}
{Hirschauer}, A.~S., {Salzer}, J.~J., {Haurberg}, N., {Gronwall}, C., \& {Janowiecki}, S. 2022, \apj, 925, 131, \dodoi{10.3847/1538-4357/ac402a}

\bibitem[{{Hirschauer} {et~al.}(2016){Hirschauer}, {Salzer}, {Skillman}, {Berg}, {McQuinn}, {Cannon}, {Gordon}, {Haynes}, {Giovanelli}, {Adams}, {Janowiecki}, {Rhode}, {Pogge}, {Croxall}, \& {Aver}}]{2016ApJ...822..108H}
{Hirschauer}, A.~S., {Salzer}, J.~J., {Skillman}, E.~D., {et~al.} 2016, \apj, 822, 108, \dodoi{10.3847/0004-637X/822/2/108}

\bibitem[{{Hook} {et~al.}(2004){Hook}, {J{\o}rgensen}, {Allington-Smith}, {Davies}, {Metcalfe}, {Murowinski}, \& {Crampton}}]{2004PASP..116..425H}
{Hook}, I.~M., {J{\o}rgensen}, I., {Allington-Smith}, J.~R., {et~al.} 2004, \pasp, 116, 425, \dodoi{10.1086/383624}

\bibitem[{{Hoyle} {et~al.}(2012){Hoyle}, {Vogeley}, \& {Pan}}]{2012MNRAS.426.3041H}
{Hoyle}, F., {Vogeley}, M.~S., \& {Pan}, D. 2012, \mnras, 426, 3041, \dodoi{10.1111/j.1365-2966.2012.21943.x}

\bibitem[{{Hoyos} {et~al.}(2012){Hoyos}, {Arag{\'o}n-Salamanca}, {Gray}, {Maltby}, {Bell}, {Barazza}, {B{\"o}hm}, {H{\"a}u{\ss}ler}, {Jahnke}, {Jogee}, {Lane}, {McIntosh}, \& {Wolf}}]{2012MNRAS.419.2703H}
{Hoyos}, C., {Arag{\'o}n-Salamanca}, A., {Gray}, M.~E., {et~al.} 2012, \mnras, 419, 2703, \dodoi{10.1111/j.1365-2966.2011.19918.x}

\bibitem[{{Hsyu} {et~al.}(2018){Hsyu}, {Cooke}, {Prochaska}, \& {Bolte}}]{2018ApJ...863..134H}
{Hsyu}, T., {Cooke}, R.~J., {Prochaska}, J.~X., \& {Bolte}, M. 2018, \apj, 863, 134, \dodoi{10.3847/1538-4357/aad18a}

\bibitem[{{Izotov} {et~al.}(2009){Izotov}, {Guseva}, {Fricke}, \& {Papaderos}}]{2009A&A...503...61I}
{Izotov}, Y.~I., {Guseva}, N.~G., {Fricke}, K.~J., \& {Papaderos}, P. 2009, \aap, 503, 61, \dodoi{10.1051/0004-6361/200911965}

\bibitem[{{Izotov} {et~al.}(2006){Izotov}, {Stasi{\'n}ska}, {Meynet}, {Guseva}, \& {Thuan}}]{2006A&A...448..955I}
{Izotov}, Y.~I., {Stasi{\'n}ska}, G., {Meynet}, G., {Guseva}, N.~G., \& {Thuan}, T.~X. 2006, \aap, 448, 955, \dodoi{10.1051/0004-6361:20053763}

\bibitem[{{Izotov} \& {Thuan}(1999)}]{1999ApJ...511..639I}
{Izotov}, Y.~I., \& {Thuan}, T.~X. 1999, \apj, 511, 639, \dodoi{10.1086/306708}

\bibitem[{{Izotov} {et~al.}(2012){Izotov}, {Thuan}, \& {Guseva}}]{2012A&A...546A.122I}
{Izotov}, Y.~I., {Thuan}, T.~X., \& {Guseva}, N.~G. 2012, \aap, 546, A122, \dodoi{10.1051/0004-6361/201219733}

\bibitem[{{Izotov} {et~al.}(2019){Izotov}, {Thuan}, \& {Guseva}}]{2019MNRAS.483.5491I}
---. 2019, \mnras, 483, 5491, \dodoi{10.1093/mnras/sty3472}

\bibitem[{{Izotov} {et~al.}(2024{\natexlab{a}}){Izotov}, {Thuan}, \& {Guseva}}]{2024MNRAS.527.3486I}
---. 2024{\natexlab{a}}, \mnras, 527, 3486, \dodoi{10.1093/mnras/stad3421}

\bibitem[{{Izotov} {et~al.}(2018){Izotov}, {Thuan}, {Guseva}, \& {Liss}}]{2018MNRAS.473.1956I}
{Izotov}, Y.~I., {Thuan}, T.~X., {Guseva}, N.~G., \& {Liss}, S.~E. 2018, \mnras, 473, 1956, \dodoi{10.1093/mnras/stx2478}

\bibitem[{{Izotov} {et~al.}(2024{\natexlab{b}}){Izotov}, {Thuan}, {Guseva}, {Schaerer}, {Worseck}, \& {Verhamme}}]{2024MNRAS.527..281I}
{Izotov}, Y.~I., {Thuan}, T.~X., {Guseva}, N.~G., {et~al.} 2024{\natexlab{b}}, \mnras, 527, 281, \dodoi{10.1093/mnras/stad3151}

\bibitem[{{James} {et~al.}(2015){James}, {Koposov}, {Stark}, {Belokurov}, {Pettini}, \& {Olszewski}}]{2015MNRAS.448.2687J}
{James}, B.~L., {Koposov}, S., {Stark}, D.~P., {et~al.} 2015, \mnras, 448, 2687, \dodoi{10.1093/mnras/stv175}

\bibitem[{{James} {et~al.}(2017){James}, {Koposov}, {Stark}, {Belokurov}, {Pettini}, {Olszewski}, \& {McQuinn}}]{2017MNRAS.465.3977J}
{James}, B.~L., {Koposov}, S.~E., {Stark}, D.~P., {et~al.} 2017, \mnras, 465, 3977, \dodoi{10.1093/mnras/stw2962}

\bibitem[{{Ju} {et~al.}(2022){Ju}, {Yin}, {Liu}, {Hao}, {Shao}, {Feng}, {Riffel}, {Liu}, {Stark}, {Shen}, {Telles}, {Fern{\'a}ndez-Trincado}, {Wang}, {Xu}, {Bizyaev}, \& {Rong}}]{2022ApJ...938...96J}
{Ju}, M., {Yin}, J., {Liu}, R., {et~al.} 2022, \apj, 938, 96, \dodoi{10.3847/1538-4357/ac9056}

\bibitem[{{Kauffmann} {et~al.}(2003){Kauffmann}, {Heckman}, {Tremonti}, {Brinchmann}, {Charlot}, {White}, {Ridgway}, {Brinkmann}, {Fukugita}, {Hall}, {Ivezi{\'c}}, {Richards}, \& {Schneider}}]{2003MNRAS.346.1055K}
{Kauffmann}, G., {Heckman}, T.~M., {Tremonti}, C., {et~al.} 2003, \mnras, 346, 1055, \dodoi{10.1111/j.1365-2966.2003.07154.x}

\bibitem[{{Kennicutt} \& {Evans}(2012)}]{2012ARA&A..50..531K}
{Kennicutt}, R.~C., \& {Evans}, N.~J. 2012, \araa, 50, 531, \dodoi{10.1146/annurev-astro-081811-125610}

\bibitem[{{Kewley} {et~al.}(2001){Kewley}, {Dopita}, {Sutherland}, {Heisler}, \& {Trevena}}]{2001ApJ...556..121K}
{Kewley}, L.~J., {Dopita}, M.~A., {Sutherland}, R.~S., {Heisler}, C.~A., \& {Trevena}, J. 2001, \apj, 556, 121, \dodoi{10.1086/321545}

\bibitem[{{Kingsburgh} \& {Barlow}(1994)}]{1994MNRAS.271..257K}
{Kingsburgh}, R.~L., \& {Barlow}, M.~J. 1994, \mnras, 271, 257, \dodoi{10.1093/mnras/271.2.257}

\bibitem[{{Kirby} {et~al.}(2013){Kirby}, {Cohen}, {Guhathakurta}, {Cheng}, {Bullock}, \& {Gallazzi}}]{2013ApJ...779..102K}
{Kirby}, E.~N., {Cohen}, J.~G., {Guhathakurta}, P., {et~al.} 2013, \apj, 779, 102, \dodoi{10.1088/0004-637X/779/2/102}

\bibitem[{{Kojima} {et~al.}(2020){Kojima}, {Ouchi}, {Rauch}, {Ono}, {Nakajima}, {Isobe}, {Fujimoto}, {Harikane}, {Hashimoto}, {Hayashi}, {Komiyama}, {Kusakabe}, {Kim}, {Lee}, {Mukae}, {Nagao}, {Onodera}, {Shibuya}, {Sugahara}, {Umemura}, \& {Yabe}}]{2020ApJ...898..142K}
{Kojima}, T., {Ouchi}, M., {Rauch}, M., {et~al.} 2020, \apj, 898, 142, \dodoi{10.3847/1538-4357/aba047}

\bibitem[{{K{\"o}ppen} \& {Hensler}(2005)}]{2005A&A...434..531K}
{K{\"o}ppen}, J., \& {Hensler}, G. 2005, \aap, 434, 531, \dodoi{10.1051/0004-6361:20042266}

\bibitem[{{Krabbe} {et~al.}(2024){Krabbe}, {Hernandez-Jimenez}, {Mendes de Oliveira}, {Jaffe}, {Oliveira}, {Cardoso}, {Smith Castelli}, {Dors}, {Cortesi}, \& {Crossett}}]{2024MNRAS.528.1125K}
{Krabbe}, A.~C., {Hernandez-Jimenez}, J.~A., {Mendes de Oliveira}, C., {et~al.} 2024, \mnras, 528, 1125, \dodoi{10.1093/mnras/stad3881}

\bibitem[{{Kroupa}(2001)}]{2001MNRAS.322..231K}
{Kroupa}, P. 2001, \mnras, 322, 231, \dodoi{10.1046/j.1365-8711.2001.04022.x}

\bibitem[{{Kunth} \& {{\"O}stlin}(2000)}]{2000A&ARv..10....1K}
{Kunth}, D., \& {{\"O}stlin}, G. 2000, \aapr, 10, 1, \dodoi{10.1007/s001590000005}

\bibitem[{{Labrie} {et~al.}(2023){Labrie}, {Simpson}, {Cardenes}, {Turner}, {Soraisam}, {Quint}, {Oberdorf}, {Placco}, {Berke}, {Smirnova}, {Conseil}, {Vacca}, \& {Thomas-Osip}}]{2023RNAAS...7..214L}
{Labrie}, K., {Simpson}, C., {Cardenes}, R., {et~al.} 2023, Research Notes of the American Astronomical Society, 7, 214, \dodoi{10.3847/2515-5172/ad0044}

\bibitem[{{Lara-L{\'o}pez} {et~al.}(2010){Lara-L{\'o}pez}, {Cepa}, {Bongiovanni}, {P{\'e}rez Garc{\'\i}a}, {Ederoclite}, {Casta{\~n}eda}, {Fern{\'a}ndez Lorenzo}, {Povi{\'c}}, \& {S{\'a}nchez-Portal}}]{2010A&A...521L..53L}
{Lara-L{\'o}pez}, M.~A., {Cepa}, J., {Bongiovanni}, A., {et~al.} 2010, \aap, 521, L53, \dodoi{10.1051/0004-6361/201014803}

\bibitem[{{Lelli} {et~al.}(2012){Lelli}, {Verheijen}, {Fraternali}, \& {Sancisi}}]{2012A&A...537A..72L}
{Lelli}, F., {Verheijen}, M., {Fraternali}, F., \& {Sancisi}, R. 2012, \aap, 537, A72, \dodoi{10.1051/0004-6361/201117867}

\bibitem[{{Lequeux} {et~al.}(1979){Lequeux}, {Peimbert}, {Rayo}, {Serrano}, \& {Torres-Peimbert}}]{1979A&A....80..155L}
{Lequeux}, J., {Peimbert}, M., {Rayo}, J.~F., {Serrano}, A., \& {Torres-Peimbert}, S. 1979, \aap, 80, 155

\bibitem[{{Lilly} {et~al.}(2013){Lilly}, {Carollo}, {Pipino}, {Renzini}, \& {Peng}}]{2013ApJ...772..119L}
{Lilly}, S.~J., {Carollo}, C.~M., {Pipino}, A., {Renzini}, A., \& {Peng}, Y. 2013, \apj, 772, 119, \dodoi{10.1088/0004-637X/772/2/119}

\bibitem[{{Lima} {et~al.}(2022){Lima}, {Sodr{\'e}}, {Bom}, {Teixeira}, {Nakazono}, {Buzzo}, {Queiroz}, {Herpich}, {Castellon}, {Dantas}, {Dors}, {Souza}, {Akras}, {Jim{\'e}nez-Teja}, {Kanaan}, {Ribeiro}, \& {Schoennell}}]{2022A&C....3800510L}
{Lima}, E.~V.~R., {Sodr{\'e}}, L., {Bom}, C.~R., {et~al.} 2022, Astronomy and Computing, 38, 100510, \dodoi{10.1016/j.ascom.2021.100510}

\bibitem[{{Lima-Dias} {et~al.}(2021){Lima-Dias}, {Monachesi}, {Torres-Flores}, {Cortesi}, {Hern{\'a}ndez-Lang}, {Barbosa}, {Mendes de Oliveira}, {Olave-Rojas}, {Pallero}, {Sampedro}, {Molino}, {Herpich}, {Jaff{\'e}}, {Amor{\'\i}n}, {Chies-Santos}, {Dimauro}, {Telles}, {Lopes}, {Alvarez-Candal}, {Ferrari}, {Kanaan}, {Ribeiro}, \& {Schoenell}}]{2021MNRAS.500.1323L}
{Lima-Dias}, C., {Monachesi}, A., {Torres-Flores}, S., {et~al.} 2021, \mnras, 500, 1323, \dodoi{10.1093/mnras/staa3326}

\bibitem[{{Lin} {et~al.}(2023){Lin}, {Scarlata}, {Mehta}, {Skillman}, {Hayes}, {McQuinn}, {Fortson}, {Chworowsky}, \& {Clarke}}]{2023ApJ...951..138L}
{Lin}, Y.-H., {Scarlata}, C., {Mehta}, V., {et~al.} 2023, \apj, 951, 138, \dodoi{10.3847/1538-4357/acd181}

\bibitem[{{Logro{\~n}o-Garc{\'\i}a} {et~al.}(2019){Logro{\~n}o-Garc{\'\i}a}, {Vilella-Rojo}, {L{\'o}pez-Sanjuan}, {Varela}, {Viironen}, {Muniesa}, {Cenarro}, {Crist{\'o}bal-Hornillos}, {Ederoclite}, {Mar{\'\i}n-Franch}, {Moles}, {V{\'a}zquez Rami{\'o}}, {Bonoli}, {D{\'\i}az-Garc{\'\i}a}, {Orsi}, {San Roman}, {Akras}, {Chies-Santos}, {Coelho}, {Daflon}, {Costa-Duarte}, {Dupke}, {Galbany}, {Gonz{\'a}lez Delgado}, {Hernandez-Jimenez}, {Lopes de Oliveira}, {Mendes de Oliveira}, {Oteo}, {Gon{\c{c}}alves}, {S{\'a}nchez-Portal}, {Schmidtobreick}, \& {Sodr{\'e}}}]{2019A&A...622A.180L}
{Logro{\~n}o-Garc{\'\i}a}, R., {Vilella-Rojo}, G., {L{\'o}pez-Sanjuan}, C., {et~al.} 2019, \aap, 622, A180, \dodoi{10.1051/0004-6361/201732487}

\bibitem[{{L{\'o}pez-S{\'a}nchez} {et~al.}(2007){L{\'o}pez-S{\'a}nchez}, {Esteban}, {Garc{\'\i}a-Rojas}, {Peimbert}, \& {Rodr{\'\i}guez}}]{2007ApJ...656..168L}
{L{\'o}pez-S{\'a}nchez}, {\'A}.~R., {Esteban}, C., {Garc{\'\i}a-Rojas}, J., {Peimbert}, M., \& {Rodr{\'\i}guez}, M. 2007, \apj, 656, 168, \dodoi{10.1086/510112}

\bibitem[{{Lotz} {et~al.}(2008){Lotz}, {Jonsson}, {Cox}, \& {Primack}}]{2008MNRAS.391.1137L}
{Lotz}, J.~M., {Jonsson}, P., {Cox}, T.~J., \& {Primack}, J.~R. 2008, \mnras, 391, 1137, \dodoi{10.1111/j.1365-2966.2008.14004.x}

\bibitem[{{Lotz} {et~al.}(2004){Lotz}, {Primack}, \& {Madau}}]{2004AJ....128..163L}
{Lotz}, J.~M., {Primack}, J., \& {Madau}, P. 2004, \aj, 128, 163, \dodoi{10.1086/421849}

\bibitem[{{Luo} {et~al.}(2021){Luo}, {Heckman}, {Hwang}, {Rowlands}, {S{\'a}nchez-Menguiano}, {Riffel}, {Bizyaev}, {Andrews}, {Fern{\'a}ndez-Trincado}, {Drory}, {S{\'a}nchez Almeida}, {Maiolino}, {Lane}, \& {Argudo-Fern{\'a}ndez}}]{2021ApJ...908..183L}
{Luo}, Y., {Heckman}, T., {Hwang}, H.-C., {et~al.} 2021, \apj, 908, 183, \dodoi{10.3847/1538-4357/abd1df}

\bibitem[{{Maeder} \& {Meynet}(2001)}]{2001A&A...373..555M}
{Maeder}, A., \& {Meynet}, G. 2001, \aap, 373, 555, \dodoi{10.1051/0004-6361:20010596}

\bibitem[{{Magrini} \& {Gon{\c{c}}alves}(2009)}]{2009MNRAS.398..280M}
{Magrini}, L., \& {Gon{\c{c}}alves}, D.~R. 2009, \mnras, 398, 280, \dodoi{10.1111/j.1365-2966.2009.15124.x}

\bibitem[{{Maiolino} \& {Mannucci}(2019)}]{2019A&ARv..27....3M}
{Maiolino}, R., \& {Mannucci}, F. 2019, \aapr, 27, 3, \dodoi{10.1007/s00159-018-0112-2}

\bibitem[{{Mannucci} {et~al.}(2010){Mannucci}, {Cresci}, {Maiolino}, {Marconi}, \& {Gnerucci}}]{2010MNRAS.408.2115M}
{Mannucci}, F., {Cresci}, G., {Maiolino}, R., {Marconi}, A., \& {Gnerucci}, A. 2010, \mnras, 408, 2115, \dodoi{10.1111/j.1365-2966.2010.17291.x}

\bibitem[{{Martin} {et~al.}(2005){Martin}, {Fanson}, {Schiminovich}, {Morrissey}, {Friedman}, {Barlow}, {Conrow}, {Grange}, {Jelinsky}, {Milliard}, {Siegmund}, {Bianchi}, {Byun}, {Donas}, {Forster}, {Heckman}, {Lee}, {Madore}, {Malina}, {Neff}, {Rich}, {Small}, {Surber}, {Szalay}, {Welsh}, \& {Wyder}}]{2005ApJ...619L...1M}
{Martin}, D.~C., {Fanson}, J., {Schiminovich}, D., {et~al.} 2005, \apjl, 619, L1, \dodoi{10.1086/426387}

\bibitem[{{McQuinn} {et~al.}(2019){McQuinn}, {van Zee}, \& {Skillman}}]{2019ApJ...886...74M}
{McQuinn}, K. B.~W., {van Zee}, L., \& {Skillman}, E.~D. 2019, \apj, 886, 74, \dodoi{10.3847/1538-4357/ab4c37}

\bibitem[{{McQuinn} {et~al.}(2015){McQuinn}, {Skillman}, {Dolphin}, {Cannon}, {Salzer}, {Rhode}, {Adams}, {Berg}, {Giovanelli}, \& {Haynes}}]{2015ApJ...815L..17M}
{McQuinn}, K. B.~W., {Skillman}, E.~D., {Dolphin}, A., {et~al.} 2015, \apjl, 815, L17, \dodoi{10.1088/2041-8205/815/2/L17}

\bibitem[{{McQuinn} {et~al.}(2020){McQuinn}, {Berg}, {Skillman}, {Adams}, {Cannon}, {Dolphin}, {Salzer}, {Giovanelli}, {Haynes}, {Hirschauer}, {Janoweicki}, {Klapkowski}, \& {Rhode}}]{2020ApJ...891..181M}
{McQuinn}, K. B.~W., {Berg}, D.~A., {Skillman}, E.~D., {et~al.} 2020, \apj, 891, 181, \dodoi{10.3847/1538-4357/ab7447}

\bibitem[{{Mej{\'\i}a-Narv{\'a}ez} {et~al.}(2017){Mej{\'\i}a-Narv{\'a}ez}, {Bruzual}, {Magris}, {Alcaniz}, {Ben{\'\i}tez}, {Carneiro}, {Cenarro}, {Crist{\'o}bal-Hornillos}, {Dupke}, {Ederoclite}, {Mar{\'\i}n-Franch}, {Mendes de Oliveira}, {Moles}, {Sodre}, {Taylor}, {Varela}, \& {V{\'a}zquez Rami{\'o}}}]{2017MNRAS.471.4722M}
{Mej{\'\i}a-Narv{\'a}ez}, A., {Bruzual}, G., {Magris}, C.~G., {et~al.} 2017, \mnras, 471, 4722, \dodoi{10.1093/mnras/stx1758}

\bibitem[{{Mendes de Oliveira} {et~al.}(2019){Mendes de Oliveira}, {Ribeiro}, {Schoenell}, {Kanaan}, {Overzier}, {Molino}, {Sampedro}, {Coelho}, {Barbosa}, {Cortesi}, {Costa-Duarte}, {Herpich}, {Hernandez-Jimenez}, {Placco}, {Xavier}, {Abramo}, {Saito}, {Chies-Santos}, {Ederoclite}, {Lopes de Oliveira}, {Gon{\c{c}}alves}, {Akras}, {Almeida}, {Almeida-Fernandes}, {Beers}, {Bonatto}, {Bonoli}, {Cypriano}, {Vinicius-Lima}, {de Souza}, {Fabiano de Souza}, {Ferrari}, {Gon{\c{c}}alves}, {Gonzalez}, {Guti{\'e}rrez-Soto}, {Hartmann}, {Jaffe}, {Kerber}, {Lima-Dias}, {Lopes}, {Menendez-Delmestre}, {Nakazono}, {Novais}, {Ortega-Minakata}, {Pereira}, {Perottoni}, {Queiroz}, {Reis}, {Santos}, {Santos-Silva}, {Santucci}, {Barbosa}, {Siffert}, {Sodr{\'e}}, {Torres-Flores}, {Westera}, {Whitten}, {Alcaniz}, {Alonso-Garc{\'\i}a}, {Alencar}, {Alvarez-Candal}, {Amram}, {Azanha}, {Barb{\'a}}, {Bernardinelli}, {Borges Fernandes}, {Branco}, {Brito-Silva}, {Buzzo}, {Caffer}, {Campillay}, {Cano}, {Carvano}, {Castejon}, {Cid
  Fernandes}, {Dantas}, {Daflon}, {Damke}, {de la Reza}, {de Melo de Azevedo}, {De Paula}, {Diem}, {Donnerstein}, {Dors}, {Dupke}, {Eikenberry}, {Escudero}, {Faifer}, {Far{\'\i}as}, {Fernandes}, {Fernandes}, {Fontes}, {Galarza}, {Hirata}, {Katena}, {Gregorio-Hetem}, {Hern{\'a}ndez-Fern{\'a}ndez}, {Izzo}, {Jaque Arancibia}, {Jatenco-Pereira}, {Jim{\'e}nez-Teja}, {Kann}, {Krabbe}, {Labayru}, {Lazzaro}, {Lima Neto}, {Lopes}, {Magalh{\~a}es}, {Makler}, {de Menezes}, {Miralda-Escud{\'e}}, {Monteiro-Oliveira}, {Montero-Dorta}, {Mu{\~n}oz-Elgueta}, {Nemmen}, {Nilo Castell{\'o}n}, {Oliveira}, {Ort{\'\i}z}, {Pattaro}, {Pereira}, {Quint}, {Riguccini}, {Rocha Pinto}, {Rodrigues}, {Roig}, {Rossi}, {Saha}, {Santos}, {Schnorr M{\"u}ller}, {Sesto}, {Silva}, {Smith Castelli}, {Teixeira}, {Telles}, {Thom de Souza}, {Th{\"o}ne}, {Trevisan}, {de Ugarte Postigo}, {Urrutia-Viscarra}, {Veiga}, {Vika}, {Vitorelli}, {Werle}, {Werner}, \& {Zaritsky}}]{2019MNRAS.489..241M}
{Mendes de Oliveira}, C., {Ribeiro}, T., {Schoenell}, W., {et~al.} 2019, \mnras, 489, 241, \dodoi{10.1093/mnras/stz1985}

\bibitem[{{Meyer} {et~al.}(2014){Meyer}, {Lisker}, {Janz}, \& {Papaderos}}]{2014A&A...562A..49M}
{Meyer}, H.~T., {Lisker}, T., {Janz}, J., \& {Papaderos}, P. 2014, \aap, 562, A49, \dodoi{10.1051/0004-6361/201220700}

\bibitem[{{Meynet} \& {Maeder}(2002)}]{2002A&A...381L..25M}
{Meynet}, G., \& {Maeder}, A. 2002, \aap, 381, L25, \dodoi{10.1051/0004-6361:20011554}

\bibitem[{{Molino} {et~al.}(2020){Molino}, {Costa-Duarte}, {Sampedro}, {Herpich}, {Sodr{\'e}}, {Mendes de Oliveira}, {Schoenell}, {Barbosa}, {Queiroz}, {Lima}, {Azanha}, {Mu{\~n}oz-Elgueta}, {Ribeiro}, {Kanaan}, {Hernandez-Jimenez}, {Cortesi}, {Akras}, {Lopes de Oliveira}, {Torres-Flores}, {Lima-Dias}, {Castellon}, {Damke}, {Alvarez-Candal}, {Jim{\'e}nez-Teja}, {Coelho}, {Pereira}, {Montero-Dorta}, {Ben{\'\i}tez}, {Gon{\c{c}}alves}, {Santana-Silva}, {Werner}, {Almeida}, {Lopes}, {Chies-Santos}, {Telles}, {de Souza}, {C}, {Gon{\c{c}}alves}, {de Souza}, {Makler}, {Buzzo}, {Placco}, {Nakazono}, {Saito}, {Overzier}, \& {Abramo}}]{2020MNRAS.499.3884M}
{Molino}, A., {Costa-Duarte}, M.~V., {Sampedro}, L., {et~al.} 2020, \mnras, 499, 3884, \dodoi{10.1093/mnras/staa1586}

\bibitem[{{Nakazono} {et~al.}(2024){Nakazono}, {R Valen{\c{c}}a}, {Soares}, {Izbicki}, {Ivezi{\'c}}, {R Lima}, {T Hirata}, {Sodr{\'e}}, {Overzier}, {Almeida-Fernandes}, {Oliveira Schwarz}, {Schoenell}, {Kanaan}, {Ribeiro}, \& {Mendes de Oliveira}}]{2024MNRAS.531..327N}
{Nakazono}, L., {R Valen{\c{c}}a}, R., {Soares}, G., {et~al.} 2024, \mnras, 531, 327, \dodoi{10.1093/mnras/stae971}

\bibitem[{{Nomoto} {et~al.}(2013){Nomoto}, {Kobayashi}, \& {Tominaga}}]{2013ARA&A..51..457N}
{Nomoto}, K., {Kobayashi}, C., \& {Tominaga}, N. 2013, \araa, 51, 457, \dodoi{10.1146/annurev-astro-082812-140956}

\bibitem[{{Osterbrock} \& {Ferland}(2006)}]{2006agna.book.....O}
{Osterbrock}, D.~E., \& {Ferland}, G.~J. 2006, {Astrophysics of gaseous nebulae and active galactic nuclei}

\bibitem[{{Pagel}(1985)}]{1985ESOC...21..155P}
{Pagel}, B.~E.~J. 1985, in European Southern Observatory Conference and Workshop Proceedings, Vol.~21, European Southern Observatory Conference and Workshop Proceedings, 155--170

\bibitem[{{Pagel} {et~al.}(1992){Pagel}, {Simonson}, {Terlevich}, \& {Edmunds}}]{1992MNRAS.255..325P}
{Pagel}, B.~E.~J., {Simonson}, E.~A., {Terlevich}, R.~J., \& {Edmunds}, M.~G. 1992, \mnras, 255, 325, \dodoi{10.1093/mnras/255.2.325}

\bibitem[{{Peng} \& {Maiolino}(2014)}]{2014MNRAS.438..262P}
{Peng}, Y.-j., \& {Maiolino}, R. 2014, \mnras, 438, 262, \dodoi{10.1093/mnras/stt2175}

\bibitem[{{Pilyugin} \& {Thuan}(2005)}]{2005ApJ...631..231P}
{Pilyugin}, L.~S., \& {Thuan}, T.~X. 2005, \apj, 631, 231, \dodoi{10.1086/432408}

\bibitem[{{Pilyugin} {et~al.}(2012){Pilyugin}, {V{\'\i}lchez}, {Mattsson}, \& {Thuan}}]{2012MNRAS.421.1624P}
{Pilyugin}, L.~S., {V{\'\i}lchez}, J.~M., {Mattsson}, L., \& {Thuan}, T.~X. 2012, \mnras, 421, 1624, \dodoi{10.1111/j.1365-2966.2012.20420.x}

\bibitem[{{Prevot} {et~al.}(1984){Prevot}, {Lequeux}, {Maurice}, {Prevot}, \& {Rocca-Volmerange}}]{1984A&A...132..389P}
{Prevot}, M.~L., {Lequeux}, J., {Maurice}, E., {Prevot}, L., \& {Rocca-Volmerange}, B. 1984, \aap, 132, 389

\bibitem[{{Pucha} {et~al.}(2024){Pucha}, {Juneau}, {Dey}, {Siudek}, {Mezcua}, {Moustakas}, {BenZvi}, {Hainline}, {Hviding}, {Mao}, {Alexander}, {Alfarsy}, {Circosta}, {Guo}, {Manwadkar}, {Martini}, {Weaver}, {Aguilar}, {Ahlen}, {Bianchi}, {Brooks}, {Canning}, {Claybaugh}, {Dawson}, {de la Macorra}, {Dey}, {Doel}, {Font-Ribera}, {Forero-Romero}, {Gazta{\~n}aga}, {Gontcho}, {Gutierrez}, {Honscheid}, {Kehoe}, {Koposov}, {Lambert}, {Landriau}, {Le Guillou}, {Meisner}, {Miquel}, {Prada}, {Rossi}, {Sanchez}, {Schlegel}, {Schubnell}, {Seo}, {Sprayberry}, {Tarl{\'e}}, \& {Zou}}]{2024arXiv241100091P}
{Pucha}, R., {Juneau}, S., {Dey}, A., {et~al.} 2024, arXiv e-prints, arXiv:2411.00091, \dodoi{10.48550/arXiv.2411.00091}

\bibitem[{{Reines} {et~al.}(2013){Reines}, {Greene}, \& {Geha}}]{2013ApJ...775..116R}
{Reines}, A.~E., {Greene}, J.~E., \& {Geha}, M. 2013, \apj, 775, 116, \dodoi{10.1088/0004-637X/775/2/116}

\bibitem[{{Reines} \& {Volonteri}(2015)}]{2015ApJ...813...82R}
{Reines}, A.~E., \& {Volonteri}, M. 2015, \apj, 813, 82, \dodoi{10.1088/0004-637X/813/2/82}

\bibitem[{{Renzini} \& {Voli}(1981)}]{1981A&A....94..175R}
{Renzini}, A., \& {Voli}, M. 1981, \aap, 94, 175

\bibitem[{{Sacchi} {et~al.}(2024){Sacchi}, {Bogd{\'a}n}, {Chadayammuri}, \& {Ricarte}}]{2024ApJ...974...14S}
{Sacchi}, A., {Bogd{\'a}n}, {\'A}., {Chadayammuri}, U., \& {Ricarte}, A. 2024, \apj, 974, 14, \dodoi{10.3847/1538-4357/ad684e}

\bibitem[{{San Roman} {et~al.}(2019){San Roman}, {S{\'a}nchez-Bl{\'a}zquez}, {Cenarro}, {D{\'\i}az-Garc{\'\i}a}, {L{\'o}pez-Sanjuan}, {Varela}, {Vilella-Rojo}, {Akras}, {Bonoli}, {Chies Santos}, {Coelho}, {Cortesi}, {Ederoclite}, {Jim{\'e}nez-Teja}, {Logro{\~n}o-Garc{\'\i}a}, {Lopes de Oliveira}, {Nogueira-Cavalcante}, {Orsi}, {V{\'a}zquez Rami{\'o}}, {Viironen}, {Crist{\'o}bal-Hornillos}, {Dupke}, {Mar{\'\i}n-Franch}, {Mendes de Oliveira}, {Moles}, \& {Sodr{\'e}}}]{2019A&A...622A.181S}
{San Roman}, I., {S{\'a}nchez-Bl{\'a}zquez}, P., {Cenarro}, A.~J., {et~al.} 2019, \aap, 622, A181, \dodoi{10.1051/0004-6361/201832894}

\bibitem[{{S{\'a}nchez} {et~al.}(2017){S{\'a}nchez}, {Barrera-Ballesteros}, {S{\'a}nchez-Menguiano}, {Walcher}, {Marino}, {Galbany}, {Bland-Hawthorn}, {Cano-D{\'\i}az}, {Garc{\'\i}a-Benito}, {L{\'o}pez-Cob{\'a}}, {Zibetti}, {Vilchez}, {Igl{\'e}sias-P{\'a}ramo}, {Kehrig}, {L{\'o}pez S{\'a}nchez}, {Duarte Puertas}, \& {Ziegler}}]{2017MNRAS.469.2121S}
{S{\'a}nchez}, S.~F., {Barrera-Ballesteros}, J.~K., {S{\'a}nchez-Menguiano}, L., {et~al.} 2017, \mnras, 469, 2121, \dodoi{10.1093/mnras/stx808}

\bibitem[{{S{\'a}nchez Almeida} {et~al.}(2016{\natexlab{a}}){S{\'a}nchez Almeida}, {Elmegreen}, {Mu{\~n}oz-Tu{\~n}{\'o}n}, \& {Elmegreen}}]{2016IAUS..308..390S}
{S{\'a}nchez Almeida}, J., {Elmegreen}, B.~G., {Mu{\~n}oz-Tu{\~n}{\'o}n}, C., \& {Elmegreen}, D.~M. 2016{\natexlab{a}}, in The Zeldovich Universe: Genesis and Growth of the Cosmic Web, ed. R.~{van de Weygaert}, S.~{Shandarin}, E.~{Saar}, \& J.~{Einasto}, Vol. 308, 390--393, \dodoi{10.1017/S1743921316010231}

\bibitem[{{S{\'a}nchez Almeida} {et~al.}(2016{\natexlab{b}}){S{\'a}nchez Almeida}, {P{\'e}rez-Montero}, {Morales-Luis}, {Mu{\~n}oz-Tu{\~n}{\'o}n}, {Garc{\'\i}a-Benito}, {Nuza}, \& {Kitaura}}]{2016ApJ...819..110S}
{S{\'a}nchez Almeida}, J., {P{\'e}rez-Montero}, E., {Morales-Luis}, A.~B., {et~al.} 2016{\natexlab{b}}, \apj, 819, 110, \dodoi{10.3847/0004-637X/819/2/110}

\bibitem[{{Scarlata} {et~al.}(2024){Scarlata}, {Hayes}, {Panagia}, {Mehta}, {Haardt}, \& {Bagley}}]{2024arXiv240409015S}
{Scarlata}, C., {Hayes}, M., {Panagia}, N., {et~al.} 2024, arXiv e-prints, arXiv:2404.09015, \dodoi{10.48550/arXiv.2404.09015}

\bibitem[{{Schiminovich} {et~al.}(2007){Schiminovich}, {Wyder}, {Martin}, {Johnson}, {Salim}, {Seibert}, {Treyer}, {Budav{\'a}ri}, {Hoopes}, {Zamojski}, {Barlow}, {Forster}, {Friedman}, {Morrissey}, {Neff}, {Small}, {Bianchi}, {Donas}, {Heckman}, {Lee}, {Madore}, {Milliard}, {Rich}, {Szalay}, {Welsh}, \& {Yi}}]{2007ApJS..173..315S}
{Schiminovich}, D., {Wyder}, T.~K., {Martin}, D.~C., {et~al.} 2007, \apjs, 173, 315, \dodoi{10.1086/524659}

\bibitem[{{Skillman} {et~al.}(2013){Skillman}, {Salzer}, {Berg}, {Pogge}, {Haurberg}, {Cannon}, {Aver}, {Olive}, {Giovanelli}, {Haynes}, {Adams}, {McQuinn}, \& {Rhode}}]{2013AJ....146....3S}
{Skillman}, E.~D., {Salzer}, J.~J., {Berg}, D.~A., {et~al.} 2013, \aj, 146, 3, \dodoi{10.1088/0004-6256/146/1/3}

\bibitem[{{Smith Castelli} {et~al.}(2024){Smith Castelli}, {Cortesi}, {Haack}, {Lopes}, {Thain{\'a}-Batista}, {Cid Fernandes}, {Lomel{\'\i}-N{\'u}{\~n}ez}, {Ribeiro}, {de Bom}, {Cernic}, {Sodr{\'e}}, {Zenocratti}, {De Rossi}, {Calder{\'o}n}, {Herpich}, {Telles}, {Saha}, {Lopes}, {Lopes-Silva}, {Gon{\c{c}}alves}, {Bambrila}, {Cardoso}, {Buzzo}, {Astudillo Sotomayor}, {Demarco}, {Leigh}, {Sarzi}, {Men{\'e}ndez-Delmestre}, {Faifer}, {Jim{\'e}nez-Teja}, {Grossi}, {Hern{\'a}ndez-Jim{\'e}nez}, {Krabbe}, {Guti{\'e}rrez Soto}, {Brand{\~a}o}, {Espinosa}, {Olave-Rojas}, {Oliveira Schwarz}, {Almeida-Fernandes}, {Schoenell}, {Ribeiro}, {Kanaan}, \& {Mendes de Oliveira}}]{2024MNRAS.530.3787S}
{Smith Castelli}, A.~V., {Cortesi}, A., {Haack}, R.~F., {et~al.} 2024, \mnras, 530, 3787, \dodoi{10.1093/mnras/stae840}

\bibitem[{{Telles} {et~al.}(2014){Telles}, {Thuan}, {Izotov}, \& {Carrasco}}]{2014A&A...561A..64T}
{Telles}, E., {Thuan}, T.~X., {Izotov}, Y.~I., \& {Carrasco}, E.~R. 2014, \aap, 561, A64, \dodoi{10.1051/0004-6361/201219270}

\bibitem[{{Thain{\'a}-Batista} {et~al.}(2023){Thain{\'a}-Batista}, {Cid Fernandes}, {Herpich}, {Mendes de Oliveira}, {Werle}, {Espinosa}, {Lopes}, {Smith Castelli}, {Sodr{\'e}}, {Telles}, {Kanaan}, {Ribeiro}, \& {Schoenell}}]{2023MNRAS.526.1874T}
{Thain{\'a}-Batista}, J., {Cid Fernandes}, R., {Herpich}, F.~R., {et~al.} 2023, \mnras, 526, 1874, \dodoi{10.1093/mnras/stad2698}

\bibitem[{{Thuan} \& {Martin}(1981)}]{1981ApJ...247..823T}
{Thuan}, T.~X., \& {Martin}, G.~E. 1981, \apj, 247, 823, \dodoi{10.1086/159094}

\bibitem[{{Tinsley}(1980)}]{1980FCPh....5..287T}
{Tinsley}, B.~M. 1980, \fcp, 5, 287, \dodoi{10.48550/arXiv.2203.02041}

\bibitem[{{Tremonti} {et~al.}(2004){Tremonti}, {Heckman}, {Kauffmann}, {Brinchmann}, {Charlot}, {White}, {Seibert}, {Peng}, {Schlegel}, {Uomoto}, {Fukugita}, \& {Brinkmann}}]{2004ApJ...613..898T}
{Tremonti}, C.~A., {Heckman}, T.~M., {Kauffmann}, G., {et~al.} 2004, \apj, 613, 898, \dodoi{10.1086/423264}

\bibitem[{{van Zee} \& {Haynes}(2006)}]{2006ApJ...636..214V}
{van Zee}, L., \& {Haynes}, M.~P. 2006, \apj, 636, 214, \dodoi{10.1086/498017}

\bibitem[{{Vazdekis} {et~al.}(2016){Vazdekis}, {Koleva}, {Ricciardelli}, {R{\"o}ck}, \& {Falc{\'o}n-Barroso}}]{2016MNRAS.463.3409V}
{Vazdekis}, A., {Koleva}, M., {Ricciardelli}, E., {R{\"o}ck}, B., \& {Falc{\'o}n-Barroso}, J. 2016, \mnras, 463, 3409, \dodoi{10.1093/mnras/stw2231}

\bibitem[{{Veilleux} \& {Osterbrock}(1987)}]{1987ApJS...63..295V}
{Veilleux}, S., \& {Osterbrock}, D.~E. 1987, \apjs, 63, 295, \dodoi{10.1086/191166}

\bibitem[{{Venhola} {et~al.}(2019){Venhola}, {Peletier}, {Laurikainen}, {Salo}, {Iodice}, {Mieske}, {Hilker}, {Wittmann}, {Paolillo}, {Cantiello}, {Janz}, {Spavone}, {D'Abrusco}, {van de Ven}, {Napolitano}, {Verdoes Kleijn}, {Capaccioli}, {Grado}, {Valentijn}, {Falc{\'o}n-Barroso}, \& {Limatola}}]{2019A&A...625A.143V}
{Venhola}, A., {Peletier}, R., {Laurikainen}, E., {et~al.} 2019, \aap, 625, A143, \dodoi{10.1051/0004-6361/201935231}

\bibitem[{{Vila-Costas} \& {Edmunds}(1993)}]{1993MNRAS.265..199V}
{Vila-Costas}, M.~B., \& {Edmunds}, M.~G. 1993, \mnras, 265, 199, \dodoi{10.1093/mnras/265.1.199}

\bibitem[{{Vilella-Rojo} {et~al.}(2015){Vilella-Rojo}, {Viironen}, {L{\'o}pez-Sanjuan}, {Cenarro}, {Varela}, {D{\'\i}az-Garc{\'\i}a}, {Crist{\'o}bal-Hornillos}, {Ederoclite}, {Mar{\'\i}n-Franch}, \& {Moles}}]{2015A&A...580A..47V}
{Vilella-Rojo}, G., {Viironen}, K., {L{\'o}pez-Sanjuan}, C., {et~al.} 2015, \aap, 580, A47, \dodoi{10.1051/0004-6361/201526374}

\bibitem[{{Vincenzo} {et~al.}(2016){Vincenzo}, {Belfiore}, {Maiolino}, {Matteucci}, \& {Ventura}}]{2016MNRAS.458.3466V}
{Vincenzo}, F., {Belfiore}, F., {Maiolino}, R., {Matteucci}, F., \& {Ventura}, P. 2016, \mnras, 458, 3466, \dodoi{10.1093/mnras/stw532}

\bibitem[{{Walsh} \& {Roy}(1989)}]{1989MNRAS.239..297W}
{Walsh}, J.~R., \& {Roy}, J.-R. 1989, \mnras, 239, 297, \dodoi{10.1093/mnras/239.2.297}

\bibitem[{{Watanabe} {et~al.}(2024){Watanabe}, {Ouchi}, {Nakajima}, {Isobe}, {Tominaga}, {Suzuki}, {Ishigaki}, {Nomoto}, {Takahashi}, {Harikane}, {Hatano}, {Kusakabe}, {Moriya}, {Nishigaki}, {Ono}, {Onodera}, \& {Sugahara}}]{2024ApJ...962...50W}
{Watanabe}, K., {Ouchi}, M., {Nakajima}, K., {et~al.} 2024, \apj, 962, 50, \dodoi{10.3847/1538-4357/ad13ff}

\bibitem[{{Werk} {et~al.}(2011){Werk}, {Putman}, {Meurer}, \& {Santiago-Figueroa}}]{2011ApJ...735...71W}
{Werk}, J.~K., {Putman}, M.~E., {Meurer}, G.~R., \& {Santiago-Figueroa}, N. 2011, \apj, 735, 71, \dodoi{10.1088/0004-637X/735/2/71}

\bibitem[{{Wesson} {et~al.}(2012){Wesson}, {Stock}, \& {Scicluna}}]{2012MNRAS.422.3516W}
{Wesson}, R., {Stock}, D.~J., \& {Scicluna}, P. 2012, \mnras, 422, 3516, \dodoi{10.1111/j.1365-2966.2012.20863.x}

\bibitem[{{Woosley} \& {Weaver}(1995)}]{1995ApJS..101..181W}
{Woosley}, S.~E., \& {Weaver}, T.~A. 1995, \apjs, 101, 181, \dodoi{10.1086/192237}

\bibitem[{{Yang} {et~al.}(2017){Yang}, {Malhotra}, {Rhoads}, \& {Wang}}]{2017ApJ...847...38Y}
{Yang}, H., {Malhotra}, S., {Rhoads}, J.~E., \& {Wang}, J. 2017, \apj, 847, 38, \dodoi{10.3847/1538-4357/aa8809}

\bibitem[{{Zenocratti} {et~al.}(2022){Zenocratti}, {De Rossi}, {Theuns}, \& {Lara-L{\'o}pez}}]{2022MNRAS.512.6164Z}
{Zenocratti}, L.~J., {De Rossi}, M.~E., {Theuns}, T., \& {Lara-L{\'o}pez}, M.~A. 2022, \mnras, 512, 6164, \dodoi{10.1093/mnras/stac906}

\bibitem[{{Zinchenko} {et~al.}(2024){Zinchenko}, {Sobolenko}, {V{\'\i}lchez}, \& {Kehrig}}]{2024A&A...690A..28Z}
{Zinchenko}, I.~A., {Sobolenko}, M., {V{\'\i}lchez}, J.~M., \& {Kehrig}, C. 2024, \aap, 690, A28, \dodoi{10.1051/0004-6361/202450232}

\bibitem[{{Zou} {et~al.}(2024){Zou}, {Sui}, {Saintonge}, {Scholte}, {Moustakas}, {Siudek}, {Dey}, {Juneau}, {Guo}, {Canning}, {Aguilar}, {Ahlen}, {Brooks}, {Claybaugh}, {Dawson}, {de la Macorra}, {Doel}, {Forero-Romero}, {Gontcho A Gontcho}, {Honscheid}, {Landriau}, {Le Guillou}, {Manera}, {Meisner}, {Miquel}, {Nie}, {Poppett}, {Rezaie}, {Rossi}, {Sanchez}, {Schubnell}, {Seo}, {Tarl{\'e}}, {Zhou}, \& {Zou}}]{2024ApJ...961..173Z}
{Zou}, H., {Sui}, J., {Saintonge}, A., {et~al.} 2024, \apj, 961, 173, \dodoi{10.3847/1538-4357/ad1409}

\end{thebibliography}
\bibliographystyle{aasjournal}

\end{document}